\documentclass[12pt]{article}
\makeatletter \@addtoreset{equation}{section} \makeatother
\renewcommand{\theequation}{\thesection.\arabic{equation}}
\newcommand{\ba}{\begin{array}}
\newcommand{\ea}{\end{array}}
\newcommand{\beq}{\begin{equation}}
\newcommand{\eeq}{\end{equation}}
\newcommand{\bea}{\begin{eqnarray}}
\newcommand{\eea}{\end{eqnarray}}

\def\bce{\begin{center}}
\def\ece{\end{center}}

\def\nonu{\nonumber}

\def\pa{\partial}

\def\be{\beta}

\def\ep{\epsilon}

\def\la{\lambda}

\def\eps6{{\displaystyle \mathop{\epsilon}^{6}}{}}
\def\g6{{\displaystyle \mathop{g}^{6}}{}}

\def\nab6{{\displaystyle \mathop{\nabla}^{6}}{}}

\def\0{{\sst{(0)}}}
\def\1{{\sst{(1)}}}
\def\2{{\sst{(2)}}}
\def\3{{\sst{(3)}}}
\def\4{{\sst{(4)}}}
\def\5{{\sst{(5)}}}
\def\6{{\sst{(6)}}}
\def\7{{\sst{(7)}}}
\def\8{{\sst{(8)}}}

\def\ba{\begin{array}}
\def\ea{\end{array}}
\def\beq{\begin{equation}}
\def\eeq{\end{equation}}
\def\be{\begin{equation}}
\def\ee{\end{equation}}

\def\la{\lambda}
\def\eps{\epsilon}

\def\ba{\begin{array}}
\def\ea{\end{array}}
\def\beq{\begin{equation}}
\def\eeq{\end{equation}}
\def\be{\begin{equation}}
\def\ee{\end{equation}}

\def\la{\lambda}
\def\eps{\epsilon}

\def\eps6{{\displaystyle \mathop{\epsilon}^{6}}{}}

\def\nab6{{\displaystyle \mathop{\nabla}^{6}}{}}

\newcommand{\bean}{\begin{eqnarray*}}
\newcommand{\eean}{\end{eqnarray*}}

\begin{document}
\thispagestyle{empty} \addtocounter{page}{-1}
   \begin{flushright}
\end{flushright}

\vspace*{1.3cm}
  
\centerline{ \Large \bf   
Higher Spin Currents in Orthogonal Wolf Space } 
\vspace*{1.5cm}
\centerline{{\bf Changhyun Ahn } and {\bf Jinsub Paeng}
} 
\vspace*{1.0cm} 
\centerline{\it 
Department of Physics, Kyungpook National University, Taegu
702-701, Korea} 
\vspace*{0.8cm} 
\centerline{\tt ahn@knu.ac.kr, \qquad jdp2r@knu.ac.kr 
} 
\vskip2cm

\centerline{\bf Abstract}
\vspace*{0.5cm}

For the ${\cal N}=4$ superconformal coset theory by 
$\frac{SO(N+4)}{SO(N) \times SU(2)} \times U(1)$ 
(that contains an orthogonal Wolf space) with $N=4$,  
the ${\cal N}=2$ WZW affine 
current algebra is obtained. The $16$ generators 
(or $11$ generators) of 
the large ${\cal N}=4$ linear (or nonlinear) 
superconformal algebra are described by 
these WZW affine currents explicitly. 
Along the line of large ${\cal N}=4$ holography,
the extra $16$ currents with spins 
$(2,\frac{5}{2},\frac{5}{2}, 3)$, $(\frac{5}{2}, 3, 3, 
\frac{7}{2})$, $(\frac{5}{2}, 3, 3, \frac{7}{2})$, and $(3, \frac{7}{2}, 
\frac{7}{2}, 4)$ 
are obtained in terms of  the WZW affine currents.
The lowest spin of this ${\cal N}=4$ multiplet is 
two rather than one which is for an unitary Wolf space.
The operator product expansions (OPEs) between the above $11$ currents and 
these extra $16$ higher spin currents are found explicitly. 

\baselineskip=18pt
\newpage
\renewcommand{\theequation}
{\arabic{section}\mbox{.}\arabic{equation}}

\section{Introduction}

It is known in \cite{GG1305} that 
the large ${\cal N}=4$ higher spin theory on $AdS_3$ 
based on the higher spin algebra
is dual to
the 't Hooft limit of the two dimensional 
large ${\cal N}=4$ superconformal coset theory.
This is a supersymmetric version of the original 
nonsupersymmetric duality described in \cite{GG1011,GG1205}.
The ${\cal N}=4$ higher spin algebra contains   
eight fields of spin $s =\frac{3}{2}, 2, \frac{5}{2}, \cdots$ and 
seven fields of spin $s=1$. 
The lowest spin for the fields is given by one. There is no field of 
spin $\frac{1}{2}$. 
The exceptional superalgebra 
$D(2,1|\frac{\mu}{1-\mu})$
is the largest finite dimensional subalgebra of this ${\cal N}=4$ 
higher spin algebra.
On the other hand,
the ${\cal N}=4$ coset theory \cite{ST} is described by the Wolf 
\cite{Wolf,Alek,Salamon} space `bosonic' coset 
$\frac{G}{H}=\frac{SU(N+M)_{k} \times SO(2N M)_1}{SU(N)_{k+M} 
 \times U(1)_{N M(N+M)(N+M+k)}}$ 
with $M=2$ \footnote{The `supersymmetric' version of this coset can be written
as $\frac{SU(N+M)_{k+N+M}}{SU(N)_{k+N+M} \times SU(M)_{k+N+M} 
 \times U(1)}$ where the central charge is given by
$c=\frac{3 k M N}{(k+M+N)}$. For $M=1$,
the minimal model holography was described in \cite{CHR1111}.
Moreover, the duality property has been studied in 
\cite{CG1203,HP1203,Ahn1206,CG1207,Ahn1208,Hikida1212} 
further. For general $M$, further analysis has been done in 
\cite{CV1312,CPV1408}.
\label{unitaryfoot}}. 
The first 
level $k$ is the one for $SU(2)_{+}$ affine algebra  while 
the second level for other
$SU(2)_{-}$ affine algebra is given by $N$ of
the large ${\cal N}=4$ nonlinear superconformal algebra.
The 't Hooft coupling constant $\la = \frac{N+1}{N+k+2}$ 
is identified with 
the free parameter of the large ${\cal N}=4$ nonlinear 
superconformal algebra.  
In other words, the numerator of $\la$ is given by 
the second level $N$ plus one 
and the denominator of $\la$ is given by 
the sum of first and second levels $(k+N)$ plus two. 
Furthermore, this 't Hooft parameter 
$\la$ in two dimensional CFT 
is equal to the above $\mu$ parameter, in the subalgebra 
of the ${\cal N}=4$ higher spin algebra,
which appears
in  the mass of scalar field in the $AdS_3$ bulk theory. 

For the $N=3$ (and $M=2$),  
the ${\cal N}=2$ WZW affine 
current algebra \cite{HS,RASS,AIS} was obtained in \cite{Ahn1311}. 
The $16$ generators (or $11$ 
generators)   of 
the large ${\cal N}=4$ linear 
\cite{STVplb,npb1988,Schoutensnpb,Ivanov1,Ivanov2,Ivanov3,Ivanov4} 
(nonlinear 
\cite{GS,cqg1989,npb1989,GK}) 
superconformal algebra were obtained.    
Based on the large ${\cal N}=4$ holography \cite{GG1305},
the extra $16$ higher spin currents, with spin contents 
$
(1,\frac{3}{2},  \frac{3}{2}, 2)$,  
$(\frac{3}{2}, 2, 2, \frac{5}{2})$, 
$(\frac{3}{2}, 2, 2, \frac{5}{2})$,   
$(2, \frac{5}{2}, 
\frac{5}{2}, 3)$
 described in terms of 
four ${\cal N}=2$ multiplets, 
were found and realized by the above ${\cal N}=2$ WZW affine currents. 
Then the
 operator product expansions (OPEs) between the above $11$ currents and 
these extra $16$ higher spin currents
were described explicitly. 
By calculating the OPEs 
between the above $16$ higher spin currents, 
the next $16$ 
higher spin currents of spins
$(2, \frac{5}{2}, \frac{5}{2}, 3)$, 
$(\frac{5}{2}, 3, 3, \frac{7}{2} )$, $(\frac{5}{2}, 3, 3, \frac{7}{2})$ 
and $(3, \frac{7}{2}, \frac{7}{2}, 4)$ were determined
from the right hand sides of these OPEs \cite{Ahn1408}.  
Moreover,
the composite fields consisting of both the $11$ currents in the large 
${\cal N}=4$ nonlinear superconformal algebra and the above 
$16$ lowest higher spin currents also occurred in 
the right hand sides of these
OPEs. The latter appeared quadratically (and linearly) in the fusion rules 
together with large ${\cal N}=4$
nonlinear superconformal family of the identity operator.  

The asymptotic 
symmetry of the $AdS_3$ higher spin theory \cite{GH1101} 
based on super higher spin algebra
$shs_{M=2}[\lambda]$ has been studied further and matches with 
those of the two-dimensional CFT Wolf space coset in the 't Hooft 
limit \cite{GP1403}. The extension of large ${\cal N}=4$ superconformal 
algebra which contains one ${\cal N}=4$ multiplet for each integer 
superspin $s=1,
2, \cdots$ as well as the currents of large ${\cal N}=4$ superconformal 
algebra has been studied in \cite{BCG1404}. 
For the particular level at 
the Kazama-Suzuki model \cite{KS1,KS2}, 
the ${\cal N}=3$ (enhanced) supersymmetry 
is observed in \cite{CHR1406}.  
This behavior also appeared in many previous examples of 
\cite{Ahn1211,Ahn1305,BCGG1305}.
Furthermore, the full spectrum of the tensionless 
string theory in $AdS_3 \times {\bf S}^3 \times {\bf T}^4$ can be 
reorganized in terms of representations of the ${\cal N}=4$ super 
${\cal W}_{\infty}$ algebra where the large level limit is taken 
\cite{GG1406}. 

In \cite{CHR1306}, the possible `bosonic' coset theory with 
$O(M)$ Chan-Paton factor is described by
\bea
\frac{SO(N+M)_k \times SO(NM)_1}{SO(N)_{k+M} \times SO(M)_{k+N}}.
\label{sogeneralcoset}
\eea
The central charge, by inserting the dimension of group, the level and the 
dual Coxeter number of the group into the central charge formula
of each factor group,  
is given by \footnote{One has 
$\frac{1}{2} (N+M)(N+M-1) \left[\frac{k}{k+(N+M-2)} \right] 
+\frac{1}{2}
N M (N M-1) \left[\frac{1}{1+(N M-2)} \right]
-\frac{1}{2} N (N-1) \left[\frac{(k+M)}{(k+M)+(N-2)} \right]
-\frac{1}{2} M (M-1) \left[\frac{(k+N)}{(k+N)+(M-2)} \right]=
 \frac{3k M N}{ 2(M+N+k-2)}$.
Or splitting the $k$ independent part of this expression 
as $\frac{1}{4} \left[ (N+M)(N+M-1)- N(N-1)-M(M-1)\right]$ and distributing 
it into other $k$ dependent parts,
one can also reexpress  the above 
central charge as 
$\frac{3}{2} \frac{1}{2} (N+M)(N+M-1) \left[ 1-\frac{2(N+M-2)}{3(k+M+N-2)} 
\right] -\frac{3}{2}  \frac{1}{2} N (N-1) \left[ 1- \frac{2(N-2)}{3(k+N+M-2)}
\right] -\frac{3}{2} \frac{1}{2} 
M(M-1) \left[ 1-\frac{2(M-2)}{3(k+N+M-2)} \right]$ which is the central 
charge for the `supersymmetric' 
coset theory in (\ref{supersogeneralcoset}).
}
\bea
c = \frac{3k M N}{ 2(M+N+k-2)}.
\label{generalcentral}
\eea
In the `stringy' limit where $M, N$ and $k$ go to infinity simultaneously,
this central charge is proportional to $N^2$ as appropriate for a stringy
model \cite{GG1207}. This central charge looks similar to 
the one in the unitary Wolf space coset in the footnote \ref{unitaryfoot}. 
The `supersymmetric' coset theory in the supersymmetric 
version of (\ref{sogeneralcoset}) is
\bea
\frac{SO(N+M)_{k+N+M-2} }{SO(N)_{k+N+M-2} \times SO(M)_{k+N+M-2}}.
\label{supersogeneralcoset}
\eea

Let us describe each case for this supersymmetric coset model. 

For $M=1$, the above coset (\ref{sogeneralcoset}) or 
(\ref{supersogeneralcoset}), by multiplying the appropriate
$SO(1)_{k+N-1}$ in the numerator of the coset 
(\ref{sogeneralcoset}), 
reproduces the one studied in \cite{CHR1209,AP1310} where the supersymmetry 
is equal to ${\cal N}=1$.

For $M=2$,  the above coset (\ref{sogeneralcoset}) or 
(\ref{supersogeneralcoset}) has been described in ${\cal N}=2$ 
Kazama-Suzuki model 
\cite{KS1,KS2} \footnote{It would be interesting to obtain the 
higher spin currents as done in \cite{Ahn1994,Ahn1206,Ahn1208}.}.

For $M=4$,
 the above coset (\ref{sogeneralcoset}) or 
(\ref{supersogeneralcoset}) reproduces
the orthogonal Wolf space \footnote{This orthogonal Wolf space 
appeared in \cite{ARV} previously.} by realizing that 
the $SO(4)$ group in the denominator can be interpreted as 
a product of two $SU(2)$ groups.   
In this case, the Wolf space central charge, by substituting $M=4$ into 
(\ref{generalcentral}), 
is given by
\bea
c_W = \frac{6kN}{(N+k+2)}.
\label{Wolfcosetcentral}
\eea
We will see that this central charge appears in the highest singular term
in the OPE between 
the two different fermionic spin-$\frac{3}{2}$ currents of
large ${\cal N}=4$ nonlinear superconformal algebra \footnote{It is natural to
ask what happens for the $M=3$ case. It would be interesting to 
find the corresponding CFT explicitly. We thank S.-J. Rey
for raising this issue.}. 

In this paper,
we would like to construct 
the $16$ generators of the large ${\cal N}=4$ linear 
superconformal  algebra
in the coset 
$
\mbox{Wolf} \times SU(2) \times U(1) = 
\frac{SO(N+4)}{SO(N) \times SU(2)} \times U(1)
$ 
(that is the fourth entry of the table $1$ in \cite{ST}) 
theory with $N=4$.
See also the papers \cite{BBSS1,BBSS2,BS} on the 
generalization of GKO coset construction.
By factoring out the spin-$1$ current and four spin-$\frac{1}{2}$ currents
\cite{GS},
the $11$ generators of the large ${\cal N}=4$ nonlinear algebra 
in the Wolf space coset 
are obtained \footnote{
As in the unitary Wolf space, one can combine the 
spin-$1$ current and one spin-$\frac{1}{2}$ current as
a single ${\cal N}=1$ super current. 
The superpartners of the remaining 
three spin-$\frac{1}{2}$ currents (i.e. three 
spin-$1$ currents) do not play an important role in 
the denominator subgroup. When one divides $SU(2) \times U(1)$ factor, 
then one obtains the `supersymmetric' Wolf space coset 
$\mbox{Wolf} = \frac{SO(N+4)}{SO(N) \times SU(2) \times SU(2)}$ where the 
second $SU(2)$ factor in the denominator refers only to  the above three 
spin-$\frac{1}{2}$ currents. Their superpartners (three spin-$1$ 
currents)  appear in the group $SO(N+4)$ and contribute to the $SU(2)_k$
affine algebra. 
The first $SU(2)$ contains the above ${\cal N}=1$ super current. 
}.
What are the spin contents for the lowest higher spin currents?
From the lessons in \cite{Ahn1106,GV1106,Ahn1202,CGKV1211,AP1301},
one should have the spin-$4$ current.
After putting this spin-$4$ current into the last place 
of the ${\cal N}=4$ multiplet and then 
the other spin contents can be filled 
automatically.  
Let us denote them by spin contents as follows:
\bea
&& \left(2, \frac{5}{2}, \frac{5}{2}, 3 \right): (T^{(2)}, 
T_{+}^{(\frac{5}{2})}, T_{-}^{(\frac{5}{2})}, T^{(3)}), \nonu \\
&&  \left(\frac{5}{2}, 3, 3, \frac{7}{2} \right):
 (U^{(\frac{5}{2})}, 
U_{+}^{(3)}, U_{-}^{(3)}, U^{(\frac{7}{2})}), \nonu \\
&& \left(\frac{5}{2}, 3, 3, \frac{7}{2} \right):
 (V^{(\frac{5}{2})}, 
V_{+}^{(3)}, V_{-}^{(3)}, V^{(\frac{7}{2})}),  \nonu \\
&& \left(3, \frac{7}{2}, \frac{7}{2}, 4 \right):
 (W^{(3)}, 
W_{+}^{(\frac{7}{2})}, W_{-}^{(\frac{7}{2})}, W^{(4)}).
\label{16}
\eea
We would like to
construct these $16$ currents (\ref{16}) in terms of ${\cal N}=2$ affine 
Kac-Moody currents in the above orthogonal Wolf space coset theory 
(\ref{supersogeneralcoset}) with $N=4$ and $M=4$ explicitly
by following the previous works in \cite{Ahn1311,Ahn1408}.
Furthermore, we calculate the various OPEs between the $11$ generators 
of the large ${\cal N}=4$ nonlinear superconformal algebra and the 
$16$ higher spin currents \footnote{The higher spin theory 
with extended supersymmetries where the two dimensional 
coset theory contains Wolf space was studied also in 
\cite{HLPR}.}.

In section $2$, the $16$ currents of ${\cal N}=4$ linear 
superconformal algebra are obtained in the `supersymmetric' coset $
\mbox{Wolf} \times SU(2) \times U(1) = 
\frac{SO(N+4)}{SO(N) \times SU(2)} \times U(1)
$ with $N=4$.
After that,  the $11$ currents of
 ${\cal N}=4$ nonlinear 
superconformal algebra are determined in the orthogonal Wolf space coset 
(\ref{sogeneralcoset}) or (\ref{supersogeneralcoset}) with $N=M=4$.

In section $3$, the extra $16$ currents in (\ref{16}) are obtained in the
orthogonal Wolf space coset (\ref{sogeneralcoset}) or 
(\ref{supersogeneralcoset}).
The OPEs between the $11$ currents in section $2$
and $16$ currents in section $3$ are obtained.

In section $4$, we summarize the main results of  this paper
together with future directions.

In Appendices $A$-$D$, some details which are necessary in sections 
$2$-$3$ are presented.

The mathematica package by Thielemans \cite{Thielemans} 
is used. 

\section{The large ${\cal N}=4$ linear and nonlinear
superconformal  algebras in the coset
minimal model    }

In this section, we would like to construct the 
large ${\cal N}=4$ nonlinear superconformal algebra realization.
Instead of making it directly from the ${\cal N}=2$ WZW affine currents,
we construct the large ${\cal N}=4$ linear superconformal algebra realization
and then its nonlinear version can be obtained by factoring out the four
fermion spin-$\frac{1}{2}$ currents and one spin-$1$ current.
  
\subsection{${\cal N}=2$ WZW affine current algebra}
Let us describe
the particular ${\cal N}=4$ superconformal coset theory described in
\cite{ST}
which can be written as
\bea
\mbox{Wolf} \times SU(2) \times U(1) = \frac{SO(N+4)}{SO(N) \times SU(2)} \times U(1),
\,\,\,
\mbox{where} \,\,\,
\mbox{Wolf}= \frac{SO(N+4)}{SO(N) \times [SU(2)]^2}.
\label{coset}
\eea
More explicitly, one has (\ref{supersogeneralcoset}) with $M=4$
and the levels $(k+N+M-2)$ at each factor group. 
The $SO(4)$ group is written in terms of two $SU(2)$ groups in the 
denominator of the orthogonal Wolf space coset.

The central charge of this coset model \cite{KS1} is given by
\bea
&&c_{SO(N+4)}-c_{SO(N)}-c_{SU(2)}+c_{U(1)}=\frac{6(k+1)(N+1)}{(k+N+2)},
\label{cosetcentral}
\eea
where each contribution can be obtained from the following expressions
\bea
c_{SO(N+4)} & = &
\frac{3}{2} \left[ \frac{1}{2}(N+4)(N+3)\right] \left[1-\frac{2(N+2)}{3(k+N+2)} \right],
\nonu \\
c_{SO(N)} & = & 
\frac{3}{2} \left[ \frac{1}{2}N(N-1)\right] \left[1-\frac{2(N-2)}{3(k+N+2)} \right],
\nonu \\
c_{SU(2)} & = & 
\frac{3}{2} \left[ 2^2-1\right] \left[1-\frac{2 \cdot 2}{3(k+N+2)} \right],\,\,\,\,\,c_{U(1)}= \frac{3}{2}.
\label{inter}
\eea
This central charge (\ref{cosetcentral}) 
is the same as the unitary case in \cite{ST,GG1305,Ahn1311,Ahn1408}
and will appear in the OPE between the spin-$2$ stress energy tensor
of large ${\cal N}=4$ linear superconformal algebra 
as usual.  
The two levels of two $SU(2)$ affine algebras of large 
${\cal N}=4$ linear superconformal algebra  are given by
the two positive integers $(k, N)$ of the coset theory 
as follows $k^{+} = k+1$ and $k^{-} =N+1=5$.
One can also express the central charge in terms of 
$(k^{+}, k^{-})$ as $c=\frac{6 k^{+} k^{-}}{(k^{+} +k^{-})}$. 
In the present case, because one chooses $N$ as $4$, 
the common factor $(k+N+2)=(k+6)$ will be present in all the important 
expressions later. For the previous case in \cite{Ahn1311,Ahn1408}, 
one could see the $(k+N+2)=(k+5)$ dependence because we took $N$ as $3$.

Furthermore, the central charge (\ref{Wolfcosetcentral}) 
of the orthogonal Wolf space 
(\ref{coset})
can be obtained as follows:
\bea
c_{\mbox{Wolf}}  & = &
c_{SO(N+4)} -c_{SO(N)} -2\,c_{SU(2)} =  \frac{6 k N}{(2+k+N)},
\label{Wolfcentral}
\eea
where the previous relations (\ref{inter}) are used. 
This can be written in terms of $c_{\mbox{Wolf}}   = \frac{6(k^{+}-1)(k^{-}-1)}
{(k^{+} +k^{-})}$.
We will see this central charge in the OPE between the spin-$\frac{3}{2}$
currents of large ${\cal N}=4$ nonlinear superconformal algebra.
Note that the central charge, in the usual OPE between the 
stress energy tensor in the large ${\cal N}=4$ nonlinear superconformal 
algebra, will take a different form. 

The $28$ basic ${\cal N}=2$
WZW affine currents of spin $\frac{1}{2}$ 
living in the orthogonal $SO(N+4=8)$ group 
are classified as follows:
\bea
&&\mbox{subgroup} \, SO(4) \times SU(2)  \times SU(2) 
\, \mbox{currents} :\,\,\, K^m =(K^1, \, K^2, \, K^3; \, K^4, \, K^5,\, K^6),
\nonu \\
&&\mbox{subgroup} \,  SO(4) \times SU(2) \times SU(2) \, 
\mbox{currents} :\,\,\, K^{\bar{m}} =(K^{\bar{1}}, \, K^{\bar{2}}, \, K^{\bar{3}}; \, 
K^{\bar{4}},\, K^{\bar{5}}, \, K^{\bar{6}}),
\nonu \\
&&\mbox{Wolf coset} \, \frac{SO(8)}{SO(4) \times SU(2)\times SU(2)} \, 
\mbox{ currents} :\,\,\, J^a =(J^7, \, J^8, \,\cdots, \, J^{14}),
\nonu \\
&&\mbox{Wolf coset} \, \frac{SO(8)}{SO(4) \times SU(2)\times SU(2)} \, 
\mbox{ currents} :\,\,\, J^{\bar{a}}=
( J^{\bar{7}}, \, J^{\bar{8}}, \, \cdots, \,  J^{\bar{14}}).
\label{QQ}
\eea
For $N=4$, the $28$ adjoint indices of $SO(8)$ can be divided in terms 
of $14$ unbarred indices and $14$ barred indices in the complex basis
\footnote{The adjoint representation ${\bf 28}$ of $SO(8)$ breaks into 
${\bf 28} \rightarrow ({\bf 1,3, 1,1}) \oplus 
({\bf 3,1,1,1}) \oplus 
({\bf 1,1,3,1}) \oplus ({\bf 1,1,1,3}) \oplus ({\bf 2,2,2,2})$ under 
the $SU(2) \times SU(2) \times SU(2) \times SU(2)$ \cite{Patera} 
where the $SO(4)$ is replaced with 
the first two $SU(2)$ factors. The first two representations are given by the 
currents having the indices $1,2,3$ (and their complex conjugated ones).
The next two representations  are given by the 
currents having the indices $4,5,6$ (and their complex conjugated ones). 
The remaining representations are given by the 
currents having the indices $7,8, \cdots, 14$ 
(and their complex conjugated ones).
}. 
The nonvanishing structure constants and the generators are described in 
Appendix $A$.
The $K^m(z)$ and $K^{\bar{m}}(z)$ of (\ref{QQ}) live in
the subgroup $SO(4) \times SU(2) \times SU(2)$ of $SO(8)$.
More precisely, 
the $6$ adjoint indices of  the subgroup $SO(4)$ are given by
$1,2,3$ (and their barred indices, $\bar{1}, \bar{2}, \bar{3}$).
The indices of the first $SU(2)$ are given by $4$, 
and $\bar{4}$, and some combination between $5$ and $\bar{5}$
while 
those of the second $SU(2)$ are given by $6$, 
and $\bar{6}$, and other combination between $5$ and $\bar{5}$.
On the other
hand, the currents $J^a(z)$ and $J^{\bar{a}}(z)$ live in the 
Wolf 
coset $\frac{SO(8)}{SO(4) \times SU(2)\times SU(2)}$ for which the remaining 
$8$ unbarred indices and $8$ barred indices are present.
One has similar spin-$1$ currents, denoted by
$D K^{\bar{m}}|_{\theta=\bar{\theta}=0}(z)$, 
$\overline{D} K^m|_{\theta=\bar{\theta}=0}(z)$, $D J^{\bar{a}}|_{\theta=\bar{\theta}=0}(z)$
and $\overline{D} J^a|_{\theta=\bar{\theta}=0}(z)$,
  corresponding to the superpartners of the above
$28$ spin-$\frac{1}{2}$ currents
\footnote{
It is also useful to write down the spin-$1$ affine current 
in terms of purely bosonic spin-$1$ current and spin-$\frac{1}{2}$ current. 
One has the following expressions \cite{HS}
\bea
\overline{D} K^m|_{\theta=\bar{\theta}=0}(z) & = & 
V^m(z) -\frac{1}{(k+N+2)} \left( f_{p \bar{q}}^{\,\,\,\, m} K^p K^{\bar{q}}
+ f_{a \bar{b}}^{\,\,\,\, m}  J^a J^{\bar{b}} \right)(z),
\nonu \\
D K^{\bar{m}}|_{\theta=\bar{\theta}=0}(z) & = & 
V^{\bar{m}}(z) -\frac{1}{(k+N+2)} \left( f_{\bar{p} q}^{\,\,\,\, \bar{m}} K^{\bar{p}} K^{q}
+ f_{m \bar{a}}^{\,\,\,\, \bar{b}}  J^{\bar{a}} J^{b} \right)(z),
\nonu \\
\overline{D} J^a|_{\theta=\bar{\theta}=0}(z) & = & 
V^a(z) -\frac{1}{(k+N+2)}  \left( f_{b \bar{m}}^{\,\,\,\, a} J^b K^{\bar{m}}
+f_{ \bar{a} m}^{\,\,\,\, b} K^m J^{\bar{b}}\right)(z),
\nonu \\
D J^{\bar{a}}|_{\theta=\bar{\theta}=0}(z) & = & 
V^{\bar{a}}(z) -\frac{1}{(k+N+2)}  \left(
f_{\bar{b} m}^{\,\,\,\, \bar{a}} J^{\bar{b}} K^{m} +
f_{a \bar{m}}^{\,\,\,\, \bar{b}} K^{\bar{m}} J^{b} \right)(z).
\label{bosonicexp}
\eea
The OPEs between the purely bosonic spin-$1$ currents,  which commute with 
the above spin-$\frac{1}{2}$ currents, are given in 
\cite{Ahn1408}:
$V^A(z) \, V^{B}(w)  =  
 -\frac{1}{(z-w)} \, f_{\bar{A} \bar{B}}^{\,\,\,\,\,\,\bar{C}} V^C(w)
+\cdots$,
$
V^{\bar{A}}(z) \, V^{\bar{B}}(w)  =  
 -\frac{1}{(z-w)} \, f_{A B}^{\,\,\,\,\,\,C} V^{\bar{C}}(w)
+\cdots$
and $
V^A(z) \, V^{\bar{B}}(w)  =  \frac{1}{(z-w)^2} \, k \, \delta^{A\bar{B}} 
 -\frac{1}{(z-w)} \, \left( 
f_{\bar{A} B}^{\,\,\,\,\,\,\bar{C}} V^C
+  f_{\bar{A} B}^{\,\,\,\,\,\,C} V^{\bar{C}}
\right)(w)
+\cdots$,
where the indices $A, B, \cdots$ and $\bar{A}, \bar{B}, \cdots $
stand for the indices of the group $SO(N+4=8)$ in the complex basis.
 Again the two extra terms of last two equations 
in (\ref{bosonicexp}) appear due to the nonvanishing structure constants
 $f_{ab}^{\,\,\, m}$ and $f_{\bar{a} 
\bar{b}}^{\,\,\,\bar{m}}$ which vanish in the unitary case \cite{Ahn1311}.
 }.

The complete OPEs for the ${\cal N}=2$ WZW affine currents 
are summarized in Appendix $B$. 

\subsection{Large ${\cal N}=4$ linear superconformal algebra in terms of 
${\cal N}=2$ WZW affine currents}
We will construct the large ${\cal N}=4$ linear
superconformal algebra realization for the coset theory (\ref{coset})
using the description of previous presentation. 

\subsubsection{Construction of two spin-$\frac{3}{2}$ currents}
Let us consider the two spin-$\frac{3}{2}$ currents
of large ${\cal N}=4$ linear superconformal algebra,
which appear in the last components of
${\cal N}=2$ multiplets
$(\frac{1}{2}, 1,1, \frac{3}{2})$ and $(\frac{1}{2}, 1,1, \frac{3}{2})$
\cite{Ahn1311}.
Let us 
denote the first one as $G_{11}(z)$ and the second one as
$G_{22}(z)$.  
As in the unitary case \cite{Ahn1311}, one should consider 
the spin-$\frac{1}{2}$ current with the Wolf space subgroup index.
Among the two $SU(2)$ subgroups, 
the second $SU(2)$ group corresponds to the above spin-$\frac{3}{2}$ 
currents. It turns out that the spin-$\frac{1}{2}$ current with
the index $6$ provides the correct 
spin-$\frac{3}{2}$ current 
\bea
G_{11}(z) =
\sqrt{2} i \left(-\frac{1}{2} \left[ D, \overline{D} \right] K^6-
\frac{(k-4)}{2 (6+k)} \pa K^6 \right)(z).
\label{g11}
\eea
Note that $K^6(w)$ and $K^{\bar{6}}(w)$ are primary under the 
$T_{SO(8)}(z)$ and $(T_{SO(8)}-T_{SO(4)})(z)$.
The coefficient $\frac{(k-4)}{(6+k)}$ can be written as
$\frac{(k^{+}-k^{-})}{(k^{+} + k^{-})}$ with 
\bea
k^{+} = k+1, \qquad  \mbox{and} \qquad  k^{-}=N+1=5.
\nonu
\eea
The second term in (\ref{g11}) is necessary for the primary field condition
on $G_{11}(w)$  
under the $T_{SO(8)}(z)$. 

The second spin-$\frac{3}{2}$ current can be obtained as follows:
\bea
G_{22}(z) =
\sqrt{2} i \left(-\frac{1}{2} \left[ D, \overline{D} \right] K^{\bar{6}}+
\frac{(k-4)}{2 (6+k)} \pa K^{\bar{6}} \right)(z).
\label{g22}
\eea
One can find some informations on 
other currents (or Wolf space central charge $c_{\mbox{Wolf}}$) 
by calculating the OPE between 
(\ref{g11}) and (\ref{g22}) and reading off the right hand side.

\subsubsection{Construction of three spin-$1$ currents}
Now one continues to obtain the other components in the previous 
two ${\cal N}=2$ multiplets of large ${\cal N}=4$ linear superconformal 
algebra. For example, one can construct the following spin-$1$ currents
associated with the indices $6$ and $\bar{6}$ as follows:
\bea
A_1(z) & = & \frac{1}{2} \left( -\overline{D} K^6 + D K^{\bar{6}} \right)(z),
\nonu \\
A_2(z) & = & \frac{i}{2} \left( \overline{D} K^6 + D K^{\bar{6}} \right)(z).
\label{a1a2}
\eea
Note that there are no singular terms in the OPEs 
of $\overline{D} K^6(z) \, \overline{D} K^6(w)$
and  $D K^{\bar{6}}(z) \, D K^{\bar{6}}(w)$.
It is easy to 
determine the overall constants in (\ref{a1a2})
because they are the affine currents of $SU(2)_{k^{+}(=k+1)}$ algebra.
Then one can fix the remaining current by calculating the OPE 
between $A_1(z)$ and $A_2(w)$ and reading off the first-order pole
\bea
A_3(w) & = &\frac{1-i}{2\sqrt{2}}(\overline{D} K^5-i\,D K^{\bar{5}})(w)
\nonu \\
&-&  \frac{i}{2 (6+k)} \left(\sum_{(m,\bar{m})=(5,\bar{5})}^{(6,\bar{6})} 
K^m K^{\bar{m}}+
\sum_{(a,\bar{a})=(7,\bar{7})}^{(14,\bar{14})} J^a J^{\bar{a}}\right)(w).
\label{a3}
\eea

Then it turns out that 
the OPEs between these spin-$1$ currents are the same as 
the ones in $(2.17)$ of \cite{Ahn1311}.

\subsubsection{Construction of four spin-$\frac{1}{2}$ currents and
other two spin-$\frac{3}{2}$ currents}
Let us determine the four spin-$\frac{1}{2}$ currents of 
large ${\cal N}=4$ linear superconformal algebra.
Let us focus on the OPE between $A_1(z)$ and $G_{11}(w)$
from (\ref{a1a2}) and (\ref{g11})
\footnote{
That is,
$
A_1(z) \, G_{11}(w) 
  =  \frac{1}{(z-w)^2} \, \frac{i}{2} \, 2(1-\gamma) \,
F_{21}(w) - \frac{1}{(z-w)}\, \frac{i}{2} \, G_{21}(w) +\cdots$.}.
This OPE is nontrivial and contains the second-order pole and 
first-order pole. 
The deformation parameter of large ${\cal N}=4$ linear superconformal algebra
is defined as 
\bea
\gamma = \frac{k^{-}}{(k^{+} + k^{-})} = \frac{5}{(6+k)},
\label{gamma}
\eea
and the second-order pole provides the following spin-$\frac{1}{2}$ current
as follows:
\bea
F_{21}(w) & = & \frac{(1+i) }{2} \,K^{5}(w).
\label{f21}
\eea
Moreover, the first-order pole  
determines
the following spin-$\frac{3}{2}$ current
\bea
G_{21}(w) & = &
\frac{\sqrt{2}}{(6+k)} \, \left( 
\sum_{(m,\bar{m})=(5,\bar{5})}^{(6,\bar{6})}K^m D K^{\bar{m}}
+\sum_{(a,\bar{a})=(7,\bar{7})}^{(14,\bar{14})}J^a D J^{\bar{a}}\right) (w)
+ 
\mbox{cubic terms}.
\label{g21}
\eea

Let us describe the OPE between $A_1(z)$ and $G_{22}(w)$ from (\ref{a1a2}) 
and (\ref{g22}) \footnote{
In other words,
$
A_1(z) \, G_{22}(w) 
  =  -\frac{1}{(z-w)^2} \, \frac{i}{2} \, 2(1-\gamma) \,
F_{12}(w) + \frac{1}{(z-w)}\, \frac{i}{2} \, G_{12}(w) +\cdots$,
with (\ref{gamma}).}. In this case, 
this nontrivial OPE contains the second-order pole and 
first-order pole. 
The former determines the spin-$\frac{1}{2}$ current
\bea
F_{12}(w) & = & \frac{(1-i) }{2} \, K^{\bar{5}}(w),
\label{f12}
\eea
and the latter fixes the spin-$\frac{3}{2}$ current
\bea
G_{12}(w) & = & \frac{5(1-i)}{(6+k)} \pa K^{\bar{5}}(w)
-\frac{\sqrt{2}}{6+k} \, \left( \sum_{(m,\bar{m})=(5,\bar{5})}^{(6,\bar{6})}
\overline{D} K^m K^{\bar{m}}+ \sum_{(a,\bar{a})=(7,\bar{7})}^{(14,\bar{14})}
\overline{D} J^a J^{\bar{a}}\right)(w)
\nonu \\
& + &
\mbox{cubic terms}.
\label{g12}
\eea

Let us determine the remaining spin-$\frac{1}{2}$ currents.
From the second order pole of the OPE between $A_3(z)$ and $G_{11}(w)$
\footnote{
One has 
$
A_3(z) \, G_{11}(w) 
  =  \frac{1}{(z-w)^2} \, \frac{i}{2} \, 2(1-\gamma) \,
F_{11}(w) - \frac{1}{(z-w)}\, \frac{i}{2} \, G_{11}(w) +\cdots$.}, 
one finds  the following spin-$\frac{1}{2}$ current
\bea
F_{11}(w) & = & \frac{i }{\sqrt{2}} \, K^{6}(w).
\label{f11}
\eea
Furthermore, the OPE between $A_3(z)$ and $G_{22}(w)$
provides the last spin-$\frac{1}{2}$ current as follows 
\footnote{
We have
$
A_3(z) \, G_{22}(w) 
  =  -\frac{1}{(z-w)^2} \, \frac{i}{2} \, 2(1-\gamma) \,
F_{22}(w) + \frac{1}{(z-w)}\, \frac{i}{2} \, G_{22}(w) +\cdots$.
}:
\bea
F_{22}(w) & = & -\frac{i }{\sqrt{2}} \,K^{\bar{6}}(w).
\label{f22}
\eea

Then the Wolf space coset subgroup $SU(2)$ containing  the indices 
$6$ and $\bar{6}$ has the three spin-$\frac{1}{2}$ currents,
$F_{11}(z)$, $F_{22}(z)$ and $(F_{12}+F_{21})(z)$.  
The nontrivial OPEs between these currents are given in $(2.29)$ of 
\cite{Ahn1311}.
The above ${\cal N}=2$ multiplets, $(\frac{1}{2}, 1, 1, \frac{3}{2})$
and  $(\frac{1}{2}, 1, 1, \frac{3}{2})$, are determined (except 
the second component of first multiplet 
and third component of second multiplet) completely (see also $(2.44)$ in 
\cite{Ahn1311}).   
Furthermore, the second and third components of the other 
${\cal N}=2$ multiplets
$(1, \frac{3}{2}, \frac{3}{2}, 2)$ and $(0, \frac{1}{2}, \frac{1}{2}, 1)$
are determined.
Because the four spin-$\frac{3}{2}$ currents are known, 
one calculates their OPEs between themselves and obtains other informations 
on the other undetermined currents. 

\subsubsection{Construction of other three spin-$1$ currents}
Let us determine the other spin-$1$ currents.
Let us focus on the OPE between $F_{11}(z)$ and $G_{21}(w)$
(and similarly  the OPE between $F_{12}(z)$ and $G_{22}(w)$)
from (\ref{f11}), (\ref{g21}), (\ref{f12}) and (\ref{g22}).  
By combining these \footnote{Explicitly
one has
$
F_{11}(z) \, G_{21}(w)  =  \frac{1}{(z-w)} \left( -i B_1 - B_2\right)(w)+
\cdots $ and 
$ F_{12}(z) \, G_{22}(w)  =  \frac{1}{(z-w)} \left( -i B_1 + B_2\right)(w)+
\cdots $.
}, then the two spin-$1$ currents 
are fixed as follows:
\bea
B_1(w) & = &-\frac{1}{2 (6+k)}( B^{'}+B^{''})(w)-\frac{(1+i)}{2\sqrt{2} (6+k)}( K^5 K^6-iK^{\bar{5}} K^{\bar{6}})(w),
\nonu \\
B_2(w) & = &
-\frac{i}{2 (6+k)}( B^{'}-B^{''})(w)+\frac{(1-i)}{2\sqrt{2} (6+k)}( K^5 K^6+iK^{\bar{5}} K^{\bar{6}})(w),
\nonu \\
B^{'}(w)& \equiv & J^{7} J^{14}(w)+J^{8} J^{13}(w)-J^{9} J^{12}(w)-J^{10} J^{11}(w),
\nonu \\
B^{''}(w)& \equiv & J^{\bar{7}} J^{\bar{14}}(w)+J^{\bar{8}} J^{\bar{13}}(w)-J^{\bar{9}} J^{\bar{12}}(w)-J^{\bar{10}} J^{\bar{11}}(w).
\label{b1b2}
\eea
After that, the remaining spin-$1$ current can be read off from the 
OPE between $B_1(z)$ and $B_2(w)$ as follows:
\bea
B_3(w) & = &
 -  \frac{i}{2 (6+k)}
 \left( \sum_{(m,\bar{m})=(5,\bar{5})}^{(6,\bar{6})}K^m K^{\bar{m}}
+\sum_{(a,\bar{a})=(7,\bar{7})}^{(14,\bar{14})}J^a J^{\bar{a}}\right) (w).
\label{b3}
\eea

The three spin-$1$ currents consist of 
the affine $SU(2)_{k^{-}=5}$ algebra. 
Note that these currents are made of quadratic fermions 
and contain the Wolf space coset indices as well as 
the Wolf space subgroup indices.
In the next subsection, we will see that the latter will disappear
in the new basis.

\subsubsection{Construction of other spin-$1$ current}
For the spin-$1$ current, one can use the OPE between 
$F_{11}(z)$ and $G_{22}(w)$ \footnote{One has
$
F_{11}(z) \, G_{22}(w)   = 
\frac{1}{(z-w)} \, \left( -i A_3 -i B_3 + U\right)(w) + \cdots $.}
and the first-order pole 
determines the following spin-$1$ current 
\bea
U(z)  & = &
\frac{(1+i)}{2 \sqrt{2}} (\overline{D} K^5+i\,D K^{\bar{5}})(z).
\label{uu}
\eea
This satisfies the correct OPE between $U(z)$ and $U(w)$
and one can also rewrite this spin-$1$ current from (\ref{f21}) 
and (\ref{f12}) as follows:
\bea
U(z) = \frac{1}{\sqrt{2}} \left(  \overline{D} F_{21} - D F_{12} \right)(z).
\label{Usuperpartner}
\eea
One can construct ${\cal N}=1$ superfield as the sum of 
$(F_{21}-F_{12})(z)$ and $U(z)$ and they correspond to the first 
$SU(2)$ subgroup of the Wolf space coset.

\subsubsection{Construction of spin-$2$ current}
So far, we have found the $15$ currents of 
large ${\cal N}=4$ linear superconformal algebra and 
should determine the last unknown spin-$2$ stress energy tensor. 
For example, from the OPE between $G_{11}(z)$ and $G_{22}(w)$, one can read off
the stress energy tensor as follows
\footnote{ One can obtain the following OPEs
\bea
T_{SO(8)}(z) \, T_{SO(8)}(w) & = & \frac{1}{(z-w)^4} \, \frac{c_{SO(8)}}{2} +
\frac{1}{(z-w)^2} \, 2T_{SO(8)}(w) + \frac{1}{(z-w)} \, \pa T_{SO(8)}(w) +
\cdots,
\nonu \\
T_{SO(8)}(z) \, T_{SO(4)}(w) & = & \frac{1}{(z-w)^4} \, \frac{c_{SO(4)}}{2} +
\frac{1}{(z-w)^2} \, 2T_{SO(4)}(w) + \frac{1}{(z-w)} \, \pa T_{SO(4)}(w) +
\cdots,
\nonu \\
T_{SO(8)}(z) \, T^{'}(w) & = & \frac{1}{(z-w)^4} \, \frac{c^{'}}{2} +
\frac{1}{(z-w)^2} \, 2T^{'}(w) + \frac{1}{(z-w)} \, \pa T^{'}(w) +
\cdots,
\nonu \\
T_{SO(4)}(w) \, T_{SO(4)}(w) & = & \frac{1}{(z-w)^4} \, \frac{c_{SO(4)}}{2} +
\frac{1}{(z-w)^2} \, 2T_{SO(4)}(w) + \frac{1}{(z-w)} \, \pa T_{SO(4)}(w) +
\cdots,
\nonu \\
T^{'}(z) \,T^{'}(w) & = & \frac{1}{(z-w)^4} \, \frac{c^{'}}{2} +
\frac{1}{(z-w)^2} \, 2T^{'}(w) + \frac{1}{(z-w)} \, \pa T^{'}+
\cdots,
\nonu \\
T_{SO(4)}(z) \,T^{'}(w) & = & +
\cdots,
\label{ttopes}
\eea
where the central charges are given by
\bea
c_{SO(8)} & = &  \frac{42(2+k)}{(6+k)}, \qquad
c_{SO(4)} = \frac{3(14+3k)}{(6+k)},
\qquad
c^{'}  =  
c_{SU(2)}-c_{U(1)} = \frac{3(14+3k)}{2(6+k)}-\frac{3}{2}=
\frac{3 (4 + k)}{(6+k)},
\nonu \\
c & = & c_{SO(8)}-c_{SO(4)}-c^{'}= \frac{30(1+k)}{(6+k)}.
\label{central}
\eea
The above OPEs can be used to calculate the OPE in (\ref{opett}).}:
\bea
T(z) = T_{SO(8)}(z) - T_{SO(4)}-T^{'}(z),\,\,\,\mbox{where}\,\,\,T^{'}(z)=T_{SU(2)}-T_{U(1)}.
\label{tstress}
\eea
Then the standard OPE between the stress tensor $T(z)$, which can be
obtained from (\ref{ttopes}), has the following
form
\bea
T(z) \, T(w) & = & \frac{1}{(z-w)^4} \, \frac{c}{2} +
\frac{1}{(z-w)^2} \, 2T(w) + \frac{1}{(z-w)} \, \pa T(w) +
\cdots,
\label{opett}
\eea
where the central charge is given in (\ref{central}).
Furthermore, one has
\bea
(T_{SO(4)}+T{'})(z) \, \Phi(w) & = & + \cdots,
\label{Tphi}
\eea
where $ \Phi(w)$ in (\ref{Tphi}) 
is the $16$ currents of large ${\cal N}=4$ linear superconformal
algebra.
Also the OPEs between $K^m(z)$ with 
$m=1, 2, 3$ (or $K^{\bar{m}}(z)$ with $\bar{m} =\bar{1}, 
\bar{2}, \bar{3}$) and these $16$ currents 
do not have singular terms. 

Summarizing this subsection, the large ${\cal N}=4$ linear 
superconformal algebra has four spin-$\frac{1}{2}$ currents 
with (\ref{f11}), (\ref{f12}), (\ref{f21}) and (\ref{f22}),
seven spin-$1$ currents with (\ref{a1a2}), (\ref{a3}), (\ref{b1b2}), 
(\ref{b3}) and (\ref{uu}), four spin-$\frac{3}{2}$ currents with 
(\ref{g11}), (\ref{g12}), (\ref{g21}) and (\ref{g22}), and spin-$2$ current
with (\ref{tstress}).  
Among spin-$1$ currents, 
the three currents corresponding to 
$SU(2)_{k^{+}}$ affine algebra and 
other three  currents corresponding to 
$SU(2)_{k^{-}}$ affine algebra contain both Wolf space subgroup index and 
the Wolf space coset index.  
In particular, the spin-$\frac{1}{2}$ current $F_{11}(z)$, 
the spin-$1$ current $A_{+}(z) (\equiv (A_1 +i A_2)(z))$ and the 
spin-$\frac{3}{2}$ current $G_{11}(z)$ are located at the same 
${\cal N}=2$ multiplet. Similarly,
 the spin-$\frac{1}{2}$ current $F_{22}(z)$, 
the spin-$1$ current $A_{-}(z) (\equiv (A_1 -i A_2)(z))$ and the 
spin-$\frac{3}{2}$ current $G_{22}(z)$ are located at the same 
other ${\cal N}=2$ multiplet.
The relevant affine spin-$1$ 
currents for these live in the second $SU(2)$ subgroup 
of the
Wolf space coset. 

\subsection{Large ${\cal N}=4$ nonlinear
superconformal  algebra realization   }
As described before, one  
should construct the large ${\cal N}=4$ nonlinear superconformal algebra
from its linear version
by decoupling four spin-$\frac{1}{2}$ currents, 
$F_{11}(z)$, $F_{12}(z)$, $F_{21}(z)$ and $F_{22}(z)$ 
and one spin-$1$ current $U(z)$.

\subsubsection{Construction of spin-$2$ stress tensor}
Let us consider the 
stress energy tensor $\hat{T}(z)$ of large 
${\cal N}=4$ nonlinear superconformal algebra \cite{GS,cqg1989,npb1989}.
By definition, this should satisfy the following regular conditions 
with (\ref{uu}) (or (\ref{Usuperpartner})), 
(\ref{f11}), (\ref{f12}), (\ref{f21}) and (\ref{f22}),
\bea
U(z) \, \hat{T}(w)  & = &  +\cdots, \qquad
F_{a}(z) \, \hat{T}(w)   =   +\cdots, \qquad a=11,12,21,22.
\label{hatTregular}
\eea
For $N=4$, the stress energy tensor of the large ${\cal N}=4$
nonlinear algebra can be written as  \cite{GS,cqg1989,npb1989}
\bea
\hat{T}(z) & = & T(z) + \frac{1}{(6+k)} U U(z) +\frac{1}{(6+k)}
 \pa F^{a} F_{a}(z)
\nonu
\\
& = &
 T_{SO(8)}(z)-T_{SO(4)}(z)-T_{SU(2)}(z)+T_{U(1)}(z)
\nonu \\
& + & \frac{1}{(6+k)} \left( U U + \pa F^{a} F_{a}\right)(z).
\label{stressnonlinear}
\eea
The relative coefficients in the extra terms
are fixed from the conditions (\ref{hatTregular}).
The stress energy tensor 
for general $N$ case can be obtained by putting $(6+k) \rightarrow
(k+N+2)$ and $T \rightarrow  T_{SO(N+4)}-T_{SO(N)}-T_{SU(2)}+T_{U(1)}$
in (\ref{stressnonlinear}) similarly.
One can read off the corresponding central charge appearing in the Virasoro
algebra by calculating the OPE $\hat{T}(z) \, \hat{T}(w)$.
Therefore, the total central charge is given by
\bea
\hat{c} =  \frac{6(k+1)(N+1)}{(k+N+2)} -3
=\frac{3(k+N+2k N)}{(k+N+2)} \rightarrow \frac{3(4+9k)}{(6+k)},
\label{chat}
\eea
where the $N=4$ is substituted in  the last stage.
The remaining $10$ currents will be primary fields under this 
stress energy tensor.

\subsubsection{Construction of six spin-$1$ currents}
Let us determine the other currents.
For the spin-$1$ current, one can add $6$  additional terms
coming from four fermions, $F_{11}(z)$, $F_{12}(z)$, $F_{21}(z)$ and
$F_{22}(z)$.
By substituting (\ref{a1a2}), (\ref{a3}), (\ref{f11}), (\ref{f12}),
(\ref{f21}) and (\ref{f22}) into the equation
$(3.6)$ (where $(5+k)$ is replaced with $(6+k)$) of \cite{Ahn1311},
one obtains
\bea
\hat{A}_1(z) & = &
-\frac{(1+i)\sqrt{2}}{4(6+k)}(K^{5} K^{\bar{6}}+iK^{6} K^{\bar{5}})(z)-\frac{1}{2} (\overline{D} K^{6}- D K^{\bar{6}})(z),
\nonu \\
\hat{A}_2(z) & = &
-\frac{(-1+i)\sqrt{2}}{4(6+k)}(K^{5} K^{\bar{6}}-iK^{6} K^{\bar{5}})(z)+\frac{i}{2} (\overline{D} K^{6}+ D K^{\bar{6}})(z),
\label{nona1a2a3}
 \\
\hat{A}_3(z) & = &
-\frac{i}{2(6+k)} \left(
\sum_{(a,\bar{a})=(7,\bar{7})}^{(14,\bar{14})} J^{a} J^{\bar{a}}+2 K^{6} K^{\bar{6}}
\right)(z)
-\frac{(1+i)\sqrt{2}}{4} \left(i\overline{D} K^{5}+ D K^{\bar{5}} \right)(z).
\nonu
\eea
One can express these in terms of 
purely bosonic spin-$1$ currents appeared in (\ref{bosonicexp}).
That is,
\bea
\hat{A}_{+}(z) & = & -V^6(z), \qquad
\hat{A}_{-}(z)  =  V^{\bar{6}}(z), \qquad 
\hat{A}_3(z) = \frac{(1-i)\sqrt{2}}{4} \left( V^5- i V^{\bar{5}}\right)(z),
\label{a1a2a3non}
\eea
where we introduce $\hat{A}_{\pm}(z) \equiv (\hat{A}_1 \pm i \hat{A}_2)(z)$.
Compared to the previous expression (\ref{nona1a2a3}), all the quadratic terms
are disappeared.
These three currents consist of $SU(2)_{\hat{k}^{+}}$ affine algebra
with $\hat{k}^{+}=k^{+}-1= k$.
Let us look at the second subgroup $SU(2)$ in the Wolf space coset.
The indices of 
spin-$\frac{1}{2}$ currents, $F_{11}(z)$, $F_{22}(z)$ and 
$(F_{12}+F_{21})(z)$ stand for the adjoint representation of this 
second subgroup $SU(2)$. All the higher spin currents including 
the $11$ currents of large ${\cal N}=4$ nonlinear superconformal algebra should 
commute with these three fermionic currents.  
The superpartners of these fermionic currents correspond to 
the above three spin-$1$ currents (\ref{a1a2a3non}) plus other terms 
\footnote{For general $N$, it is nontrivial to express the above three 
spin-$1$ currents. One expects that they can be written in terms of 
sum of  some tensors with indices $a$ multiplied by $V^a(z)$
and   other tensors with indices ${\bar{a}}$ multiplied by 
$V^{\bar{a}}(z)$.  }.  

Let us move on the other type of spin-$1$ currents.
As done before, the spin-$1$ currents, $B_1(z)$, $B_2(z)$ and
$B_3(z)$
can  be modified under the factoring out the fermions and spin-$1$ current.
By substituting (\ref{b1b2}), (\ref{b3}), (\ref{f11}), (\ref{f12}),
(\ref{f21}) and (\ref{f22}) into the equation $(3.10)$ with a replacement 
of $(6+k)$ 
 in \cite{Ahn1311},
one has
\bea
\hat{B}_1(z) & = &
-\frac{1}{2 (6+k)} \left(B^{'} +B^{''}\right)(z),
\nonu \\
\hat{B}_2(z) & = &
-\frac{i}{2 (6+k)} \left(B^{'} -B^{''}\right)(z),
\nonu \\
\hat{B}_3(z) & = &
-\frac{i}{2 (6+k)} \sum_{(a,\bar{a})=(7,\bar{7})}^{(14,\bar{14})} J^{a} J^{\bar{a}}(z),
\label{b1b2b3nonlinear}
\eea
where $B^{'}(z)$ and $B^{''}(z)$ 
are given in (\ref{b1b2}).
All the $K^m(z)$ and $K^{\bar{m}}(z)$ dependences are 
disappeared because the additional quadratic terms in the fermions 
of $\hat{B}_i(z)$ cancel those terms.
Note that these three currents have only the Wolf space coset indices
and satisfy the $SU(2)_{\hat{k}^{-}}$ affine algebra with 
$\hat{k}^{-} (=k^{-}-1=N=4)$.
Because the purely bosonic currents $V^m(z)$ or $V^{\bar{m}}(z)$ 
in  (\ref{a1a2a3non})
commute with fermions $J^a(z)$ or $J^{\bar{a}}(z)$ in (\ref{b1b2b3nonlinear}),
one sees that there are no singular terms in the OPEs between 
$\hat{A}_i(z)$ and $\hat{B}_{j}(w)$.
In this respect, the expression in (\ref{a1a2a3non}) is more 
useful  than the one in (\ref{nona1a2a3}) 
\footnote{One expects that 
for general $N$, the three spin-$1$ currents, $\hat{B}_{+}(z)$, 
$\hat{B}_{-}(z)$ and $\hat{B}_3(z)$ can be written as 
some tensors (whose values are $1$ or $-1$) 
with indices $a$ and $b$ multiplied by $J^a J^{b}(z)$,
some tensors (whose values are $1$ or $-1$) 
with indices $\bar{a}$ and $\bar{b}$ multiplied by $J^{\bar{a}} 
J^{\bar{b}}(z)$ and other tensors (whose values are $1$) 
with indices $a$ and $\bar{b}$ 
multiplied by $J^a J^{\bar{b}}(z)$ respectively.}.  

\subsubsection{ Construction of four spin-$\frac{3}{2}$ currents}
Let us construct four spin-$\frac{3}{2}$ currents $\hat{G}_{11}(z), \hat{G}_{22}(z), \hat{G}_{12}(z)$, and $\hat{G}_{21}(z)$  \cite{GS,cqg1989,npb1989}.
By substituting the expressions (\ref{f11}), (\ref{f12}), (\ref{f21}),
(\ref{f22}), (\ref{uu}), (\ref{g11}), (\ref{g12}), (\ref{g21}), (\ref{g22}),
(\ref{nona1a2a3}) and (\ref{b1b2b3nonlinear}) 
into the equation (3.13) of \cite{Ahn1311},
the following results can be obtained
\bea
\hat{G}_{11}(z) & = &  \frac{i \sqrt{2}}{(6+k)} \,
 \left( 
J^7 \overline{D}J^{14}+J^8 \overline{D}J^{13}-
J^9 \overline{D}J^{12}-J^{10} \overline{D}J^{11}
 +J^{11} \overline{D}J^{10}
\right.
\nonu \\
&+& \left. 
J^{12}\overline{D} J^{9}- J^{13}\overline{D} J^{8}-J^{14}\overline{D} J^{7} 
\right)(z)
+
\mbox{cubic terms},
\label{nong11}
\\
\hat{G}_{22}(z) & = &   \frac{i \sqrt{2}}{(6+k)} \,
 \left( 
J^{\bar{7}} DJ^{\bar{14}}+J^{\bar{8}} D J^{\bar{13}}-J^{\bar{9}} D J^{\bar{12}}-
J^{\bar{10}} D J^{\bar{11}}
 +J^{\bar{11}} D J^{\bar{10}}
\right.
\nonu \\
&+& \left. 
J^{\bar{12}} D J^{\bar{9}}- J^{\bar{13}} D J^{\bar{8}}-J^{\bar{14}} D J^{\bar{7}} 
\right)(z)
+
\mbox{cubic terms},
\label{nong22}
\\
\hat{G}_{12}(z) & = &
\frac{4(1-i)}{ (6+k)} \pa  K^{\bar{5}}(z) -\frac{\sqrt{2}}{ (6+k)}
 \sum_{(a,\bar{a})=(7,\bar{7})}^{(14,\bar{14})} \overline{D}J^a J^{\bar{a}}(z)
+\mbox{cubic terms}, 
\label{nong12}
\\
\hat{G}_{21}(z) & = &
\frac{\sqrt{2}}{ (6+k)}
\sum_{(a,\bar{a})=(7,\bar{7})}^{(14,\bar{14})} J^a DJ^{\bar{a}}(z)
+
\mbox{other cubic terms}.
\label{nong21}
\eea
In this case, one also reexpresses
them using the purely bosonic spin-$1$ currents
\footnote{
One has alternative form
as follows:
\bea
\hat{G}_{11}(z) & = &  -\frac{i \sqrt{2}}{(6+k)} \,
 \left( 
J^7 V^{14}+J^8 V^{13}-J^9 V^{12}-J^{10} V^{11}
 +J^{11} V^{10}
+
J^{12} V^{9}- J^{13} V^{8}-J^{14} V^{7} 
\right)(z),
\nonu
\\
\hat{G}_{22}(z) & = &   -\frac{i \sqrt{2}}{(6+k)} \,
 \left( 
J^{\bar{7}} V^{\bar{14}}+J^{\bar{8}} V^{\bar{13}}-J^{\bar{9}} V^{\bar{12}}-
J^{\bar{10}} V^{\bar{11}}
 +J^{\bar{11}} V^{\bar{10}}
+ 
J^{\bar{12}} V^{\bar{9}}- J^{\bar{13}} V^{\bar{8}}-J^{\bar{14}} V^{\bar{7}} 
\right)(z),
\nonu
\\
\hat{G}_{12}(z) & = &
-\frac{\sqrt{2}}{ (6+k)}
 \sum_{(a,\bar{a})=(7,\bar{7})}^{(14,\bar{14})}  J^{\bar{a}} V^a(z), 
\qquad
\hat{G}_{21}(z)  = 
\frac{\sqrt{2}}{ (6+k)}
\sum_{(a,\bar{a})=(7,\bar{7})}^{(14,\bar{14})} J^a V^{\bar{a}}(z).
\label{newbasis}
\eea
It is obvious that the last two fermions in (\ref{newbasis}) 
living in the ${\cal N}=2$
stress energy tensor can be written  
as some tensors (whose values are $1$) with indices $\bar{a}$ and $b$ 
multiplied by $J^{\bar{a}} V^b(z)$ and 
other tensors (whose values are $1$) with indices $a$ and $\bar{b}$ 
multiplied by $J^{a} V^{\bar{b}}(z)$ respectively.
For the first two in (\ref{newbasis}), one expects that 
they can be realized by  some tensors (whose values are either $1$ or $-1$) 
with indices $a$ and $b$ 
multiplied by $J^{a} V^b(z)$
and other tensors (whose values are either $1$ or $-1$) 
with indices $\bar{a}$ and $\bar{b}$ 
multiplied by $J^{\bar{a}} V^{\bar{b}}(z)$ respectively.
}. 
All the cubic terms appearing in (\ref{nong11}), (\ref{nong12}), 
(\ref{nong21}) or (\ref{nong22}) are disappeared in (\ref{newbasis}).
All the indices appearing in these four spin-$\frac{3}{2}$ currents 
are coming from 
the Wolf space coset indices.
The various OPEs between these spin-$\frac{3}{2}$ currents 
are given in the equation $(A.3)$ of \cite{Ahn1311} 
where one should replace $(5+k)$ with $(6+k)$
and the Wolf space central charge $c_{\mbox{Wolf}}$ is given in 
(\ref{Wolfcentral}) (or the third equation in footnote \ref{nonfoot}). 

\subsubsection{$U(1)$ charges of the $11$ currents of
 large ${\cal N}=4$ nonlinear superconformal algebra}
As in the $U(1)$ current of $(3.19)$ in \cite{Ahn1311},
one can describe the $U(1)$ current here.
By realizing that the $U(1)$ current appears in the second-order pole
of the OPE
$\hat{G}_{21}(z) \, \hat{G}_{12}(w)$,
the $U(1)$ current is identified with
$\left(- 2 i \gamma_A \hat{A}_3 -2 i \gamma_B \hat{B}_3\right)(z)$
from the OPE (3.18) of \cite{Ahn1311}.
Here 
\bea
\gamma_A = \frac{\hat{k}^{-}}{(\hat{k}^{+} +\hat{k}^{-} +2)} =
\frac{4}{(6+k)}, \qquad
\gamma_B = \frac{\hat{k}^{+}}{(\hat{k}^{+} +\hat{k}^{-} +2)} =
\frac{k}{(6+k)}.
\label{gammaab}
\eea
Then one obtains the following first-order poles for
the $11$ currents as follows:
\bea
\left(- 2 i \gamma_A \hat{A}_3 -2 i \gamma_B \hat{B}_3\right)(z) \,
\hat{A}_{\pm}(w)|_{\frac{1}{(z-w)}} & = &
\frac{1}{(z-w)}
\left[ \mp \frac{8}{(6+k)}\right]
\hat{A}_{\pm}(w),
\nonu \\
\left(- 2 i \gamma_A \hat{A}_3 -2 i \gamma_B \hat{B}_3\right)(z) \,
\hat{B}_{\pm}(w)|_{\frac{1}{(z-w)}} & = &
\frac{1}{(z-w)}
\left[ \mp \frac{2k}{(6+k)}\right]
\hat{B}_{\pm}(w),
\nonu \\
\left(- 2 i \gamma_A \hat{A}_3 -2 i \gamma_B \hat{B}_3\right)(z) \,
\hat{A}_3(w)|_{\frac{1}{(z-w)}} & = &
0,
\nonu \\
\left(- 2 i \gamma_A \hat{A}_3 -2 i \gamma_B \hat{B}_3\right)(z) \,
\hat{B}_3(w)|_{\frac{1}{(z-w)}} & = &
0,
\nonu \\
\left(- 2 i \gamma_A \hat{A}_3 -2 i \gamma_B \hat{B}_3\right)(z) \,
\left(
\begin{array}{c}
\hat{G}_{11} \\
\hat{G}_{22} \\
\end{array} \right) (w)|_{\frac{1}{(z-w)}} & = &
\frac{1}{(z-w)}
\left[ \pm \frac{(-4+k)}{(6+k)}\right]
\left(
\begin{array}{c}
\hat{G}_{11} \\
\hat{G}_{22} \\
\end{array} \right) (w),
\nonu \\
\left(- 2 i \gamma_A \hat{A}_3 -2 i \gamma_B \hat{B}_3\right)(z) \,
\left(
\begin{array}{c}
\hat{G}_{12} \\
\hat{G}_{21} \\
\end{array} \right)  (w)|_{\frac{1}{(z-w)}} & = &
\frac{1}{(z-w)}
\left[ \mp \frac{(4+k)}{(6+k)}\right]
\left(
\begin{array}{c}
\hat{G}_{12} \\
\hat{G}_{21} \\
\end{array} \right)(w),
\nonu \\
\left(- 2 i \gamma_A \hat{A}_3 -2 i \gamma_B \hat{B}_3\right)(z) \,
\hat{T}(w)|_{\frac{1}{(z-w)}} & = &
0.
\label{u1first}
\eea
We present the $U(1)$ charges for the $11$ currents in Table $1$ (the relevant
Tables $2$-$10$ will appear later).

\begin{table}[ht]
\centering 
\begin{tabular}{|c||c| } 
\hline 
$U(1)$ charge & $11$ currents of large ${\cal N}=4$ nonlinear
superconformal algebra  \\ [0.5ex] 
\hline \hline 
$\frac{2k}{(6+k)}$  & $\hat{B}_{-}(\equiv\hat{B}_1 - i \hat{B}_2) $
\\ 
\hline
$\frac{(4+k)}{(6+k)}$ &  $\hat{G}_{21}$ \\
\hline
$\frac{(-4+k)}{(6+k)}$ & $\hat{G}_{11}$  \\
\hline
$\frac{8}{(6+k)} $ & $\hat{A}_{-}(\equiv\hat{A}_1 - i \hat{A}_2)$  \\
\hline
$0$ & $\hat{A}_3, \quad \hat{B}_3, \quad \hat{T}$  \\
\hline
$-\frac{8}{(6+k)}$ & $ \hat{A}_{+}(\equiv\hat{A}_1 + i \hat{A}_2)$  \\
\hline
$-\frac{(-4+k)}{(6+k)}$ & $\hat{G}_{22}$  \\
\hline
$-\frac{(4+k)}{(6+k)} $ & $\hat{G}_{12}$  \\
\hline
$-\frac{2k}{(6+k)}$ & $\hat{B}_{+}(\equiv\hat{B}_1 + i \hat{B}_2)$  \\
[1ex] 
\hline 
\end{tabular}
\caption{The $U(1)$ charges for the $11$ currents from the equation
(\ref{u1first}).
The first four currents with positive
($ k > 4 $)
$U(1)$ charges have their conjugated  currents with each
opposite (negative) $U(1)$ charge.
  } 
\end{table}

For the affine currents, $K^m(z)$ where $m=1,2, \cdots, 6$
has vanishing $U(1)$ charge (and their complex conjugated ones also).
The $J^a(z)$ where $a=7,8, \cdots, 14$ has the $U(1)$ charge $\frac{k}{(6+k)}$
while the complex conjugated ones has the $U(1)$ charge $-\frac{k}{(6+k)}$.
For the spin-$1$ currents, $\overline{D} K^m(z)$with $m=1,2, \cdots, 5$ 
has vanishing $U(1)$ charge and   $D K^{\bar{m}}(z)$with $\bar{m}= \bar{1},
\bar{2}, \cdots, \bar{5}$ 
has also vanishing $U(1)$ charge.  

Summarizing this subsection, 
the large ${\cal N}=4$ nonlinear superconformal algebra is constructed.  
There are six spin-$1$ currents with (\ref{nona1a2a3}) and 
(\ref{b1b2b3nonlinear}),
four spin-$\frac{3}{2}$ currents with (\ref{newbasis}) and spin-$2$ stress
energy tensor (\ref{stressnonlinear}).
The three spin-$1$ currents corresponding to $SU(2)_{\hat{k}^{+}}$ affine algebra
live in the second subgroup $SU(2)$ of Wolf space coset and 
the other three spin-$1$ currents corresponding to
$SU(2)_{\hat{k}^{-}}$ affine algebra live in the first subgroup 
$SU(2)$ of   
Wolf space coset. 
Compared to the linear version of previous subsection,
the quadratic terms in $A_i(z)$ are disappeared  and also 
the cubic terms in $G_a(z)$ are disappeared.
Furthermore, some of the quadratic terms in $B_i(z)$
are disappeared
\footnote{
\label{nonfoot}
The large ${\cal N}=4$ nonlinear superconformal algebra 
\cite{GS,cqg1989,npb1989,GK} is 
\bea
\hat{T}(z) \, \hat{T}(w) & = & \frac{1}{(z-w)^4} \, \frac{\hat{c}}{2} +
\frac{1}{(z-w)^2} \, 2 \hat{T}(w) + \frac{1}{(z-w)} \, 
\pa \hat{T}(w) +
\cdots,
\nonu \\
\hat{T}(z) \, \Phi(w) & = & 
\frac{1}{(z-w)^2} \, h \, \Phi(w) + \frac{1}{(z-w)} \, 
\pa \Phi(w) +
\cdots, 
\nonu \\
\hat{G}_{a}(z) \, \hat{G}_{b}(w) & = &
\frac{1}{(z-w)^3} \, \frac{2}{3} c_{\mbox{Wolf}} \, \hat{\delta}_{ab} -
\frac{1}{(z-w)^2} \frac{16}{(6+k)} 
\left[ \hat{k}^{-} \, \alpha_{ab}^{+i} \,  \hat{A}_i + \hat{k}^{+} \, 
\alpha_{ab}^{-i} \, 
\hat{B}_i \right](w)\nonu \\
 & + & 
\frac{1}{(z-w)} \left[ 2 \hat{T} \,  \hat{\delta}_{ab}
-\frac{1}{2}  \,    \frac{16}{(6+k)} 
 ( \hat{k}^{-} \, 
\alpha_{ab}^{+i} \,  \pa \hat{A}_i + \hat{k}^{+} \, 
\alpha_{ab}^{-i} \, \pa 
\hat{B}_i ) +
\right. \nonu \\
& - & \left. \frac{16}{(6+k)} (  
\alpha^{+i} \,  \hat{A}_i -
\alpha^{-i} \, 
\hat{B}_i)_{c(a }  (  
\alpha^{+j} \,  \hat{A}_j -
\alpha^{-j} \, 
\hat{B}_j )_{b)}^{\,\,c}   \right](w) +  \cdots,
\nonu \\
\hat{A}_i(z) \, \hat{G}_{a}(w) 
 & = &  \frac{1}{(z-w)} \, \alpha_{ab}^{+ i} \, \hat{G}^b(w) +\cdots,  
\qquad
\hat{B}_i(z) \, \hat{G}_{a}(w)  =  
\frac{1}{(z-w)} \, \alpha_{ab}^{-i} \, \hat{G}^{b}(w) +\cdots,
\nonu \\
\hat{A}_i(z) \, \hat{A}_j(w) & = & 
-\frac{1}{(z-w)^2} \, \frac{1}{2} \hat{k}^{+} \, \hat{\delta}_{ij} +
 \frac{1}{(z-w)} \, \ep_{ijk} \, \hat{A}_k(w)
 +\cdots,
\nonu \\
\hat{B}_i(z) \, 
\hat{B}_j(w) & = & -\frac{1}{(z-w)^2} \, \frac{1}{2} \, \hat{k}^{-} 
\, \hat{\delta}_{ij} +
\frac{1}{(z-w)} \, \ep_{ijk} \, \hat{B}_k(w) 
+\cdots,
\nonu 
\eea
where $\hat{c}$ is given by (\ref{chat}), 
$c_{\mbox{Wolf}}$ is given by (\ref{Wolfcentral}),
$h =\frac{3}{2}$ for $\Phi=\hat{G}_a$, $h =1$ for 
$\Phi=\hat{A}_i, \hat{B}_i$ and $\hat{\delta}_{ab}$ is normalized to be $1$.
}. 

\section{The $16$ 
higher spin currents  in the orthogonal Wolf space coset 
$\frac{SO(8)}{SO(4) \times SU(2) \times SU(2)}$}
This section will consider the particular supersymmetric Wolf space
coset minimal model
(\ref{coset}) and the following higher spin currents  introduced in section
$1$ will be
determined:
\bea
\left(2, \frac{5}{2}, \frac{5}{2}, 3 \right)
& : & (T^{(2)}, T_{+}^{(\frac{5}{2})}, T_{-}^{(\frac{5}{2})}, T^{(3)}),
\nonu \\
 \left(\frac{5}{2}, 3, 3, \frac{7}{2} \right) & : &
(U^{(\frac{5}{2})}, U_{+}^{(3)}, U_{-}^{(3)}, U^{(\frac{7}{2})} ), \nonu \\
\left(\frac{5}{2}, 3, 3, \frac{7}{2} \right) & : &
(V^{(\frac{5}{2})}, V^{(3)}_{+}, V^{(3)}_{-}, V^{(\frac{7}{2})}),  \nonu \\
\left(3, \frac{7}{2}, \frac{7}{2}, 4 \right) & : &
 (W^{(3)}, W_{+}^{(\frac{7}{2})}, W_{-}^{(\frac{7}{2})}, W^{(4)}).
\label{new16comp}
\eea
The highest spin-$4$ current is located at the last component 
of the last ${\cal N}=2$ multiplet.
One can interpret the other $15$ currents as the ${\cal N}=4$ superpartner
of this spin-$4$ current.
As in the unitary case \cite{Ahn1311}, 
it is important to obtain the lowest higher spin-$2$ current in the beginning.
Then all the other component fields can be determined using the 
currents of large ${\cal N}=4$ nonlinear superconformal algebra.
Note that the field contents of (\ref{new16comp}) appear in the unitary 
case \cite{GG1305,Ahn1408}.

\subsection{Construction of higher spin currents of spins
$\left(2, \frac{5}{2}, \frac{5}{2}, 3 \right)$}
The $28$ ${\cal N}=2$ WZW affine currents  of spin-$\frac{1}{2}$
are given by $K^m(z)$, $K^{\bar{m}}(z)$, $J^a(z)$ and $J^{\bar{a}}(z)$.
Let us denote them as $Q^A(z)$ and $Q^{\bar{A}}(z)$ where $A=(m,a)$ and $\bar{A}=
(\bar{m},\bar{a})$.
Now let us  determine the spin-$2$ current in (\ref{new16comp}).
We write down the most general spin-$2$ currents
(with $14596$  coefficient functions), which depend on
the level $k$,  as
\bea
T^{(2)}(z) & = & \sum_{A>B=1}^{14}  \sum_{\bar{C}>\bar{D}=\bar{1}}^{\bar{14}}c_{A,B,\bar{C},\bar{D}} Q^A Q^B Q^{\bar{C}} Q^{\bar{D}}(z)
\nonu \\
& + &
  \sum_{A,C=1}^{14}  \sum_{\bar{B}=\bar{1}}^{\bar{14}}c_{A,\bar{B},C} Q^A  Q^{\bar{B}}\overline{D}Q^{C}(z)
  + \sum_{A=1}^{14}  \sum_{\bar{B},\bar{C}=\bar{1}}^{\bar{14}}c_{A,\bar{B},\bar{C}} Q^A  Q^{\bar{B}}D Q^{\bar{C}}(z)
\nonu \\
&+&\sum_{A=1}^{14}  \sum_{\bar{B}=\bar{1}}^{\bar{14}}c_{A,\bar{B}} \pa Q^A  Q^{\bar{B}}(z)
  + \sum_{A=1}^{14}  \sum_{\bar{B}=\bar{1}}^{\bar{14}}c_{A,\bar{B}}^{'} Q^A \pa Q^{\bar{B}}(z)
\nonu \\
&+&\sum_{A,B=1}^{14} c_{A,B} \overline{D} Q^A  \overline{D} Q^{B}(z)
  + \sum_{A=1}^{14}  \sum_{\bar{B}=\bar{1}}^{\bar{14}}c_{A,\bar{B}}^{''} \overline{D} Q^AD Q^{\bar{B}}(z)
\nonu \\
&+&\sum_{\bar{A},\bar{B}=1}^{\bar{14}} c_{\bar{A},\bar{B}} D Q^{\bar{A}}  D Q^{\bar{B}}(z)
  + \sum_{A=1}^{14}  c_A \pa \overline{D} Q^A(z)+ \sum_{\bar{A}=\bar{1}}^{\bar{14}}  c_{\bar{A}} \pa D Q^{\bar{A}}(z).
\label{generalt2}
\eea
The above quartic, cubic, quadratic and linear terms 
have the equal number of unbarred quantity and barred quantity.
We would like to determine the coefficient functions appearing in
(\ref{generalt2})
explicitly.
Because 
the regularity conditions between the spin-$1$ current $U(z)$ (and the
spin-$\frac{1}{2}$ currents $F_{a}(z)$)  and the spin-$2$ current
$T^{(2)}(w)$
are preserved in this extended nonlinear
algebra,
the following relations, together with (\ref{uu}), (\ref{f11}), (\ref{f12}),
(\ref{f21}), (\ref{f22}) and (\ref{generalt2}), should be satisfied
\bea
U(z) \, T^{(2)}(w) & = & +\cdots,
\qquad
F_{a}(z) \, T^{(2)}(w)   =  +\cdots, \qquad a=11,12,21,22.
\label{uft2}
\eea
Then after (\ref{uft2}) the remaining undetermined coefficient functions
can be fixed by the following primary field condition under the stress
tensor (\ref{stressnonlinear})
\bea
\hat{T}(z) \, T^{(2)}(w) & = & \frac{1}{(z-w)^2} \, 2T^{(2)}(w) +
\frac{1}{(z-w)} \, \pa T^{(2)}(w) + \cdots.
\label{t2primary}
\eea
All the first-order singular terms between
the above spin-$2$ current and other six spin-$1$ currents
can be obtained from the defining equations
(\ref{nona1a2a3}), (\ref{b1b2b3nonlinear}) and (\ref{generalt2})
and by requiring that the commutators between the zero mode $T_0^{(2)}$ and six
spin-$1$ currents should vanish \cite{GG1305}
\bea
T^{(2)}(z) \, \hat{A}_i(w)|_{\frac{1}{(z-w)}} & = & 0, \qquad
T^{(2)}(z) \, \hat{B}_i(w)|_{\frac{1}{(z-w)}} =0.\qquad
\label{t2sixpole1}
\eea
All the remaining coefficient functions, after using (\ref{t2primary}) and 
(\ref{t2sixpole1}),  
are completely determined except
overall constants \footnote{
After computing the OPE $T^{(2)}(z) \, T^{(2)}(w)$, 
one finds that three undetermined coefficients are fixed
and, with an appropriate normalization constant,
one has
\bea
T^{(2)}(z) \, T^{(2)}(w)&=& -\frac{1}{(z-w)^4}\, 
\left[\frac{8 k (6+k)^3 (4+5 k) (16+11 k)}{(2+k) (8+k)}\right]
\nonu \\
&-& \frac{1}{(z-w)^2} \left[ \frac{36 i k (6+k)^2}{(8+k)} T^{(2)} 
+ \frac{16 (6+k)^4 (4+5 k)}{(8+k)} \hat{T} 
+\frac{16 (6+k)^4 (4+5 k)}{(2+k) (8+k)} \hat{A}_3 \hat{A}_3
\right.
\nonu \\
 & + &   \frac{16 (6+k)^4 (4+5 k)}{(2+k) (8+k)} \hat{A}_{+} \hat{A}_{-}
+  \frac{8 (6+k)^4 (4+5 k)}{3 (8+k)} 
\hat{B}_{3} \hat{B}_{3}+  
\frac{8 (6+k)^4 (4+5 k)}{3 (8+k)}
\hat{B}_{+} \hat{B}_{-} \nonu \\
& 
+ & \left.  \frac{16 i (6+k)^4 (4+5 k)}{(2+k) (8+k)} 
\pa \hat{A}_3 +  \frac{8 i (6+k)^4 (4+5 k)}{3 (8+k)}
\pa \hat{B}_3 \right](w)
+ \frac{1}{(z-w)}\frac{1}{2} \pa \{ T^{(2)} \, T^{(2)} \}_{-2}(w)+\cdots.
\nonu
\eea
There exist other six different solutions for $T^{(2)}(z)$. In other words,
these six different expressions for $T^{(2)}(z)$ are independent each other.
All the OPEs in Appendix $C$ and $D$ do not depend on the explicit seven 
different forms for
$T^{(2)}(z)$.}.
It turns out that the lowest higher spin-$2$
current (\ref{generalt2}) can be obtained as follows :
\bea
T^{(2)}(z) & = & \frac{2i}{(6+k)}\left[
K^1 J^8 K^{\bar{1}} J^{\bar{8}}
+K^1 J^8 K^{\bar{2}} J^{\bar{7}}
+K^1 J^{10} K^{\bar{1}} J^{\bar{10}}
+K^1 J^{12} K^{\bar{1}} J^{\bar{12}}\right](z)
\nonu \\
&+& \mbox{other 341 terms},
\label{t2}
\eea
where each term has vanishing $U(1)$ charge.

Let us calculate the OPE between the spin-$\frac{3}{2}$ current 
$\hat{G}_{21}(z)$ given in (\ref{nong21}) and 
the spin-$2$ current $T^{(2)}(w)$ given in (\ref{t2}).
This particular OPE 
is motivated by the fact that 
the OPE between the second component of ${\cal N}=2$
super stress tensor and the first component of 
any ${\cal N}=2$ super primary current provides 
its second component.
It turns out that
\bea
\hat{G}_{21}(z) \, T^{(2)}(w)  & = &
\frac{1}{(z-w)}
T_{+}^{(\frac{5}{2})}(w) +
\cdots,
\label{g21t2}
\eea
where the spin-$\frac{5}{2}$ current is given by
\bea
T_{+}^{(\frac{5}{2})}(z) & = &
 -\frac{2i \sqrt{2}(16+11k)}{(2+k)(6+k)(8+k)}\left[
K^{1} K^{4} J^{8} K^{\bar{4}}J^{\bar{11}}
+K^{1} K^{4} J^{10} K^{\bar{4}}J^{\bar{13}}
\right. \nonu \\
&-& \left.
K^{1} K^{4} J^{12} K^{\bar{4}}J^{\bar{7}}
-K^{1} K^{4} J^{14} K^{\bar{4}}J^{\bar{9}}\right](z)
+ \mbox{other 636 terms}.
\label{t+}
\eea
This is a primary field under the stress energy tensor (\ref{stressnonlinear}).
One can check that both sides in (\ref{g21t2}) have correct $U(1)$ charge
defined in (\ref{u1first}). 
It turns out that the corresponding $U(1)$ charge of $T^{(2)}(w)$
is equal to zero and the $U(1)$ charge of $T_{+}^{(\frac{5}{2})}(w)$
is given by the one for $\hat{G}_{21}(z)$, $\frac{(4+k)}{(6+k)}$.
We will come to this issue in subsection $3.6$ later.
The OPE (\ref{g21t2}) 
has very simple form compared to the unitary case in \cite{Ahn1408} where 
the corresponding OPE is given by $\hat{G}_{21}(z) \, P^{(2)}(w)$ and there 
are many composite fields in the first-order pole, for example.
According to Table $4$, there are also other five composite fields 
satisfying the above $U(1)$ charge but they do not appear in the first order 
pole
in (\ref{g21t2}).

Recall that the OPE between the third component of ${\cal N}=2$
super stress tensor and the first component of 
any ${\cal N}=2$ super primary current provides 
its third component.
Then one describes the following OPE
between 
 the spin-$\frac{3}{2}$ current 
$\hat{G}_{12}(z)$ in (\ref{nong12}) and 
the spin-$2$ current $T^{(2)}(w)$ in (\ref{t2})
\bea
\hat{G}_{12}(z) \, T^{(2)}(w)  & = &
\frac{1}{(z-w)}
T_{-}^{(\frac{5}{2})} (w) +
\cdots.
\label{g12t2}
\eea
In the right hand side of (\ref{g12t2}), the spin-$\frac{5}{2}$ current appears
as follows:
\bea
T_{-}^{(\frac{5}{2})}(z) & = &
 \frac{2i \sqrt{2}(16+11k)}{(2+k)(6+k)(8+k)}\left[
K^{4} J^{7} K^{\bar{1}} K^{\bar{4}}J^{\bar{12}}
+K^{4} J^{8} K^{\bar{2}} K^{\bar{4}}J^{\bar{12}}
\right. \nonu \\
&+& \left.
K^{4} J^{9} K^{\bar{1}} K^{\bar{4}}J^{\bar{14}}
-K^{4} J^{11} K^{\bar{1}} K^{\bar{4}}J^{\bar{8}}\right](z)
+ \mbox{other 643 terms},
\label{t-}
\eea
which transforms as a primary field under the stress energy tensor
(\ref{stressnonlinear}).
The $U(1)$ charge conservation implies that the $U(1)$ charge of
this spin-$\frac{5}{2}$ current is given by $-\frac{(4+k)}{(6+k)}$
which is the $U(1)$ charge of $\hat{G}_{12}(z)$ in the left hand side of
(\ref{g12t2}).
In this case also, 
the OPE (\ref{g12t2}) 
has simple form compared to the unitary case in \cite{Ahn1408} where 
the corresponding OPE is given by $\hat{G}_{12}(z) \, P^{(2)}(w)$ 
as before.
Moreover,  
there are no other five composite fields  in (\ref{g12t2}).

Once again,
 the OPE between the second component of ${\cal N}=2$
super stress tensor and the third component of 
any ${\cal N}=2$ super primary current provides 
its first and fourth components.
It turns out that the OPE between the spin-$\frac{3}{2}$ current
$\hat{G}_{21}(z)$ and the spin-$\frac{5}{2}$ current $T_{-}^{(\frac{5}{2})}(w)$
(\ref{t-}) reads as
\bea
\hat{G}_{21}(z) \, T_{-}^{(\frac{5}{2})}(w) & = &
\frac{1}{(z-w)^2} 4T^{(2)} (w)  
 +  \frac{1}{(z-w)}  \left[
 \frac{1}{4} 4\pa
   T^{(2)} +T^{(3)} \right](w) +  \cdots.
\label{g21t-}
\eea
In the second-order pole of (\ref{g21t-}),
one sees that the first component of the first ${\cal N}=2$ multiplet
in (\ref{new16comp}) and, in the first-order pole,
the following spin-$3$ current occurs
\bea
T^{(3)}(z) & = &
 \frac{2i \sqrt{2}(16+11k)}{(2+k)(6+k)^2(8+k)}\left[
K^{1} K^{4} J^{8} K^{\bar{4}}K^{\bar{5}}J^{\bar{11}}
-K^{1} K^{4} J^{10} K^{\bar{3}}K^{\bar{4}}J^{\bar{13}}
\right. \nonu \\
&+& \left.
K^{1} K^{4} J^{12} K^{\bar{3}}K^{\bar{4}}J^{\bar{7}}
-K^{1} K^{4} J^{12} K^{\bar{4}}K^{\bar{5}}J^{\bar{7}}\right](z)
+ \mbox{other 4941 terms}.
\label{t3}
\eea
This is a primary field under the stress energy tensor 
(\ref{stressnonlinear}) and has a vanishing $U(1)$ charge.
The relative coefficient $\frac{1}{4}$ above is fixed by 
the spins of the left hand side of (\ref{g21t-}) and the spin of $T^{(2)}(w)$. 
The OPE (\ref{g21t-}) 
has very simple form compared to the unitary case in \cite{Ahn1408} where 
the corresponding OPE is given by $\hat{G}_{21}(z) \, P_{-}^{(\frac{5}{2})}(w)$
and three kinds of quasiprimary fields due to the presence of
spin-$1$ currents with vanishing $U(1)$ charges appear in the first-order 
pole.
In other words, there are no other composite fields (appearing in Table $5$)
in the first order pole of (\ref{g21t-}). 

In this subsection, the first ${\cal N}=2$ multiplet given in the 
first one in (\ref{new16comp}) is determined 
completely.

\subsection{Construction of higher spin currents of spins
$\left(\frac{5}{2}, 3, 3, \frac{7}{2} \right)$}
Now let us move on the other ${\cal N}=2$ multiplet in (\ref{new16comp}).
According to the behavior in the unitary case \cite{Ahn1311},
the spin-$\frac{3}{2}$ current $\hat{G}_{11}(z)$ provides
the second ${\cal N}=2$ multiplet after calculating the OPE between 
$\hat{G}_{11}(z)$ and the first ${\cal N}=2$ multiplet.
One obtains, from (\ref{nong11}) and (\ref{t2}),  the following OPE   
\bea
\hat{G}_{11}(z) \, T^{(2)}(w)  & = &
\frac{1}{(z-w)}U^{(\frac{5}{2})} (w) +
\cdots.
\label{g11t2}
\eea
In the right hand side of (\ref{g11t2}), 
the following spin-$\frac{5}{2}$ current, which is the lowest component 
of the second ${\cal N}=2$ multiplet, occurs
\bea
U^{(\frac{5}{2})}(z)& = &
 \frac{2(1-i) }{(2+k)(6+k)}\left[
K^{1} K^{6} J^{8} K^{\bar{5}}J^{\bar{11}}
-K^{1} K^{6} J^{10} K^{\bar{5}}J^{\bar{13}}
\right. \nonu \\
&-& \left.
K^{1} K^{6} J^{12} K^{\bar{5}}J^{\bar{7}}
+K^{1} K^{6} J^{14} K^{\bar{5}}J^{\bar{9}}\right](z)
+ \mbox{other 540 terms}.
\label{u5half}
\eea
This is a primary field under the stress energy tensor 
(\ref{stressnonlinear}) and the $U(1)$ charge is given by 
$\frac{(-4+k)}{(6+k)}$.
There are no other five composite fields in (\ref{g11t2}).

Let us describe the following OPE
between the above spin-$\frac{3}{2}$ current $\hat{G}_{11}(z)$ (\ref{nong11})
and the next component spin-$\frac{5}{2}$ current 
$T^{(\frac{5}{2})}_{+}(w)$ (\ref{t+}) of the second ${\cal N}=2$ multiplet 
\footnote{
The normalization with sign for the spin-$3$ current comes from 
the following OPE result 
\bea
\hat{G}_{21}(z) \, U^{(\frac{5}{2})}(w)  & = &
\frac{1}{(z-w)}  U^{(3)}_{+} (w) +
\cdots.
\label{g21u5half}
\eea
The $U(1)$ charge for the spin-$3$ current $U_{+}^{(3)}(w)$ 
in (\ref{g21u5half})
is given by the sum of
the $U(1)$ charge of $\hat{G}_{21}(z)$ and the one of
$U^{(
\frac{5}{2})}(w)$:$\frac{2k}{(6+k)}$.
This spin-$3$ current appears in (\ref{g11t+}).
The OPE (\ref{g21u5half}) 
has very simple form compared to the unitary case in \cite{Ahn1408} where 
the corresponding OPE is given by $\hat{G}_{21}(z) \, Q^{(\frac{5}{2})}(w)$
and the first-order pole contains the quasiprimary field due to the 
spin-$1$ current.
}
\bea
\hat{G}_{11}(z) \, T^{(\frac{5}{2})}_{+}(w)  & = &
\frac{1}{(z-w)} \left[ -U^{(3)}_{+} \right](w) +
\cdots.
\label{g11t+}
\eea
Here the explicit form for the spin-$3$ current is given by
\bea
U^{(3)}_{+}(z)& = &
 \frac{8(16+11k)}{(2+k)(6+k)^2(8+k)}\left[
K^{1} K^{4} J^{8} J^{12} K^{\bar{3}}K^{\bar{5}}
+K^{1} K^{4} J^{8} J^{13} K^{\bar{1}}K^{\bar{4}}
\right. \nonu \\
&-& \left.
K^{1} K^{4} J^{9} J^{12} K^{\bar{1}}K^{\bar{4}}
+K^{1} K^{4} J^{10} J^{11} K^{\bar{1}}K^{\bar{4}}\right](z)
+ \mbox{other 2086 terms}.
\label{u3+}
\eea
According to Table $5$, there are also other composite fields 
with the 
correct $U(1)$ charge, but they do not appear in the first order pole of
(\ref{g11t+}).

Let us describe the following OPE
between the above spin-$\frac{3}{2}$ current $\hat{G}_{11}(z)$ (\ref{nong11})
and the third component spin-$\frac{5}{2}$ current 
$T^{(\frac{5}{2})}_{-}(w)$ (\ref{t-}) of the second ${\cal N}=2$ multiplet 
\footnote{
In this case the normalization for the spin-$3$ current 
comes from the following OPE 
\bea
\hat{G}_{12}(z) \, U^{(\frac{5}{2})}(w)  & = &
\frac{1}{(z-w)} U^{(3)}_{-} (w) +
\cdots.
\label{g12u5half}
\eea
The $U(1)$ charge for the spin-$3$ current $U_{-}^{(3)}(w)$ 
in (\ref{g12u5half}) is given by the sum of
the $U(1)$ charge of $\hat{G}_{12}(z)$ and the one of
$U^{(
\frac{5}{2})}(w)$:$-\frac{8}{(6+k)}$.
This spin-$3$ current appears in the OPE (\ref{g11t-}).
The OPE (\ref{g12u5half}) 
has very simple form compared to the unitary case in \cite{Ahn1408} where 
the corresponding OPE is given by $\hat{G}_{12}(z) \, Q^{(\frac{5}{2})}(w)$
and the first-order pole contains the quasiprimary field due to the 
spin-$1$ current.
}
\bea
\hat{G}_{11}(z) \, T^{(\frac{5}{2})}_{-}(w)  & = &
\frac{1}{(z-w)} \left[ -U^{(3)}_{-} \right](w) +
\cdots,
\label{g11t-}
\eea
where the spin-$3$ current in the first-order pole 
is given by
\bea
U^{(3)}_{-}(z) & = &
 -\frac{2(1-i)\sqrt{2}}{(2+k)(6+k)^2}\left[
K^{1} K^{6} J^{10} K^{\bar{1}}K^{\bar{5}}J^{\bar{10}}
-K^{1} K^{6} J^{10} K^{\bar{4}}K^{\bar{5}}J^{\bar{7}}
\right. \nonu \\
&+& \left.
K^{1} K^{6} J^{12} K^{\bar{1}}K^{\bar{5}}J^{\bar{12}}
+K^{1} K^{6} J^{14} K^{\bar{1}}K^{\bar{5}}J^{\bar{14}}\right](z)
+ \mbox{other 2068 terms}.
\label{u3-}
\eea
There are no other composite fields 
(with the correct $U(1)$ charge) in the first order pole of
(\ref{g11t-}).

Let us consider the following OPE between (\ref{nong21}) and (\ref{u3-}), 
as in (\ref{g21t-})
\bea
\hat{G}_{21}(z) \, U_{-}^{(3)}(w) & = &
\frac{1}{(z-w)^2} \frac{2(11+3k)}{(6+k)} U^{(\frac{5}{2})}(w)
\nonu \\
& + & \frac{1}{(z-w)}  \left[ \frac{1}{5}  \frac{2(11+3k)}{(6+k)} \pa U^{(\frac{5}{2})}+U^{(\frac{7}{2})}
 \right](w)+ \cdots.
\label{g21u-}
\eea
The relative coefficient $\frac{1}{5}$ is fixed by the spins 
of $\hat{G}_{21}(z)$, $U_{-}^{(3)}(w)$ and $U^{(\frac{5}{2})}(w)$
in (\ref{g21u-}).
The spin-$\frac{7}{2}$ current is given by
\bea
U^{(\frac{7}{2})}(z) & = &
 -\frac{8\sqrt{2}(16+11k)}{(2+k)(6+k)^3(8+k)}\left[
K^{1} K^{4} K^{6}J^{7} K^{\bar{1}}K^{\bar{4}}J^{\bar{7}}
+K^{1} K^{4} K^{6}J^{8} K^{\bar{3}}K^{\bar{5}}J^{\bar{9}}
\right. \nonu \\
&-& \left.
K^{1} K^{4} K^{6}J^{9} K^{\bar{1}}K^{\bar{4}}J^{\bar{9}}
+K^{1} K^{4} K^{6}J^{10} K^{\bar{1}}K^{\bar{4}}J^{\bar{10}}\right](z)
+ \mbox{ 13191 terms}.
\label{u7half}
\eea
The $U(1)$ charge of this current is the same as the one of the 
spin-$\frac{5}{2}$ current $U^{(\frac{5}{2})}(z)$.
The OPE (\ref{g21u-}) 
has very simple form compared to the unitary case in \cite{Ahn1408} where 
the corresponding OPE is given by $\hat{G}_{21}(z) \, Q_{-}^{(3)}(w)$
and the first-order pole contains the quasiprimary fields due to the 
spin-$\frac{3}{2}$ currents.
In this case also, there are no other possible composite fields 
(appearing in Table $6$) in the first order pole of (\ref{g21u-}).

In this subsection, the second ${\cal N}=2$ multiplet  
in (\ref{new16comp}) is determined 
completely.

\subsection{Construction of higher spin currents of spins
$\left(\frac{5}{2}, 3, 3, \frac{7}{2} \right)$}
Now let us move on the other ${\cal N}=2$ multiplet in (\ref{new16comp}).
According to the behavior in the unitary case \cite{Ahn1311},
the spin-$\frac{3}{2}$ current $\hat{G}_{22}(z)$ provides
the third ${\cal N}=2$ multiplet after calculating the OPE between 
$\hat{G}_{22}(z)$ and the first ${\cal N}=2$ multiplet.
One obtains, from (\ref{nong22}) and (\ref{t2}), the following OPE 
\bea
\hat{G}_{22}(z) \, T^{(2)}(w)  & = &
\frac{1}{(z-w)}  V^{(\frac{5}{2})} (w) +
\cdots.
\label{g22t2}
\eea
In the right hand side of (\ref{g22t2}), 
the following spin-$\frac{5}{2}$ current, which is the lowest component 
of the third ${\cal N}=2$ multiplet, occurs
\bea
V^{(\frac{5}{2})}(z) & = &
 \frac{\sqrt{2}(16+11k)}{(2+k)(6+k)}\left[
K^{1} J^{7}J^{\bar{7}}J^{\bar{11}} J^{\bar{13}}
-K^{1} J^{9}J^{\bar{9}}J^{\bar{11}} J^{\bar{13}}
\right. \nonu \\
&-& \left.
K^{1} J^{11}J^{\bar{7}}J^{\bar{9}} J^{\bar{11}}
+K^{1} J^{13}J^{\bar{7}}J^{\bar{9}} J^{\bar{13}}\right](z)
+ \mbox{other 542 terms}.
\label{v5half}
\eea
This primary spin-$\frac{5}{2}$ current has the $U(1)$ charge of
$\frac{(4-k)}{(6+k)}$.

Let us describe the following OPE
between the above spin-$\frac{3}{2}$ current $\hat{G}_{22}(z)$
(\ref{nong22}) and the third component spin-$\frac{5}{2}$ current 
$T^{(\frac{5}{2})}_{+}(w)$ (\ref{t+}) of the first ${\cal N}=2$ multiplet 
\footnote{
The normalization with sign for the spin-$3$ current comes from 
the following OPE result 
\bea
\hat{G}_{21}(z) \, V^{(\frac{5}{2})}(w)  & = &
\frac{1}{(z-w)} V^{(3)}_{+}(w) +
\cdots.
\label{g21v5half}
\eea
The $U(1)$ charge for the spin-$3$ current $V_{+}^{(3)}(w)$ 
in (\ref{g21v5half}) is given by the sum of
the $U(1)$ charge of $\hat{G}_{21}(z)$ and the one of
$V^{(
\frac{5}{2})}(w)$:$\frac{8}{(6+k)}$.
The OPE (\ref{g21v5half}) 
has very simple form compared to the unitary case in \cite{Ahn1408} where 
the corresponding OPE is given by $\hat{G}_{21}(z) \, R^{(\frac{5}{2})}(w)$
and the first-order pole contains the quasiprimary field due to the 
spin-$1$ current.
}
\bea
\hat{G}_{22}(z) \, T^{(\frac{5}{2})}_{+}(w)  & = &
\frac{1}{(z-w)} \left[ -V^{(3)}_{+} \right](w) +
\cdots,
\label{g22t+}
\eea
where the spin-$3$ current in the first-order pole of (\ref{g22t+})
is given by
\bea
V^{(3)}_{+}(z) & = &
 \frac{2(16+11k)}{(2+k)(6+k)^2}\left[
K^{1} K^{2} J^{7} J^{\bar{7}}J^{\bar{9}}J^{\bar{11}}
-K^{1} K^{2} J^{8} J^{\bar{8}}J^{\bar{9}}J^{\bar{11}}
\right. \nonu \\
&+& \left.
K^{1} K^{2} J^{13} J^{\bar{9}}J^{\bar{11}}J^{\bar{13}}
-K^{1} K^{2} J^{14} J^{\bar{9}}J^{\bar{11}}J^{\bar{14}}\right](z)
+ \mbox{other 2114 terms}.
\label{v3+}
\eea

Let us describe the following OPE
between the above spin-$\frac{3}{2}$ current $\hat{G}_{22}(z)$
(\ref{nong22}) and the next component spin-$\frac{5}{2}$ current 
$T^{(\frac{5}{2})}_{-}(w)$ (\ref{t-}) of the first ${\cal N}=2$ multiplet 
\footnote{
The normalization with sign for the spin-$3$ current comes from 
the following OPE result 
\bea
\hat{G}_{12}(z) \, V^{(\frac{5}{2})}(w)  & = &
\frac{1}{(z-w)}  V^{(3)}_{-}(w) +
\cdots.
\label{g12v5half}
\eea
The $U(1)$ charge for the spin-$3$ current $V_{-}^{(3)}(w)$ 
in (\ref{g12v5half}) is given by the sum of
the $U(1)$ charge of $\hat{G}_{12}(z)$ and the one of
$V^{(
\frac{5}{2})}(w)$:$-\frac{2k}{(6+k)}$.
This spin-$3$ current appears in (\ref{g22t-}).
The OPE (\ref{g12v5half}) 
has very simple form compared to the unitary case in \cite{Ahn1408} where 
the corresponding OPE is given by $\hat{G}_{12}(z) \, R^{(\frac{5}{2})}(w)$
and the first-order pole contains the quasiprimary field due to the 
spin-$1$ current.
}
\bea
\hat{G}_{22}(z) \, T^{(\frac{5}{2})}_{-}(w)  & = &
\frac{1}{(z-w)} \left[ -V^{(3)}_{-} \right](w) +
\cdots.
\label{g22t-}
\eea
In this case, the spin-$3$ current is given by
\bea
V^{(3)}_{-}(z) & = &
 \frac{8(16+11k)}{(2+k)(6+k)^2(8+k)}\left[
K^{1} K^{4} K^{\bar{1}} K^{\bar{4}}J^{\bar{7}}J^{\bar{14}}
-K^{1} K^{4} K^{\bar{1}} K^{\bar{4}}J^{\bar{8}}J^{\bar{13}}
\right. \nonu \\
&+& \left.
K^{1} K^{4} K^{\bar{1}} K^{\bar{4}}J^{\bar{9}}J^{\bar{12}}
-K^{1} K^{4} K^{\bar{1}} K^{\bar{4}}J^{\bar{10}}J^{\bar{11}}\right](z)
+ \mbox{other 2204 terms}.
\label{v3-}
\eea

Let us consider the following OPE between (\ref{nong21}) and (\ref{v3-}), 
as in (\ref{g21t-})
\bea
\hat{G}_{21}(z) \, V_{-}^{(3)}(w) & = &
\frac{1}{(z-w)^2}\frac{(30+4k)}{(6+k)} V^{(\frac{5}{2})}(w)
\nonu \\
& + & \frac{1}{(z-w)}  \left[ \frac{1}{5}  \frac{(30+4k)}{(6+k)}\pa V^{(\frac{5}{2})}+V^{(\frac{7}{2})}
 \right](w)+ \cdots.
\label{g21v-}
\eea
The relative coefficient $\frac{1}{5}$ is fixed by the spins 
of $\hat{G}_{21}(z)$, $V_{-}^{(3)}(w)$ and $V^{(\frac{5}{2})}(w)$
in (\ref{g21v-}).
The spin-$\frac{7}{2}$ current is given by
\bea
V^{(\frac{7}{2})}(z)  & = &
 -\frac{2\sqrt{2}(16+11k)}{(2+k)(6+k)^3}\left[
K^{1} K^{2} J^{7} K^{\bar{1}}J^{\bar{7}}J^{\bar{8}}J^{\bar{9}}
+K^{1} K^{2} J^{7} K^{\bar{1}}J^{\bar{7}}J^{\bar{11}}J^{\bar{14}}
\right. \nonu \\
&+& \left.
K^{1} K^{2} J^{7} K^{\bar{1}}J^{\bar{9}}J^{\bar{11}}J^{\bar{12}}
+K^{1} K^{2} J^{7} K^{\bar{2}}J^{\bar{7}}J^{\bar{11}}J^{\bar{13}}\right](z)
\nonu \\
&+& \mbox{other 13109 terms}.
\label{v7half}
\eea
The $U(1)$ charge of this current is the same as the one of the 
spin-$\frac{5}{2}$ current $V^{(\frac{5}{2})}(z)$.
The OPE (\ref{g21v-}) 
has very simple form compared to the unitary case in \cite{Ahn1408} where 
the corresponding OPE is given by $\hat{G}_{21}(z) \, R_{-}^{(3)}(w)$
and the first-order pole contains the quasiprimary fields due to the 
spin-$\frac{3}{2}$ currents.

In this subsection, the third ${\cal N}=2$ multiplet  
in (\ref{new16comp}) is determined 
completely.

\subsection{Construction of higher spin currents of spins
$\left(3, \frac{7}{2}, \frac{7}{2}, 4 \right)$}
Let us consider 
the following OPE between (\ref{nong22}) and (\ref{u5half})
\bea
\hat{G}_{22}(z) \, U^{(\frac{5}{2})}(w) & = &
\frac{1}{(z-w)^2}4T^{(2)}(w)
+  \frac{1}{(z-w)}  \left[ \frac{1}{4} 4 \pa T^{(2)}+W^{(3)}
 \right](w)+ \cdots.
\label{g22u5half}
\eea
The spin-$3$ current in (\ref{g22u5half}) is given by
\bea
W^{(3)}(z) & = &
 \frac{8i(16+11k)}{(2+k)(6+k)^2(8+k)}\left[
-K^{1} K^{4} K^{6} K^{\bar{4}}J^{\bar{7}}J^{\bar{9}}
+K^{1} K^{4} K^{6} K^{\bar{4}}J^{\bar{11}}J^{\bar{13}}
\right. \nonu \\
&+& \left.
K^{1} K^{4} J^{9} K^{\bar{1}}K^{\bar{4}}J^{\bar{9}}
-K^{1} K^{4} J^{11} K^{\bar{1}}K^{\bar{4}}J^{\bar{11}}\right](z)
+ \mbox{other 5367 terms}.
\label{w3}
\eea
The OPE (\ref{g22u5half}) 
has very simple form compared to the unitary case in \cite{Ahn1408} where 
the corresponding OPE is given by $\hat{G}_{22}(z) \, Q^{(\frac{5}{2})}(w)$
and the first-order pole contains the quasiprimary fields due to the 
spin-$1$ currents.
In other words, there are no other possible composite fields in 
(\ref{g22u5half}).

The following OPE between (\ref{nong21}) and (\ref{w3}) 
provides the spin-$\frac{7}{2}$ current
\bea
\hat{G}_{21}(z) \,W^{(3)}(w) & = &
\frac{1}{(z-w)^2} \frac{(4-k)}{(6+k)} T_{+}^{(\frac{5}{2})}(w)
\nonu \\
& + & \frac{1}{(z-w)}  \left[ \frac{1}{5}   \frac{(4-k)}{(6+k)} \pa T_{+}^{(\frac{5}{2})}+W_{+}^{(\frac{7}{2})}
 \right](w)+ \cdots,
\label{g21w3}
\eea
where the spin-$\frac{7}{2}$ current is
\bea
W_{+}^{(\frac{7}{2})}(z) & = &
 \frac{8i\sqrt{2}(16+11k)}{(2+k)(6+k)^3(8+k)}\left[
K^{1} K^{2} K^{4} J^{7}K^{\bar{1}}K^{\bar{4}}J^{\bar{11}}
-K^{1} K^{2} K^{4} J^{8}K^{\bar{2}}K^{\bar{4}}J^{\bar{11}}
\right. \nonu \\
&-& \left.
K^{1} K^{2} K^{4} J^{10}K^{\bar{1}}K^{\bar{4}}J^{\bar{14}}
+K^{1} K^{2} K^{4} J^{10}K^{\bar{2}}K^{\bar{4}}J^{\bar{13}}\right](z)
+ \mbox{16536 terms}.
\label{w7+}
\eea
Note that there is the $(k-4)$ factor in the $T_{+}^{(\frac{5}{2})}(w)$
and its descendant field in (\ref{g21w3}). 
When $k=4$ (or $\hat{k}^{+} = \hat{k}^{-}$), these terms vanish.
Also the large ${\cal N}=4$ nonlinear superonformal algebra 
with $\hat{k}^{+} = \hat{k}^{-}$
becomes the $SO({\cal N}=4)$ Knizhnik-Bershadsky algebra 
\cite{Knizhnik,Bershadsky}. 
The OPE (\ref{g21w3}) 
has very simple form compared to the unitary case in \cite{Ahn1408} where 
the corresponding OPE is given by $\hat{G}_{21}(z) \, S^{(3)}(w)$
and the first-order pole contains the quasiprimary fields due to the 
spin-$\frac{3}{2}$ currents.

Let us describe the following OPE
between (\ref{g12}) and (\ref{w3})
\bea
\hat{G}_{12}(z) \,W^{(3)}(w) & = &
\frac{1}{(z-w)^2} \frac{(-4+k)}{(6+k)} T_{-}^{(\frac{5}{2})}
(w)
\nonu \\
& + & \frac{1}{(z-w)}  \left[ \frac{1}{5}   \frac{(-4+k)}{(6+k)} \pa T_{-}^{(\frac{5}{2})}+W_{-}^{(\frac{7}{2})}
 \right](w)+ \cdots.
\label{g12w3}
\eea
The spin-$\frac{7}{2}$ current is given by
\bea
W_{-}^{(\frac{7}{2})}(z) & = &
 \frac{16i\sqrt{2}(16+11k)}{(2+k)(6+k)^3(8+k)}\left[
K^{1} K^{4} K^{6} K^{\bar{1}}K^{\bar{4}}J^{\bar{7}}J^{\bar{14}}
-K^{1} K^{4} K^{6} K^{\bar{1}}K^{\bar{4}}J^{\bar{8}}J^{\bar{13}}
\right. \nonu \\
&+& \left.
K^{1} K^{4} K^{6} K^{\bar{1}}K^{\bar{4}}J^{\bar{9}}J^{\bar{12}}
-K^{1} K^{4} K^{6} K^{\bar{1}}K^{\bar{4}}J^{\bar{10}}J^{\bar{11}}\right](z)
+ \mbox{ 16662 terms}.
\label{w7-}
\eea
There is the $(k-4)$ factor in the $T_{-}^{(\frac{5}{2})}(w)$
and its descendant field in (\ref{g12w3}). 
The OPE (\ref{g12w3}) 
has very simple form compared to the unitary case in \cite{Ahn1408} where 
the corresponding OPE is given by $\hat{G}_{12}(z) \, S^{(3)}(w)$
and the first-order pole contains the quasiprimary fields due to the 
spin-$\frac{3}{2}$ currents.

From the following OPE between (\ref{nong21}) and (\ref{w7-})
\bea
\hat{G}_{21}(z) \,W_{-}^{(\frac{7}{2})}(w) & = &
\frac{1}{(z-w)^3} \left[ -\frac{48(-4+k)}{5(6+k)}T^{(2)}
 \right](w)
\nonu \\
& + & \frac{1}{(z-w)^2}  \left[ -\frac{6(-4+k)}{5(6+k)}T^{(3)}+\frac{2(16+3k)}{(6+k)}W^{(3)}
+\frac{16i}{(6+k)}(\hat{A_3}
-\hat{B_3})T^{(2)}
 \right](w)
 \nonu \\
 &+&\frac{1}{(z-w)} \left[ \frac{1}{6}\pa \{\hat{G}_{21} \,W_{-}^{(\frac{7}{2})}\}_{-2}
 \right. \nonu \\
& - & \left. \frac{144(-4+k)}{(324+179k)}(\hat{T}T^{(2)}-\frac{3}{10} \pa^2 T^{(2)})
 + W^{(4)}
 \right](w)
 + \cdots,
\label{g21w7mhalf}
\eea
one obtains, from (\ref{g21w7mhalf}),  the following spin-$4$ current
\bea
W^{(4)}(z) & = &
 \frac{16i(16+11k)}{(2+k)(6+k)^4(8+k)}\left[
K^{1} K^{2} K^{4} J^{7}K^{\bar{1}}K^{\bar{2}}K^{\bar{4}}J^{\bar{7}}
+K^{1} K^{2} K^{4} J^{8}K^{\bar{1}}K^{\bar{2}}K^{\bar{4}}J^{\bar{8}}
\right. \nonu \\
&+& \left.
2K^{1} K^{2} K^{4} J^{10}K^{\bar{1}}K^{\bar{2}}K^{\bar{4}}J^{\bar{10}}
-2K^{1} K^{2} K^{4} J^{12}K^{\bar{1}}K^{\bar{2}}K^{\bar{4}}J^{\bar{12}}\right](z)
\nonu \\
&+& \mbox{other 98358 terms}.
\label{w4}
\eea
There is no descendant field for the spin-$2$ field (appearing in the 
third order pole) in the second order pole  in (\ref{g21w7mhalf})
because the spin difference of the left hand side 
is given by $\frac{3}{2}-\frac{7}{2} =-2$
which is equal to the minus spin of the above spin-$2$ current.
Each term in the second order pole of (\ref{g21w7mhalf}) 
is a primary field under the stress 
energy tensor.
The spin-$4$ current appears in the OPEs, (\ref{g22u7half}), (\ref{g11v7half}),
(\ref{g12w+7half}) and (\ref{g21w-7half}).
Note that the quasiprimary field in the first-order pole in 
(\ref{g21w7mhalf}) has the factor $(k-4)$.
The OPE (\ref{g21w7mhalf}) 
has very simple form compared to the unitary case in \cite{Ahn1408} where 
the corresponding OPE is given by $\hat{G}_{21}(z) \, S_{-}^{(\frac{7}{2})}(w)$
and the first-order pole contains many quasiprimary fields.

Therefore,
the $16$ higher spin currents are found explicitly.
We have also checked that these $16$ currents as well as the $11$
currents of large ${\cal N}=4$ nonlinear superconformal algebra
have regular OPEs with the spin $\frac{1}{2}$ currents $K^m(w)$ and 
$K^{\bar{m}}(w)$
with $m=1,2,3, \bar{m}=\bar{1}, \bar{2},\bar{3}$ 
(and their superpartner spin-$1$ currents $\overline{D} K^m(w)$ 
and $D K^{\bar{m}}(w)$). 
These six spin-$\frac{1}{2}$ currents (and six spin-$1$ currents)
live in the subgroup $SO(4)$ of Wolf space coset.
Furthermore, the OPEs between the above $(11+16)$ currents and 
 the spin $\frac{1}{2}$ currents $K^4(w)$ and 
$K^{\bar{4}}(w)$ 
(and their superpartner spin-$1$ currents $\overline{D} K^4(w)$ 
and $D K^{\bar{4}}(w)$) have regular terms. 
These two spin-$\frac{1}{2}$ currents (and two spin-$1$ currents)
live in the first subgroup $SU(2)$ of Wolf space coset.
By construction, the remaining currents, spin-$\frac{1}{2}$ current 
$\frac{(1+i)}{2} 
(K^5+i K^{\bar{5}})(z)$ which is $(F_{12}-F_{21})(z)$
and its superpartner spin-$1$ current $\frac{(1+i)}{2\sqrt{2}} 
(\overline{D} K^5 + i D K^{\bar{5}})(z)$ which is 
$U(z)$ (both living in  the first subgroup $SU(2)$ of Wolf space coset)
do not have any singular terms with the above $(11+16)$ currents. 
Furthermore, the three spin-$\frac{1}{2}$ currents
$\frac{(1+i)}{2} 
(K^5-i K^{\bar{5}})(z)(=(F_{12}+F_{21})(z))$, 
$ F_{11}(z)= \frac{i}{\sqrt{2}} 
K^6(z)$ and $F_{22}(z)=-\frac{i}{\sqrt{2}} 
K^{\bar{6}}(z)$, which live in the second $SU(2)$ subgroup 
of the Wolf space coset, do not have any singular terms with 
the above $(11+16)$ currents \footnote{The three spin-$1$ currents,
$( \overline{D} K^5-i D K^{\bar{5}})(z)$, $ \overline{D} K^6(z)$ and $
D K^{\bar{6}}(z)$, do have the singular terms with the above $(11+16)$ 
currents. }.

\subsection{The OPEs between the $11$ currents of 
large ${\cal N}=4$ nonlinear superconformal algebra and $16$ 
higher spin currents    }
In section $2$, the $11$ currents of
large ${\cal N}=4$ nonlinear superconformal algebra
were constructed and in section $3$, the $16$
higher spin currents were found.
For the extension of  large ${\cal N}=4$ nonlinear
superconformal  algebra, one should calculate
the OPEs between the $11$ currents in section $2$ and
the $16$ currents in section $3$
as follows:
\bea
\left(
\begin{array}{c}
 \hat{T} \\
\hat{A}_1, \hat{A}_2, \hat{A}_3 \\
\hat{B}_1, \hat{B}_2, \hat{B}_3 \\
\hat{G}_{11}, \hat{G}_{12}, \hat{G}_{21}, \hat{G}_{22}
 \\
\end{array}
\right)(z)  \,
\left(
\begin{array}{cccc}
T^{(2)}, T_{+}^{(\frac{5}{2})}, T_{-}^{(\frac{5}{2})},
T^{(3)} \\
U^{(\frac{5}{2})}, U_{+}^{(3)}, U_{-}^{(3)}, U^{(\frac{7}{2})}
 \\
V^{(\frac{5}{2})}, V_{+}^{(3)},
V_{-}^{(3)}, V^{(\frac{7}{2})}
\\
W^{(3)}, W_{+}^{(\frac{7}{2})},
W_{-}^{(\frac{7}{2})}, W^{(4)} \\
\end{array} \right)(w).
\label{opediagram}
\eea
In other words, the OPEs between $11$ currents in the left hand side of
(\ref{opediagram}) and the $16$ currents in the right hand side of
(\ref{opediagram}) are needed.

\subsection{$U(1)$ charges of higher spin currents}
It is straightforward to calculate the various $U(1)$ charges under
the $U(1)$ current introduced in the previous section. The result is
as follows with (\ref{gammaab}):
\bea
\left(- 2 i \gamma_A \hat{A}_3 -2 i \gamma_B \hat{B}_3\right)(z) \,
T^{(2)}(w)|_{\frac{1}{(z-w)}} & = &
0, \nonu \\
\left(- 2 i \gamma_A \hat{A}_3 -2 i \gamma_B \hat{B}_3\right)(z) \,
T_{\pm}^{(\frac{5}{2})} (w)|_{\frac{1}{(z-w)}} & = &
\frac{1}{(z-w)}
\left[  \pm \frac{(4+k)}{(6+k)}\right]
T_{\pm}^{(\frac{5}{2})} (w),
\nonu \\
\left(- 2 i \gamma_A \hat{A}_3 -2 i \gamma_B \hat{B}_3\right)(z) \,
T^{(3)}(w)|_{\frac{1}{(z-w)}} & = &
0,
\nonu \\
\left(- 2 i \gamma_A \hat{A}_3 -2 i \gamma_B \hat{B}_3\right)(z) \,
\left(
\begin{array}{c}
U^{(\frac{5}{2})} \\
V^{(\frac{5}{2})} \\
\end{array} \right) (w)|_{\frac{1}{(z-w)}} & = &
\frac{1}{(z-w)}
\left[  \pm \frac{(-4+k)}{(6+k)}\right]
\left(
\begin{array}{c}
U^{(\frac{5}{2})} \\
V^{(\frac{5}{2})} \\
\end{array} \right)(w),
\nonu \\
\left(- 2 i \gamma_A \hat{A}_3 -2 i \gamma_B \hat{B}_3\right)(z) \,
\left(
\begin{array}{c}
U_{+}^{(3)} \\
V_{-}^{(3)} \\
\end{array} \right)
(w)|_{\frac{1}{(z-w)}} & = &
\frac{1}{(z-w)}
\left[  \pm \frac{2k}{(6+k)}\right]
\left(
\begin{array}{c}
U_{+}^{(3)} \\
V_{-}^{(3)} \\
\end{array} \right)
(w),
\nonu \\
\left(- 2 i \gamma_A \hat{A}_3 -2 i \gamma_B \hat{B}_3\right)(z) \,
\left(
\begin{array}{c}
U_{-}^{(3)} \\
V_{+}^{(3)} \\
\end{array} \right)
(w)|_{\frac{1}{(z-w)}} & = &
\frac{1}{(z-w)}
\left[ \mp \frac{8}{(6+k)}\right]
\left(
\begin{array}{c}
U_{-}^{(3)} \\
V_{+}^{(3)} \\
\end{array} \right)(w),
\nonu \\
\left(- 2 i \gamma_A \hat{A}_3 -2 i \gamma_B \hat{B}_3\right)(z) \,
\left(
\begin{array}{c}
U^{(\frac{7}{2})} \\
V^{(\frac{7}{2})} \\
\end{array} \right)
(w)|_{\frac{1}{(z-w)}} & = &
\frac{1}{(z-w)}
\left[ \pm  \frac{(-4+k)}{(6+k)}\right]
\left(
\begin{array}{c}
U^{(\frac{7}{2})} \\
V^{(\frac{7}{2})} \\
\end{array} \right) (w),
\nonu \\
\left(- 2 i \gamma_A \hat{A}_3 -2 i \gamma_B \hat{B}_3\right)(z) \,
W^{(3)}(w)|_{\frac{1}{(z-w)}} & = &
0,
\nonu \\
\left(- 2 i \gamma_A \hat{A}_3 -2 i \gamma_B \hat{B}_3\right)(z) \,
W_{\pm}^{(\frac{7}{2})} (w)|_{\frac{1}{(z-w)}} & = &
\frac{1}{(z-w)}
\left[  \pm \frac{(4+k)}{(6+k)}\right]
W_{\pm}^{(\frac{7}{2})} (w),
\nonu \\
\left(- 2 i \gamma_A \hat{A}_3 -2 i \gamma_B \hat{B}_3\right)(z) \,
W^{(4)}(w)|_{\frac{1}{(z-w)}} & = &
0.
\label{u1second}
\eea
We present these $U(1)$ charges in the Table $2$ explicitly \footnote{Compared to the unitary case in \cite{Ahn1311}, the $U(1)$ charges behave similarly 
in the sense that for given $U(1)$ charge each current among $16$ currents
shares their $U(1)$ charges with each current among $11$ currents.    }. 
Then the $U(1)$ charges of  all the composite fields 
coming from the $(11+16)$ currents  can be determined by 
the assignments in Table $1$ and Table $2$.
The necessary $U(1)$ assignments in this paper are presented in 
Table $3$-Table $10$.

\subsection{Structure of OPEs between the $11$ currents and $16$ currents}
The $16$ higher spin currents are primary fields under the stress energy 
tensor $\hat{T}(z)$ (\ref{stressnonlinear}).
Then the remaining nontrivial 
OPEs between the $11$ currents and the $16$ currents
are given by 1) the OPEs between the six spin-$1$ currents and the 
$16$ higher spin currents described in Appendix $C$ and 2)
the OPEs between the four spin-$\frac{3}{2}$ currents and 
the $16$ higher spin currents described in Appendix $D$.  
One should write down the correct terms, which preserve the right 
$U(1)$ charge described before via Tables $1$-$10$,  
with arbitrary coefficients at the specific 
pole (with fixed spin) in the given OPE. 

For example, in the OPEs between the spin-$1$ currents of 
large ${\cal N}=4$ nonlinear algebra and the largest spin-$4$ higher spin
current, the first-order pole contains 
the spin-$4$ fields in Appendices (\ref{A+w4})-(\ref{B3w4}).
The $U(1)$ charges of composite spin-$4$ fields are presented in Tables 
$7,8$. The $U(1)$ charge of the left hand side of Appendix 
(\ref{A+w4}) is the sum of
the $U(1)$ charge of $\hat{A}_{+}(z)$ and the one of 
$W^{(4)}(w)$. The former is given by $-\frac{8}{(6+k)}$ 
and the latter is given by $0$ from Table $2$. 
Therefore, the total $U(1)$ charge, $-\frac{8}{(6+k)}$, should appear in the 
right hand side of Appendix (\ref{A+w4}). 
Then from the first row of Table $8$, 
there exist $50$ possible composite spin-$4$ fields with the $U(1)$ charge 
$-\frac{8}{(6+k)}$.    
Because the left hand side of Appendix 
(\ref{A+w4}) is written in terms of ${\cal N}=2$
WZW affine currents explicitly, let us subtract the above $50$ terms with 
arbitrary $k$-dependent coefficients from the left hand side of Appendix
(\ref{A+w4}).
In order for these quantities to vanish, one should solve this equation.
It turns out that the unknown coefficient functions can be obtained completely
and they are given in Appendix (\ref{A+w4}) \footnote{Compared to the 
convention in \cite{BCG1404}, because the third order pole of
the OPE between the OPE $\hat{A}_3(z)$ and $W^{(4)}(w)$ (\ref{A3w4}) (similarly 
the OPE between the OPE $\hat{B}_3(z)$ and $W^{(4)}(w)$ (\ref{B3w4})) 
does not vanish, 
one can find other basis where these higher order poles should vanish. 
Their $W^{(4)}(z)$  in \cite{BCG1404} 
is not a primary field (but a quasiprimary field) 
under the stress energy tensor.
Note that the other OPEs appearing in Appendix $C$ do not contain 
the third order (and higher order) poles. Then we do not need to change 
the basis for these remaining $15$ higher spin currents.
}.  

The OPEs between the spin-$\frac{3}{2}$ currents of
large ${\cal N}=4$ nonlinear superconformal algebra and the above spin-$4$ 
higher spin current are presented in Appendix (\ref{g11w4})-(\ref{g21w4}).
One can have the composite spin-$\frac{9}{2}$ fields in the first-order pole of
Appendix (\ref{g11w4}).
The total $U(1)$ charge of this OPE should be preserved. 
The $U(1)$ charge of the left hand side is given by 
$\frac{(-4+k)}{(6+k)}$ from the Table $2$, and the possible  
composite fields with 
this $U(1)$ charge are presented in the Table $9$. 
The unknown 
coefficient functions can be obtained completely
and are given in Appendix (\ref{g11w4}) \footnote{The highest order pole 
in Appendix $D$ appears in the third order pole. As in \cite{Ahn1311}, one 
can find the new basis where these singular terms vanish.}.

The result from Appendices $C$ and $D$ shows that 
the right hand sides of all the OPEs contains the composite fields which 
can be obtained from the known $(11+16)$ currents. 
One expects that the new primary fields will appear in the 
OPEs between the $16$ currents and themselves.

\begin{table}[ht]
\centering 
\begin{tabular}{|c||c||c| } 
\hline 
$U(1)$ charge & $16$ higher spin currents & $11$ currents in Table 1
\\ [0.5ex] 
\hline \hline 
$\frac{2k}{(6+k)}$  & $ \quad U_{+}^{(3)}$ &  $\hat{B}_{-}$
\\ 
\hline
$\frac{(4+k)}{(6+k)}$ &  $ T_{+}^{(\frac{5}{2})}, \quad
W_{+}^{(\frac{7}{2})}$ &  $\hat{G}_{21}$ \\
\hline
$\frac{(-4+k)}{(6+k)}$ & $\quad U^{(\frac{5}{2})}, \quad
U^{(\frac{7}{2})}$ &  $\hat{G}_{11}$  \\
\hline
$\frac{8}{(6+k)} $ & $V_{+}^{(3)}$ &  $\hat{A}_{-}$  \\
\hline
$0$ & $ T^{(2)}, \quad T^{(3)}, \quad
W^{(3)}, \quad W^{(4)}$ & $\hat{A}_3$, \qquad $\hat{B}_3$, \qquad
$\hat{T}$  \\
\hline
$-\frac{8}{(6+k)}$ & $ U_{-}^{(3)}$ & $\hat{A}_{+}$  \\
\hline
$-\frac{(-4+k)}{(6+k)}$ & $ V^{(\frac{5}{2})}, \quad
V^{(\frac{7}{2})}$ & $\hat{G}_{22}$  \\
\hline
$-\frac{(4+k)}{(6+k)} $ & $ T_{-}^{(\frac{5}{2})}, \quad
W_{-}^{(\frac{7}{2})}$ & $\hat{G}_{12}$  \\
\hline
$-\frac{2k}{(6+k)}$ & $V_{-}^{(3)}$ &
$\hat{B}_{+}$  \\
[1ex] 
\hline 
\end{tabular}
\caption{The $U(1)$ charges for the $16$ currents from the equation
(\ref{u1second}).
For convenience, the $U(1)$ charges for the $11$ currents of
large ${\cal N}=4$ nonlinear superconformal algebra
in Table $1$ are presented also.
} 
\centering 
\end{table}

\begin{table}[ht]
\centering 
\begin{tabular}{|c||c| } 
\hline 
$U(1)$ charge & Composite fields of spin-$2$ \\ [0.5ex] 
\hline \hline 
$\frac{2k}{(6+k)}$  & $\hat{B}_{-}\hat{A}_{3},\quad \hat{B}_{-}\hat{B}_{3},\quad \pa \hat{B}_{-}$
\\ 
\hline
$\frac{4k}{(6+k)}$ &  $\hat{B}_{-}\hat{B}_{-}$\\
\hline
$\frac{2(4+k)}{(6+k)}$ &  $\hat{B}_{-}\hat{A}_{-}$  \\
\hline
$\frac{2(-4+k)}{(6+k)} $ &  $\hat{B}_{-}\hat{A}_{+}$  \\
\hline
$\frac{16}{(6+k)}$ &  $\hat{A}_{-}\hat{A}_{-}$  \\
\hline
$\frac{8}{(6+k)}$ & $\hat{A}_{-}\hat{A}_{3},\quad \hat{A}_{-}\hat{B}_{3},\quad \pa \hat{A}_{-}$  \\
\hline
$0$ & $\hat{T},\,\, T^{(2)},\,\, \hat{A}_{3}\hat{A}_{3},\,\,
 \hat{A}_{3}\hat{B}_{3},\,\, \hat{A}_{-}\hat{A}_{+},\,\, 
\hat{B}_{3}\hat{B}_{3},\,\,
 \hat{B}_{-}\hat{B}_{+},\,\, \pa \hat{A}_{3}, \,\, \pa \hat{B}_{3}$ \\
\hline
$-\frac{8}{(6+k)} $ &  $\hat{A}_{3}\hat{A}_{+},\quad \hat{B}_{3}\hat{A}_{+},\quad \pa \hat{A}_{+}$  \\
\hline
$-\frac{16}{(6+k)}$ & $\hat{A}_{+}\hat{A}_{+}$  \\
\hline
$-\frac{2(-4+k)}{(6+k)}$ & $\hat{A}_{-}\hat{B}_{+}$ \\
\hline
$-\frac{2(4+k)}{(6+k)}$ & $\hat{A}_{+}\hat{B}_{+}$  \\
\hline
$-\frac{4k}{(6+k)}$ & $\hat{B}_{+}\hat{B}_{+}$  \\
\hline
$-\frac{2k}{(6+k)}$ & $\hat{A}_{3}\hat{B}_{+},\quad \hat{B}_{3}\hat{B}_{+}, \quad \pa \hat{B}_{+}$  \\
[1ex] 
\hline 
\end{tabular}
\caption{The $U(1)$ charges for the spin-$2$ composite fields.
  } 
\end{table}

\begin{table}[ht]
\centering 
\begin{tabular}{|c||c| } 
\hline 
$U(1)$ charge & Composite fields of spin-$\frac{5}{2}$ \\ [0.5ex] 
\hline \hline 
$\frac{(4+3k)}{(6+k)}$  & $\hat{B}_{-}\hat{G}_{21}$
\\ 
\hline
$\frac{(-4+3k)}{(6+k)}$ &  $\hat{B}_{-}\hat{G}_{11}$\\
\hline
$\frac{(12+k)}{(6+k)}$ &  $\hat{G}_{21}\hat{A}_{-}$  \\
\hline
$\frac{(-12+k)}{(6+k)} $ &  $\hat{G}_{11}\hat{A}_{+}$  \\
\hline
$\frac{(4+k)}{(6+k)}$ & $T_{+}^{(\frac{5}{2})},\quad \hat{B}_{-}\hat{G}_{22},\quad \hat{G}_{11}\hat{A}_{-},\quad
\hat{G}_{21}\hat{A}_{3},\quad \hat{G}_{21}\hat{B}_{3},\quad \pa \hat{G}_{21}$  \\
\hline
$\frac{(-4+k)}{(6+k)}$ & $U^{(\frac{5}{2})},\quad \hat{B}_{-}\hat{G}_{12},\quad \hat{G}_{11}\hat{A}_{3},\quad
\hat{G}_{11}\hat{B}_{3},\quad \hat{G}_{21}\hat{A}_{+},\quad \pa \hat{G}_{11}$  \\
\hline
$-\frac{(-4+k)}{(6+k)} $ & $V^{(\frac{5}{2})},\quad \hat{A}_{3}\hat{G}_{22},\quad \hat{A}_{-}\hat{G}_{12},\quad
\hat{B}_{3}\hat{G}_{22},\quad \hat{G}_{21}\hat{B}_{+},\quad \pa \hat{G}_{22}$  \\
\hline
$-\frac{(4+k)}{(6+k)}$ & $T_{-}^{(\frac{5}{2})},\quad \hat{A}_{3}\hat{G}_{12},\quad \hat{A}_{+}\hat{G}_{22},\quad
\hat{B}_{3}\hat{G}_{12},\quad \hat{G}_{11}\hat{B}_{+},\quad \pa \hat{G}_{12}$  \\
\hline
$-\frac{(-12+k)}{(6+k)}$ & $\hat{A}_{-}\hat{G}_{22}$  \\
\hline
$-\frac{(12+k)}{(6+k)}$ & $\hat{A}_{+}\hat{G}_{12}$ \\
\hline
$-\frac{(-4+3k)}{(6+k)}$ & $\hat{G}_{22}\hat{B}_{+}$  \\
\hline
$-\frac{(4+3k)}{(6+k)}$ & $\hat{G}_{12}\hat{B}_{+}$  \\
[1ex] 
\hline 
\end{tabular}
\caption{The $U(1)$ charges for the spin-$\frac{5}{2}$ composite fields.
  } 
\end{table}

\begin{table}[ht]
\centering 
\begin{tabular}{|c||c| } 
\hline 
$U(1)$ charge & Composite fields of spin-$3$ \\ [0.5ex] 
\hline \hline 
$\frac{2k}{(6+k)}$  & $U_{+}^{(3)},\quad \hat{A}_{3} \pa \hat{B}_{-},\quad \hat{B}_{3} \pa \hat{B}_{-},
\quad \hat{B}_{-}T^{(2)}, \quad \hat{B}_{-} \hat{T},\quad \hat{B}_{-}\hat{A}_{3}\hat{A}_{3}$,\\
& $\hat{B}_{-}\hat{A}_{3}\hat{B}_{3},
\quad \hat{B}_{-}\hat{A}_{-}\hat{A}_{+},\quad \hat{B}_{-}\hat{B}_{3}\hat{B}_{3},\quad \hat{B}_{-}\hat{B}_{-}\hat{B}_{+}$,\\
& $\hat{B}_{-} \pa \hat{A}_{3}, \quad \hat{B}_{-} \pa \hat{B}_{3},\quad \hat{G}_{21} \hat{G}_{11},\quad \pa^2 \hat{B}_{-}$
\\ 
\hline
$\frac{4k}{(6+k)}$ &  $\hat{B}_{-}\hat{B}_{-}\hat{A}_{3}, \quad \hat{B}_{-}\hat{B}_{-}\hat{B}_{3},\quad \hat{B}_{-} \pa \hat{B}_{-}$\\
\hline
$\frac{2(4+k)}{(6+k)}$ &  $\hat{A}_{-} \pa \hat{B}_{-}, \quad \hat{B}_{-}\hat{A}_{-}\hat{A}_{3},\quad \hat{B}_{-}\hat{A}_{-}\hat{B}_{3},
\quad \hat{B}_{-} \pa \hat{A}_{-},\quad  \hat{G}_{21}\hat{G}_{21}$  \\
\hline
$\frac{2(-4+k)}{(6+k)} $ &  $\hat{A}_{+} \pa \hat{B}_{-}, \quad \hat{B}_{-}\hat{A}_{3}\hat{A}_{+},\quad \hat{B}_{-}\hat{B}_{3}\hat{A}_{+},
\quad \hat{B}_{-} \pa \hat{A}_{+},\quad  \hat{G}_{11}\hat{G}_{11}$  \\
\hline
$\frac{16}{(6+k)}$ & $\hat{A}_{-}\hat{A}_{-}\hat{A}_{3},\quad \hat{A}_{-}\hat{A}_{-}\hat{B}_{3},
\quad \hat{A}_{-} \pa \hat{A}_{-}$   \\
\end{tabular}
\caption{The $U(1)$ charges for the spin-$3$ composite fields.
  } 
\end{table}

\begin{table}[ht]
\centering 
\begin{tabular}{|c||c| } 
\hline 
$U(1)$ charge & Composite fields of spin-$\frac{7}{2}$ \\ [0.5ex] 
\hline \hline 
$\frac{(4+3k)}{(6+k)}$  & $\hat{B}_{-} T_{+}^{(\frac{5}{2})},\quad \hat{B}_{-}\hat{B}_{-}\hat{G}_{22},
\quad \hat{B}_{-}\hat{G}_{11} \hat{A}_{-}, \quad \hat{B}_{-} \hat{G}_{21}\hat{A}_{3},\quad \hat{B}_{-}\hat{G}_{21}\hat{B}_{3}$,\\
& $\hat{B}_{-}\pa \hat{G}_{21},\quad \hat{G}_{21} \pa \hat{B}_{-}$
\\ 
\hline
$\frac{(-4+3k)}{(6+k)}$ & $\hat{B}_{-} U^{(\frac{5}{2})},\quad \hat{B}_{-}\hat{B}_{-}\hat{G}_{12},
\quad \hat{B}_{-}\hat{G}_{11} \hat{A}_{3}, \quad \hat{B}_{-} \hat{G}_{11}\hat{B}_{3},\quad \hat{B}_{-}\hat{G}_{21}\hat{A}_{+}$,\\
& $\hat{B}_{-}\pa \hat{G}_{11},\quad \hat{G}_{11} \pa \hat{B}_{-}$
\\
\hline
$\frac{(12+k)}{(6+k)}$ &  $\hat{A}_{-} \pa \hat{G}_{21},\quad \hat{B}_{-}\hat{A}_{-}\hat{G}_{22},
\quad \hat{G}_{11}\hat{A}_{-} \hat{A}_{-}, \quad \hat{G}_{21} \hat{A}_{-}\hat{A}_{3},\quad \hat{G}_{21}\hat{A}_{-}\hat{B}_{3}$,\\
& $\hat{G}_{21}\pa \hat{A}_{-},\quad T_{+}^{(\frac{5}{2})} \hat{A}_{-}$
\\
\hline
$\frac{(-12+k)}{(6+k)} $ &  $\hat{A}_{+} \pa \hat{G}_{11},\quad \hat{B}_{-}\hat{A}_{+}\hat{G}_{12},
\quad \hat{G}_{11}\hat{A}_{3} \hat{A}_{+}, \quad \hat{G}_{11} \hat{B}_{3}\hat{A}_{+},\quad \hat{G}_{11} \pa \hat{A}_{+}$,\\
\end{tabular}
\caption{The $U(1)$ charges for the spin-$\frac{7}{2}$ composite fields.
  } 
\end{table}

\begin{table}[ht]
\centering 
\begin{tabular}{|c||c| } 
\hline 
$U(1)$ charge & Composite fields of spin-$4$
\\ [0.5ex] 
\hline \hline 
$\frac{2k}{(6+k)}$  & $
\hat{A}_3  \hat{A}_{3} \pa \hat{B}_{-},
 \hat{A}_3\hat{B}_{3}\pa  \hat{B}_{-},
\hat{A}_{3} \pa^2 \hat{B}_{-},
\hat{A}_{-} \hat{A}_{+} \pa \hat{B}_{-},
 \hat{B}_{3} \hat{B}_{3}\pa  \hat{B}_{-},
\hat{B}_{3} \pa^2 \hat{B}_{-},
\hat{B}_{-} T^{(3)} $, \\
&  $
\hat{B}_{-}  W^{(3)},
 \hat{B}_{-}\hat{A}_{3} T^{(2)},
\hat{B}_{-} \hat{A}_{3} \hat{T},
\hat{B}_{-} \hat{A}_{3} \hat{A}_3  \hat{A}_{3},
\hat{B}_{-} \hat{A}_{3} \hat{A}_3  \hat{B}_{3},
\hat{B}_{-} \hat{A}_{3} \hat{B}_3  \hat{B}_{3},
\hat{B}_{-} \hat{A}_{3} \pa \hat{A}_3  $, \\
&  $
\hat{B}_{-}  \hat{A}_{3} \pa \hat{B}_{3},
\hat{B}_{-} \hat{A}_{-} \hat{A}_3  \hat{A}_{+},
\hat{B}_{-} \hat{A}_{-} \hat{B}_3  \hat{A}_{+},
\hat{B}_{-} \hat{A}_{-}\pa  \hat{A}_{+},
\hat{B}_{-} \hat{A}_{+} \pa \hat{A}_{-},
\hat{B}_{-} \hat{B}_{3} T^{(2)},
\hat{B}_{-} \hat{B}_{3} \hat{T}  $, \\
&  $
\hat{B}_{-}  \hat{B}_{3} \hat{B}_3 \hat{B}_{3},
\hat{B}_{-} \hat{B}_{3} \pa \hat{A}_3,
\hat{B}_{-} \hat{B}_{3} \pa \hat{B}_3,
\hat{B}_{-} \hat{B}_{-}\hat{A}_3  \hat{B}_{+},
\hat{B}_{-} \hat{B}_{-}\hat{B}_3  \hat{B}_{+},
\hat{B}_{-} \hat{B}_{-} \pa \hat{B}_{+} $, \\
&  $
\hat{B}_{-} \hat{B}_{+} \pa \hat{B}_{-},
\hat{B}_{-} \hat{G}_{11}  \hat{G}_{22},
\hat{B}_{-} \hat{G}_{21}  \hat{G}_{12},
\hat{B}_{-} \pa T^{(2)},
\hat{B}_{-} \pa \hat{T},
\hat{B}_{-} \pa^2 \hat{A}_3,
\hat{B}_{-} \pa^2 \hat{B}_3,
\hat{G}_{11} \hat{G}_{11}  \hat{A}_{-}  $, \\
&  $
\hat{G}_{11} \pa \hat{G}_{21},
\hat{G}_{21}U^{(\frac{5}{2})},
\hat{G}_{21} \hat{G}_{11}  \hat{A}_{3},
\hat{G}_{21} \hat{G}_{11}  \hat{B}_{3},
\hat{G}_{21} \hat{G}_{21}  \hat{A}_{+},
\hat{G}_{21} \pa \hat{G}_{11},
T^{(2)} \pa \hat{B}_{-}$, \\
\hline
\end{tabular}
\caption{The $U(1)$ charges for the spin-$4$ composite fields.
 } 
\end{table}

\begin{table}[ht]
\centering 
\begin{tabular}{|c||c| } 
\hline 
$U(1)$ charge & Composite fields of spin-$4$
\\ [0.5ex] 
\hline \hline 
$-\frac{8}{(8+k)}$
&
$\hat{A}_{3} U_{-}^{(3)},
\hat{A}_{3} \hat{A}_{3} \hat{A}_{3} \hat{A}_{+},
\hat{A}_{3} \hat{A}_{3} \hat{B}_{3} \hat{A}_{+},
\hat{A}_{3} \hat{A}_{3}\pa \hat{A}_{+},
\hat{A}_{3} \hat{A}_{+} \pa \hat{A}_{3}  ,
\hat{A}_{3} \hat{A}_{+} \pa \hat{B}_{3}$,\\
&
$\hat{A}_{3} \hat{B}_{3} \hat{B}_{3} \hat{A}_{+},
\hat{A}_{3} \hat{B}_{3}\pa \hat{A}_{+},
\hat{A}_{3} T^{(2)} \hat{A}_{+},
\hat{A}_{3} \hat{T} \hat{A}_{+},
\hat{A}_{3} \pa^2 \hat{A}_{+},
\hat{A}_{-} \hat{A}_{3} \hat{A}_{+} \hat{A}_{+}$,\\
&
$\hat{A}_{-} \hat{A}_{+} \pa \hat{A}_{+},
\hat{A}_{-} \hat{B}_{3}\hat{A}_{+} \hat{A}_{+},
\hat{A}_{+} \hat{A}_{+} \pa \hat{A}_{-},
\hat{A}_{+} \hat{B}_{+} \pa \hat{B}_{-},
\hat{A}_{+} \pa T^{(2)},
\hat{A}_{+} \pa \hat{T}$,\\
&
$\hat{A}_{+} \pa^2 \hat{A}_{3},
\hat{A}_{+} \pa^2 \hat{B}_{3},
\hat{B}_3 U_{-}^{(3)},
\hat{B}_{3} \hat{A}_{+} \pa \hat{A}_{3},
\hat{B}_{3} \hat{A}_{+} \pa \hat{B}_{3},
\hat{B}_{3} \hat{B}_{3} \hat{B}_{3} \hat{A}_{+},
\hat{B}_{3} \hat{B}_{3} \pa \hat{A}_{+}$,\\
&
$\hat{B}_{3} T^{(2)} \hat{A}_{+},
\hat{B}_{3} \hat{T} \hat{A}_{+},
\hat{B}_3 \pa^2 \hat{A}_{+},
\hat{B}_{-} \hat{A}_{3} \hat{A}_{+}\hat{B}_{+},
\hat{B}_{-} \hat{A}_{+} \pa \hat{B}_{+},
\hat{B}_{-} \hat{B}_{3} \hat{A}_{+} \hat{B}_{+}$,\\
&
$\hat{B}_{-} \hat{B}_{+} \pa \hat{A}_{+},
\hat{B}_{-} \hat{G}_{12} \hat{G}_{12},
\hat{G}_{11} T_{-}^{(\frac{5}{2})},
\hat{G}_{11} \hat{A}_{3} \hat{G}_{12},
\hat{G}_{11} \hat{A}_{+} \hat{G}_{22},
\hat{G}_{11} \hat{B}_{3} \hat{G}_{12}$,\\
\end{tabular}
\caption{The $U(1)$ charges for the spin-$4$ composite fields.
 } 
\end{table}

\begin{table}[ht]
\centering 
\begin{tabular}{|c||c| } 
\hline 
$U(1)$ charge & Composite fields of spin-$\frac{9}{2}$
\\ [0.5ex] 
\hline \hline 
$\frac{(4+k)}{(6+k)}$
&
$
\hat{A}_3  \hat{A}_{3} \pa \hat{G}_{21},
\hat{A}_3  \hat{B}_{3} \pa \hat{G}_{21},
\hat{A}_3  \hat{G}_{22} \pa \hat{B}_{-},
\hat{A}_{3} \pa T_{+}^{(\frac{5}{2})},
 \hat{A}_{3} \pa^2 \hat{G}_{21},
\hat{A}_{-}\hat{A}_{3} \pa \hat{G}_{11} $, \\
&
$
\hat{A}_{-}  \hat{A}_{+} \pa \hat{G}_{21},
\hat{A}_{-}  \hat{B}_{3} \pa \hat{G}_{11},
\hat{A}_{-}  \hat{G}_{12} \pa \hat{B}_{-},
\hat{A}_{-} \pa U^{(\frac{5}{2})},
 \hat{A}_{-} \pa^2 \hat{G}_{11},
\hat{B}_{3}\hat{B}_{3} \pa \hat{G}_{21} $, \\
&
$
\hat{B}_{3}  \hat{G}_{22} \pa \hat{B}_{-},
\hat{B}_{3} \pa T_{+}^{(\frac{5}{2})},
\hat{B}_{3}  \pa^2 \hat{G}_{21},
\hat{B}_{-} V^{(\frac{7}{2})},
 \hat{B}_{-} \hat{A}_{3}V^{(\frac{5}{2})},
\hat{B}_{-}\hat{A}_{3} \hat{A}_{3} \hat{G}_{22},
 \hat{B}_{-}\hat{A}_{3} \hat{B}_{3} \hat{G}_{22}$, \\
 &
$
\hat{B}_{-}  \hat{A}_{3} \pa \hat{G}_{22},
\hat{B}_{-} \hat{A}_{-} T_{-}^{(\frac{5}{2})},
 \hat{B}_{-}\hat{A}_{-} \hat{A}_{3} \hat{G}_{12},
 \hat{B}_{-}\hat{A}_{-} \hat{A}_{+} \hat{G}_{22},
 \hat{B}_{-}\hat{A}_{-} \hat{B}_{3} \hat{G}_{12}$, \\
&
$
\hat{B}_{-}  \hat{A}_{+} \pa \hat{G}_{12},
\hat{B}_{-} \hat{B}_{3} V^{(\frac{5}{2})},
\hat{B}_{-}\hat{B}_{3} \hat{B}_{3} \hat{G}_{22},
\hat{B}_{-} \hat{B}_{3} \pa \hat{G}_{22},
\hat{B}_{-}\hat{B}_{-} \hat{G}_{22} \hat{B}_{+},
\hat{B}_{-}\hat{B}_{+} \pa \hat{G}_{21}$, \\
&
$
\hat{B}_{-}  \hat{G}_{11}\hat{A}_{-} \hat{B}_{+},
\hat{B}_{-} \hat{G}_{12} \pa \hat{A}_{-},
\hat{B}_{-}\hat{G}_{21} \hat{A}_{3} \hat{B}_{+},
\hat{B}_{-}\hat{G}_{21} \hat{B}_{3} \hat{B}_{+},
\hat{B}_{-}\hat{G}_{21} \pa\hat{B}_{+},
\hat{B}_{-}\hat{G}_{22} \pa \hat{A}_{3}$, \\
\hline
\end{tabular}
\caption{The $U(1)$ charges for the spin-$\frac{9}{2}$ composite fields.
 } 
\end{table}

\begin{table}[ht]
\centering 
\begin{tabular}{|c||c| } 
\hline 
$U(1)$ charge & Composite fields of spin-$\frac{9}{2}$
\\ [0.5ex] 
\hline \hline 
$-\frac{(-4+k)}{(6+k)}$
&
$
\hat{A}_3 V^{(\frac{7}{2})},
\hat{A}_3  \hat{A}_{3} V^{(\frac{5}{2})},
\hat{A}_3  \hat{A}_{3} \hat{A}_3\hat{G}_{22},
\hat{A}_3  \hat{A}_{3} \hat{B}_3\hat{G}_{22},
 \hat{A}_{3} \hat{A}_3\pa \hat{G}_{22},
\hat{A}_{3} \hat{B}_3 V^{(\frac{5}{2})},
\hat{A}_{3}\hat{B}_{3} \hat{B}_{3}\hat{G}_{22} $, \\
&
$
\hat{A}_{3}  \hat{B}_{3} \pa \hat{G}_{22},
\hat{A}_{3}  \hat{B}_{+} \pa \hat{G}_{21},
\hat{A}_{3}  \hat{G}_{12} \pa \hat{A}_{-},
\hat{A}_{3} \hat{G}_{22}\pa \hat{A}_3,
\hat{A}_{3} \hat{G}_{22}\pa \hat{B}_3,
\hat{A}_{3}T^{(2)} \hat{G}_{22} $, \\
&
$
\hat{A}_{3}  \hat{T} \hat{G}_{22},
\hat{A}_{3} \pa V^{(\frac{5}{2})},
\hat{A}_{3}  \pa^2 \hat{G}_{22},
\hat{A}_{-} W_{-}^{(\frac{7}{2})},
 \hat{A}_{-} \hat{A}_{3}T_{-}^{(\frac{5}{2})},
\hat{A}_{-}\hat{A}_{3} \hat{A}_{3} \hat{G}_{12},
 \hat{A}_{-}\hat{A}_{3} \hat{A}_{+} \hat{G}_{22}$, \\
 &
$
\hat{A}_{-}  \hat{A}_{3}\hat{B}_3 \hat{G}_{12},
\hat{A}_{-} \hat{A}_{3}\pa \hat{G}_{12},
 \hat{A}_{-}\hat{A}_{-} \hat{A}_{+} \hat{G}_{12},
 \hat{A}_{-}\hat{A}_{+} V^{(\frac{5}{2})},
 \hat{A}_{-}\hat{A}_{+} \pa \hat{G}_{22}$, \\
&
$
\hat{A}_{-}  \hat{B}_{3} T_{-}^{(\frac{5}{2})},
\hat{A}_{-} \hat{B}_{3}\hat{A}_{+} \hat{G}_{22},
\hat{A}_{-}\hat{B}_{3} \hat{B}_{3} \hat{G}_{12},
\hat{A}_{-} \hat{B}_{3} \pa \hat{G}_{12},
\hat{A}_{-}\hat{B}_{+} \pa \hat{G}_{11},
\hat{A}_{-}\hat{G}_{12} \pa \hat{A}_{3}$, \\
&
$
\hat{A}_{-}  \hat{G}_{12}\pa \hat{B}_{3},
\hat{A}_{-} \hat{G}_{22} \pa \hat{A}_{+},
\hat{A}_{-}T^{(2)} \hat{G}_{12},
\hat{A}_{-}\hat{T} \hat{G}_{12},
\hat{A}_{-}\pa T_{-}^{(\frac{5}{2})},
\hat{A}_{-}\pa^2 \hat{G}_{12} $, \\
\hline
\end{tabular}
\caption{The $U(1)$ charges for the spin-$\frac{9}{2}$ composite fields.
 } 
\end{table}

\section{Conclusions and outlook }

The $16$ higher spin currents given in (\ref{16}) and $11$ currents of 
large ${\cal N}=4$ nonlinear algebra are determined explicitly in 
the Wolf space
coset model $\frac{SO(8)}{SO(4) \times SU(2) \times SU(2)}$. 
Some  OPEs of the extended large ${\cal N}=4$ nonlinear algebra
are described in Appendices $C$ and $D$. 
As in the unitary case \cite{Ahn1311}, 
the $(11+16)$ currents commute with 
the ${\cal N}=2$
WZW affine currents
living in the ${\cal N}=1$ subgroup $SO(4) \times SU(2)$ and the bosonic 
second subgroup $SU(2)$.  
Three of four spin-$\frac{1}{2}$ currents
live in the above bosonic second subgroup 
$SU(2)$ while one of them and spin-$1$ current 
live in the ${\cal N}=1$ first subgroup $SU(2)$.
The two $SU(2)$ affine algebras of large ${\cal N}=4$ nonlinear algebra
 are embedded in the ${\cal N}=4$ 
coset theory in a nontrivial way.
That is,
the $SU(2)_4$ affine algebra is realized by the Wolf space coset currents 
(\ref{b1b2b3nonlinear}) 
while
the $SU(2)_k$ affine algebra is realized by the currents (\ref{a1a2a3non}) 
from the Wolf space 
coset subgroup. 

One should determine the 
higher spin currents for general $N$ \footnote{In order to
do this job, it is useful to obtain the higher spin currents for low $N$ 
values $N=5,
8, 9, 12, 13$ 
where the corresponding $SO(N+4)$ group has even dimensions and 
one does not introduce the extra $U(1)$ factor in order to describe the 
${\cal N}=2 $ WZW
affine currents  in the 
complex basis. One expects that the structure constants appearing all 
the OPEs we have discussed in this paper depend on $N$-dependence 
explicitly and one determines the structure constants for general $N$
from the above low $N$ values results. Furthermore, one should 
calculate each pole term by hand rather than the mathematica package. }. 
This is because 
one should have the general $N$-dependence for the higher spin currents 
in order to 
calculate the various three point functions (for example,
\cite{Ahn1111,CHR1211,MZ1211,AK1308}).
Based on the present results for $N=4$ case, one should find their 
$N$-generalization. For the large ${\cal N}=4$ nonlinear 
superconformal algebra, one can find their general $N$ expressions 
in \cite{cqg1989}. It is an open problem to construct the higher spin currents
in terms of ${\cal N}=2$ WZW affine currents with general $N$. 
For the unitary case \cite{Ahn1311}, 
the spin of lowest higher spin current is given by
one. On the other hands, for the orthogonal case, the lowest spin of
the higher spin currents is given by two.  
The immediate question is why the spin-$1$ current (can be written in terms of
spin-$\frac{1}{2}$ currents and its superpartner spin-$1$ currents with
arbitrary tensorial coefficients) for the orthogonal Wolf 
space coset (for general $N$) is not allowed in the spin contents.
Somehow the various geometrical, tensorial coefficient functions
appearing in this spin-$1$ current satisfy nontrivial identities
and it will turn out that there exists no spin-$1$ current because all the 
coefficient functions vanish.  
Similarly, for the candidate of spin-$\frac{3}{2}$ Casimir 
current one should show that 
the coefficient functions in front of this current vanish. 
However, for the unitary Wolf space coset \cite{Ahn1311}, one should expect 
that there exists a nontrivial spin-$1$ current (and all the $15$ higher 
spin currents) for general $N$ with nontrivial
tensorial coefficient functions.
Once the spin-$1$ current for general $N$ is found, it is straightforward to
determine other higher spin currents. 

As in the unitary case \cite{Ahn1311}, 
it is an open problem to construct the 
next higher spin currents given in
\bea
 \left(4, \frac{9}{2}, \frac{9}{2}, 5 \right) & : & (P^{(4)}, 
P_{+}^{(\frac{9}{2})}, P_{-}^{(\frac{9}{2})}, P^{(5)}), \nonu \\
  \left(\frac{9}{2}, 5, 5, \frac{11}{2} \right) & : &
 (Q^{(\frac{9}{2})}, 
Q_{+}^{(5)}, Q_{-}^{(5)}, Q^{(\frac{11}{2})}), \nonu \\
 \left(\frac{9}{2}, 5, 5, \frac{11}{2} \right) & : &
 (R^{(\frac{9}{2})}, 
R_{+}^{(5)}, R_{-}^{(5)}, R^{(\frac{11}{2})}),  \nonu \\
 \left(5, \frac{11}{2}, \frac{11}{2}, 6 \right) & : &
 (S^{(5)}, 
S_{+}^{(\frac{11}{2})}, S_{-}^{(\frac{11}{2})}, S^{(6)}).
\label{next16}
\eea
In order to see this structure, one should consider the OPEs 
between the higher spin currents in (\ref{new16comp}) and themselves.
One expects that the explicit calculations on these constructions 
will be rather complicated because the spins become larger and the number of
corresponding WZW affine currents become bigger and bigger. 
Even in the unitary case \cite{Ahn1408}, the particular OPE between the 
higher spin currents has very complicated expression.
For the OPE between $W^{(4)}(z) \, W^{(4)}(w)$, one should 
determine the possible composite fields of spin $s=7$.
In this paper, we have considered the composite fields up to the spin
$s=\frac{9}{2}$ and the possible 
composite fields with $s=5, \frac{11}{2}, 6, 
\frac{13}{2}$ and $7$ should be classified according to their 
$U(1)$ charges.

It would be interesting to how one can obtain 
the higher spin algebra in $AdS_3$ bulk theory side.
From the higher spin contents in (\ref{new16comp}) and (\ref{next16}),
one expects that there are 
\bea
&& \mbox{Six fields of spin}  \,\, s  =  1,3,5, \cdots, (2n-1), \cdots,
\nonu \\
&& \mbox{Four fields of spin} \,\, s  =  
\frac{3}{2}, \frac{5}{2}, \frac{7}{2}, \cdots, (n+\frac{1}{2}), \cdots,
\nonu \\
&&  \mbox{Two fields of spin} \,\, s  =  2, 4, 6, \cdots, 2n, \cdots.
\label{fieldcon}
\eea
In \cite{GG1305}, the explicit construction using the oscillator formalism 
\cite{PV9806,PV9812} for 
spin-$1$ and four spin-$\frac{3}{2}$ currents living in the 
lowest ${\cal N}=4$ multiplet (as well as the currents of large ${\cal N}=4$
superconformal algebra) was found. In the present case, 
it is nontrivial to obtain the spin-$2$ current and 
four spin-$\frac{5}{2}$ currents (living in (\ref{new16comp})) 
using the method of higher spin algebra 
in the bulk. Furthermore, it is an open problem to construct 
the $16$ higher spin currents we have constructed in this paper using the
oscillator formalism in the context of higher spin algebra.  
One expects that it is nontrivial to express them in the primary basis where
all the higher spin fields (\ref{fieldcon}) 
are primary under the stress energy tensor.

So far, we described an extension of large ${\cal N}=4$ 
nonlinear superconformal
algebra. By incorporating both 
the four spin-$\frac{1}{2}$ currents and spin-$1$
current again, it is interesting to obtain an extension of 
large ${\cal N}=4$ linear superconformal algebra.    
One of the main results in \cite{GG1406} is that 
the small ${\cal N}=4$ superconformal algebra (spin-$2$ stress energy tensor, 
four spin-$\frac{3}{2}$ currents and three spin-$1$ currents) 
plays the important role 
in CFT side and $AdS_3$ bulk side. 
Then it is natural to consider the extension of small ${\cal N}=4$ 
superconformal algebra. It is known that the small ${\cal N}=4$ superconformal
algebra can be obtained from the large ${\cal N}=4$ linear 
superconformal algebra in several ways rather than the large ${\cal N}=4$
nonlinear superconformal algebra. 
In other words, it is not straightforward to decouple the half of 
spin-$1$ currents from the OPE in the two fermionic spin-$\frac{3}{2}$ 
currents with nonlinear terms in the spin-$1$ currents.

In the classification of \cite{ST}, there exists
the following coset 
\bea
\mbox{Wolf} \times SU(2) \times U(1) = \frac{Sp(N+2)}{Sp(N)} \times U(1).
\label{spcoset}
\eea
It would be interesting to obtain the higher spin currents 
in this symplectic 
Wolf space coset model. Note that $Sp(2)$ is isomorphic to 
$SU(2)$. One expects that the spin-$\frac{1}{2}$ current and spin-$1$ current
should live in $U(1)$ inside of $Sp(N)$
and the three spin-$\frac{1}{2}$ currents should be in $SU(2)$ subgroup
of Wolf space coset.
Also this $SU(2)$ group plays the role of $SU(2)_k$ affine algebra 
of large ${\cal N}=4$ nonlinear superconformal algebra.
This coset (\ref{spcoset}) 
looks similar to the orthogonal coset but the $SU(2)$ in the 
denominator is missing.
This is the reason why the ${\cal N}=1$ subgroup corresponding to
one spin-$1$ current and three spin-$\frac{1}{2}$ currents 
should be present inside of 
$Sp(N)$. 

\vspace{.7cm}

\centerline{\bf Acknowledgments}

We would like to thank H. Kim for discussions. 
This work was supported by the Mid-career Researcher Program through
the National Research Foundation of Korea (NRF) grant 
funded by the Korean government (MEST) 
(No. 2012-045385/2013-056327/2014-051185).
CA would like to thank the participants of the focus program of 
Asia Pacific Center
for Theoretical Physics (APCTP) 
on
``Liouville, Integrability and Branes (10) Focus Program at Asia-Pacific
Center for Theoretical Physics'',
Sept. 03-14, 2014 for their feedbacks.
We appreciate APCTP for its hospitality during completion of this work.
CA acknowledges warm hospitality from 
the School of  Liberal Arts (and Institute of Convergence Fundamental
Studies), Seoul National University of Science and Technology.

\newpage

\appendix

\renewcommand{\thesection}{\large \bf \mbox{Appendix~}\Alph{section}}
\renewcommand{\theequation}{\Alph{section}\mbox{.}\arabic{equation}}

\section{ The $SO(8)$ generators in the complex
  basis (or Cartan-Weyl basis)  }
A Lie algebra is described by a set of generators $ \{X_i\}$ and their commutation relations
\bea
\left[ X_i, X_j \right] &=& f_{ij}^{\,\,\,k} X_k.
\nonu
\eea
The number of generators is the dimension of the algebra and the constants $f_{ij}^{\,\,\,k}$ are structure constants.
In the standard Cartan-Weyl basis, the commutation relations between generators are summarized as follows \cite{Georgi,francesco,cahn}:
\bea
\left[H_i,\,\,H_j \right] & =& 0, \,\,\,\,H_i = H_i^{\dagger}\,\,\,\,\,\,\,\,\,\,\,\, i= 1,2,3 \cdots, l,
\nonu \\
\left[H_i,\,\, E_{\alpha}\right] & =&  \alpha_i E_{\alpha},
\nonu \\
\left[E_{\alpha},\,\, E_{-\alpha}\right] & =&  \alpha^{i} H_i,
\nonu \\
\left[E_{\alpha},\,\, E_{\beta}\right] &=&  N_{\alpha \beta}E_{\alpha+\beta}\mbox{ if $\alpha +\beta$ is a non zero root},
\nonu \\
&=& 0  \,\,\,\,\,\,\,\,\,\,\,\,\,\,\,\,\,\,\,\,\,\mbox{ if $\alpha +\beta$ is not a root}.
\nonu
\eea
The $H_i$ are elements of the Cartan generators of  the Lie group ($l$ is the rank of the algebra) and can be diagonalized simultaneously.
They also satisfy
$
\mbox{Tr}(H_i H_j)= k_D \delta_{ij},
$
where $k_D$ is some constant that depends on the representation and on the normalization of generators \cite{Georgi}.
The vector $\alpha=( \alpha_1,\alpha_2, \cdots, \alpha_l) $ in the $l$-dimensional space is called a root. Root components can be extracted
 from the nonzero eigenvalues of the $H_i$ in the adjoint representation \cite{francesco}.
The $E_{\pm \alpha}$ are corresponding raising and lowering operators for the weights
and satisfy
$
E_{\alpha} = \left(E_{-\alpha}\right)^{\dagger}$.
To find the complex basis of a Lie algebra with even dimension, one just needs to construct generators $ \widetilde{ H}_i$  which are
linear combinations of $H_i$ and satisfy
$\left[\widetilde{ H}_i, \widetilde{ H}_j \right] = 0$ where $ 
i= 1,2,3 \cdots, l$ and one can find its complex conjugated one for each
$\widetilde{H}_i$ where $i =1, 2, \cdots, \frac{l}{2}$. 

The $SO(N)$ generator can be realized as $N \times N$ matrix with the $(i,\,j)$
components as follows:
\bea
X^{ij}_{A}=X^{ij}_{(ab)} =i(\delta^{i}_a \delta^{j}_b-\delta^{j}_{a} \delta^{i}_b),
\label{SOgenerators}
\eea
where the generators satisfy the usual commutator relation
$\left[ X_A, \, X_B \right ]= f_{AB}^{\,\,\,\,C} X_C$ and $\mbox{Tr}(X_A X_B)=2\delta_{AB}$.
From (\ref{SOgenerators}), we find the 28 generators of  $SO(8)$ as follows:
\bea
X_1 & = &
\left(
\begin{array}{cccccccccc}
0 & i &0 &0 &0 &0 &0 &0  \\
-i& 0 &0 &0 &0 &0 &0 &0 \\
0 & 0 &0 &0 &0 &0 &0 &0\\
0 & 0 &0 & 0 & 0  &0 &0 &0\\
0 & 0 &0 & 0 & 0 &0 &0 &0\\
0 & 0 &0 & 0 & 0 &0 &0 &0\\
0 & 0 &0 & 0 & 0 &0 &0 &0\\
0 & 0 &0 & 0 & 0 &0 &0 &0\\
\end{array} \right),
X_2 =
\left(
\begin{array}{cccccccccc}
0 & 0 &i &0 &0 &0 &0 &0  \\
0& 0 &0 &0 &0 &0 &0 &0 \\
-i & 0 &0 &0 &0 &0 &0 &0\\
0 & 0 &0 & 0 & 0  &0 &0 &0\\
0 & 0 &0 & 0 & 0 &0 &0 &0\\
0 & 0 &0 & 0 & 0 &0 &0 &0\\
0 & 0 &0 & 0 & 0 &0 &0 &0\\
0 & 0 &0 & 0 & 0 &0 &0 &0\\
\end{array} \right),
\nonu \\
X_3  & = &
\left(
\begin{array}{cccccccccc}
0 & 0 &0 &i &0 &0 &0 &0  \\
0& 0 &0 &0 &0 &0 &0 &0 \\
0 & 0 &0 &0 &0 &0 &0 &0\\
-i & 0 &0 & 0 & 0  &0 &0 &0\\
0 & 0 &0 & 0 & 0 &0 &0 &0\\
0 & 0 &0 & 0 & 0 &0 &0 &0\\
0 & 0 &0 & 0 & 0 &0 &0 &0\\
0 & 0 &0 & 0 & 0 &0 &0 &0\\
\end{array} \right),
X_4 =
\left(
\begin{array}{cccccccccc}
0 & 0 &0 &0 &i &0 &0 &0  \\
0& 0 &0 &0 &0 &0 &0 &0 \\
0 & 0 &0 &0 &0 &0 &0 &0\\
0 & 0 &0 & 0 & 0  &0 &0 &0\\
-i & 0 &0 & 0 & 0 &0 &0 &0\\
0 & 0 &0 & 0 & 0 &0 &0 &0\\
0 & 0 &0 & 0 & 0 &0 &0 &0\\
0 & 0 &0 & 0 & 0 &0 &0 &0\\
\end{array} \right),
\nonu \\
X_5  & = &
\left(
\begin{array}{cccccccccc}
0 & 0 &0 &0 &0 &i &0 &0  \\
0& 0 &0 &0 &0 &0 &0 &0 \\
0 & 0 &0 &0 &0 &0 &0 &0\\
0 & 0 &0 & 0 & 0  &0 &0 &0\\
0 & 0 &0 & 0 & 0 &0 &0 &0\\
-i & 0 &0 & 0 & 0 &0 &0 &0\\
0 & 0 &0 & 0 & 0 &0 &0 &0\\
0 & 0 &0 & 0 & 0 &0 &0 &0\\
\end{array} \right),
X_6  =
\left(
\begin{array}{cccccccccc}
0 & 0 &0 &0 &0 &0 &i &0  \\
0& 0 &0 &0 &0 &0 &0 &0 \\
0 & 0 &0 &0 &0 &0 &0 &0\\
0 & 0 &0 & 0 & 0  &0 &0 &0\\
0 & 0 &0 & 0 & 0 &0 &0 &0\\
0 & 0 &0 & 0 & 0 &0 &0 &0\\
-i & 0 &0 & 0 & 0 &0 &0 &0\\
0 & 0 &0 & 0 & 0 &0 &0 &0\\
\end{array} \right),
\nonu \\
X_7 & = &
\left(
\begin{array}{cccccccccc}
0 & 0 &0 &0 &0 &0 &0 &i  \\
0& 0 &0 &0 &0 &0 &0 &0 \\
0 & 0 &0 &0 &0 &0 &0 &0\\
0 & 0 &0 & 0 & 0  &0 &0 &0\\
0 & 0 &0 & 0 & 0 &0 &0 &0\\
0 & 0 &0 & 0 & 0 &0 &0 &0\\
0 & 0 &0 & 0 & 0 &0 &0 &0\\
-i & 0 &0 & 0 & 0 &0 &0 &0\\
\end{array} \right),
X_{8} =
\left(
\begin{array}{cccccccccc}
0 & 0 &0 &0 &0 &0 &0 &0  \\
0& 0 &i &0 &0 &0 &0 &0 \\
0 & -i &0 &0 &0 &0 &0 &0\\
0 & 0 &0 & 0 & 0  &0 &0 &0\\
0 & 0 &0 & 0 & 0 &0 &0 &0\\
0 & 0 &0 & 0 & 0 &0 &0 &0\\
0 & 0 &0 & 0 & 0 &0 &0 &0\\
0 & 0 &0 & 0 & 0 &0 &0 &0\\
\end{array} \right),
\nonu \\
X_9 & = &
\left(
\begin{array}{cccccccccc}
0 & 0 &0 &0 &0 &0 &0 &0  \\
0& 0 &0 &i &0 &0 &0 &0 \\
0 & 0 &0 &0 &0 &0 &0 &0\\
0 & -i &0 & 0 & 0  &0 &0 &0\\
0 & 0 &0 & 0 & 0 &0 &0 &0\\
0 & 0 &0 & 0 & 0 &0 &0 &0\\
0 & 0 &0 & 0 & 0 &0 &0 &0\\
0 & 0 &0 & 0 & 0 &0 &0 &0\\
\end{array} \right),
X_{10} =
\left(
\begin{array}{cccccccccc}
0 & 0 &0 &0 &0 &0 &0 &0  \\
0& 0 &0 &0 &i &0 &0 &0 \\
0 & 0 &0 &0 &0 &0 &0 &0\\
0 & 0 &0 & 0 & 0  &0 &0 &0\\
0 & -i &0 & 0 & 0 &0 &0 &0\\
0 & 0 &0 & 0 & 0 &0 &0 &0\\
0 & 0 &0 & 0 & 0 &0 &0 &0\\
0 & 0 &0 & 0 & 0 &0 &0 &0\\
\end{array} \right),
\nonu \\
X_{11} & = &
\left(
\begin{array}{cccccccccc}
0 & 0 &0 &0 &0 &0 &0 &0  \\
0& 0 &0 &0 &0 &i &0 &0 \\
0 & 0 &0 &0 &0 &0 &0 &0\\
0 & 0 &0 & 0 & 0  &0 &0 &0\\
0 & 0 &0 & 0 & 0 &0 &0 &0\\
0 & -i &0 & 0 & 0 &0 &0 &0\\
0 & 0 &0 & 0 & 0 &0 &0 &0\\
0 & 0 &0 & 0 & 0 &0 &0 &0\\
\end{array} \right),
X_{12} =
\left(
\begin{array}{cccccccccc}
0 & 0 &0 &0 &0 &0 &0 &0  \\
0& 0 &0 &0 &0 &0 &i &0 \\
0 & 0 &0 &0 &0 &0 &0 &0\\
0 & 0 &0 & 0 & 0  &0 &0 &0\\
0 & 0 &0 & 0 & 0 &0 &0 &0\\
0 & 0 &0 & 0 & 0 &0 &0 &0\\
0 &-i &0 & 0 & 0 &0 &0 &0\\
0 & 0 &0 & 0 & 0 &0 &0 &0\\
\end{array} \right),
\nonu \\
X_{13} & = &
\left(
\begin{array}{cccccccccc}
0 & 0 &0 &0 &0 &0 &0 &0  \\
0& 0 &0 &0 &0 &0 &0 &i \\
0 & 0 &0 &0 &0 &0 &0 &0\\
0 & 0 &0 & 0 & 0  &0 &0 &0\\
0 & 0 &0 & 0 & 0 &0 &0 &0\\
0 & 0 &0 & 0 & 0 &0 &0 &0\\
0 & 0 &0 & 0 & 0 &0 &0 &0\\
0 & -i &0 & 0 & 0 &0 &0 &0\\
\end{array} \right),
X_{14} =
\left(
\begin{array}{cccccccccc}
0 & 0 &0 &0 &0 &0 &0 &0  \\
0& 0 &0 &0 &0 &0 &0 &0 \\
0 & 0 &0 &i &0 &0 &0 &0\\
0 & 0 &-i & 0 & 0  &0 &0 &0\\
0 & 0 &0 & 0 & 0 &0 &0 &0\\
0 & 0 &0 & 0 & 0 &0 &0 &0\\
0 & 0 &0 & 0 & 0 &0 &0 &0\\
0 & 0 &0 & 0 & 0 &0 &0 &0\\
\end{array} \right),
\nonu \\
X_{15} & = &
\left(
\begin{array}{cccccccccc}
0 & 0 &0 &0 &0 &0 &0 &0  \\
0 & 0 &0 &0 &0 &0 &0 &0 \\
0 & 0 &0 &0 &i &0 &0 &0\\
0 & 0 &0 & 0 & 0  &0 &0 &0\\
0 & 0 &-i & 0 & 0 &0 &0 &0\\
0 & 0 &0 & 0 & 0 &0 &0 &0\\
0 & 0 &0 & 0 & 0 &0 &0 &0\\
0 & 0 &0 & 0 & 0 &0 &0 &0\\
\end{array} \right),
X_{16} =
\left(
\begin{array}{cccccccccc}
0 & 0 &0 &0 &0 &0 &0 &0  \\
0 & 0 &0 &0 &0 &0 &0 &0 \\
0 & 0 &0 &0 &0 &i &0 &0\\
0 & 0 &0 & 0 & 0  &0 &0 &0\\
0 & 0 &0 & 0 & 0 &0 &0 &0\\
0 & 0 &-i & 0 & 0 &0 &0 &0\\
0 & 0 &0 & 0 & 0 &0 &0 &0\\
0 & 0 &0 & 0 & 0 &0 &0 &0\\
\end{array} \right),
\nonu \\
X_{17}  & = &
\left(
\begin{array}{cccccccccc}
0 & 0 &0 &0 &0 &0 &0 &0  \\
0 & 0 &0 &0 &0 &0 &0 &0 \\
0 & 0 &0 &0 &0 &0 &i &0\\
0 & 0 &0 & 0 & 0  &0 &0 &0\\
0 & 0 &0 & 0 & 0 &0 &0 &0\\
0 & 0 &0 & 0 & 0 &0 &0 &0\\
0 & 0 &-i & 0 & 0 &0 &0 &0\\
0 & 0 &0 & 0 & 0 &0 &0 &0\\
\end{array} \right),
X_{18} =
\left(
\begin{array}{cccccccccc}
0 & 0 &0 &0 &0 &0 &0 &0  \\
0 & 0 &0 &0 &0 &0 &0 &0 \\
0 & 0 &0 &0 &0 &0 &0 &i\\
0 & 0 &0 & 0 & 0  &0 &0 &0\\
0 & 0 &0 & 0 & 0 &0 &0 &0\\
0 & 0 &0 & 0 & 0 &0 &0 &0\\
0 & 0 &0 & 0 & 0 &0 &0 &0\\
0 & 0 &-i & 0 & 0 &0 &0 &0\\
\end{array} \right),
\nonu \\
X_{19}  & = &
\left(
\begin{array}{cccccccccc}
0 & 0 &0 &0 &0 &0 &0 &0  \\
0 & 0 &0 &0 &0 &0 &0 &0 \\
0 & 0 &0 &0 &0 &0 &0 &0\\
0 & 0 &0 & 0 & i  &0 &0 &0\\
0 & 0 &0 & -i & 0 &0 &0 &0\\
0 & 0 &0 & 0 & 0 &0 &0 &0\\
0 & 0 &0 & 0 & 0 &0 &0 &0\\
0 & 0 &0 & 0 & 0 &0 &0 &0\\
\end{array} \right),
X_{20} =
\left(
\begin{array}{cccccccccc}
0 & 0 &0 &0 &0 &0 &0 &0  \\
0 & 0 &0 &0 &0 &0 &0 &0 \\
0 & 0 &0 &0 &0 &0 &0 &0\\
0 & 0 &0 & 0 & 0  &i &0 &0\\
0 & 0 &0 & 0 & 0 &0 &0 &0\\
0 & 0 &0 & -i & 0 &0 &0 &0\\
0 & 0 &0 & 0 & 0 &0 &0 &0\\
0 & 0 &0 & 0 & 0 &0 &0 &0\\
\end{array} \right),
\nonu \\
X_{21} & = &
\left(
\begin{array}{cccccccccc}
0 & 0 &0 &0 &0 &0 &0 &0  \\
0 & 0 &0 &0 &0 &0 &0 &0 \\
0 & 0 &0 &0 &0 &0 &0 &0\\
0 & 0 &0 & 0 & 0  &0 &i &0\\
0 & 0 &0 & 0 & 0 &0 &0 &0\\
0 & 0 &0 & 0 & 0 &0 &0 &0\\
0 & 0 &0 & -i & 0 &0 &0 &0\\
0 & 0 &0 & 0 & 0 &0 &0 &0\\
\end{array} \right),
X_{22} =
\left(
\begin{array}{cccccccccc}
0 & 0 &0 &0 &0 &0 &0 &0  \\
0 & 0 &0 &0 &0 &0 &0 &0 \\
0 & 0 &0 &0 &0 &0 &0 &0\\
0 & 0 &0 & 0 & 0  &0 &0 &i\\
0 & 0 &0 & 0 & 0 &0 &0 &0\\
0 & 0 &0 & 0 & 0 &0 &0 &0\\
0 & 0 &0 & 0 & 0 &0 &0 &0\\
0 & 0 &0 & -i & 0 &0 &0 &0\\
\end{array} \right),
\nonu \\
X_{23} & = &
\left(
\begin{array}{cccccccccc}
0 & 0 &0 &0 &0 &0 &0 &0  \\
0 & 0 &0 &0 &0 &0 &0 &0 \\
0 & 0 &0 &0 &0 &0 &0 &0\\
0 & 0 &0 & 0 & 0  &0 &0 &0\\
0 & 0 &0 & 0 & 0 &i &0 &0\\
0 & 0 &0 & 0 & -i &0 &0 &0\\
0 & 0 &0 & 0 & 0 &0 &0 &0\\
0 & 0 &0 & 0 & 0 &0 &0 &0\\
\end{array} \right),
X_{24} =
\left(
\begin{array}{cccccccccc}
0 & 0 &0 &0 &0 &0 &0 &0  \\
0 & 0 &0 &0 &0 &0 &0 &0 \\
0 & 0 &0 &0 &0 &0 &0 &0\\
0 & 0 &0 & 0 & 0  &0 &0 &0\\
0 & 0 &0 & 0 & 0 &0 &i &0\\
0 & 0 &0 & 0 & 0 &0 &0 &0\\
0 & 0 &0 & 0 & -i &0 &0 &0\\
0 & 0 &0 & 0 & 0 &0 &0 &0\\
\end{array} \right),
\nonu \\
X_{25} & = &
\left(
\begin{array}{cccccccccc}
0 & 0 &0 &0 &0 &0 &0 &0  \\
0 & 0 &0 &0 &0 &0 &0 &0 \\
0 & 0 &0 &0 &0 &0 &0 &0\\
0 & 0 &0 & 0 & 0  &0 &0 &0\\
0 & 0 &0 & 0 & 0 &0 &0 &i\\
0 & 0 &0 & 0 & 0 &0 &0 &0\\
0 & 0 &0 & 0 & 0 &0 &0 &0\\
0 & 0 &0 & 0 & -i &0 &0 &0\\
\end{array} \right),
X_{26} =
\left(
\begin{array}{cccccccccc}
0 & 0 &0 &0 &0 &0 &0 &0  \\
0 & 0 &0 &0 &0 &0 &0 &0 \\
0 & 0 &0 &0 &0 &0 &0 &0\\
0 & 0 &0 & 0 & 0  &0 &0 &0\\
0 & 0 &0 & 0 & 0 &0 &0 &0\\
0 & 0 &0 & 0 & 0 &0 &i &0\\
0 & 0 &0 & 0 & 0 &-i &0 &0\\
0 & 0 &0 & 0 & 0 &0 &0 &0\\
\end{array} \right),
\nonu \\
X_{27} & = &
\left(
\begin{array}{cccccccccc}
0 & 0 &0 &0 &0 &0 &0 &0  \\
0 & 0 &0 &0 &0 &0 &0 &0 \\
0 & 0 &0 &0 &0 &0 &0 &0\\
0 & 0 &0 & 0 & 0  &0 &0 &0\\
0 & 0 &0 & 0 & 0 &0 &0 &0\\
0 & 0 &0 & 0 & 0 &0 &0 &i\\
0 & 0 &0 & 0 & 0 &0 &0 &0\\
0 & 0 &0 & 0 & 0 &-i &0 &0\\
\end{array} \right),
X_{28} =
\left(
\begin{array}{cccccccccc}
0 & 0 &0 &0 &0 &0 &0 &0  \\
0 & 0 &0 &0 &0 &0 &0 &0 \\
0 & 0 &0 &0 &0 &0 &0 &0\\
0 & 0 &0 & 0 & 0  &0 &0 &0\\
0 & 0 &0 & 0 & 0 &0 &0 &0\\
0 & 0 &0 & 0 & 0 &0 &0 &0\\
0 & 0 &0 & 0 & 0 &0 &0 &i\\
0 & 0 &0 & 0 & 0 &0 &-i &0\\
\end{array} \right).
\label{SO8originalgenerators}
\eea
The $SO(8)$ generators in (\ref{SO8originalgenerators})  are not expressed in the complex basis.
To express them in the complex basis, one should find the Cartan-Weyl basis  of the $SO(8)$.
We choose $X_1, X_{14}, X_{23}$ and $X_{28}$ to form the Cartan subalgebra.\footnote{
There are 104 other possibilities to form the Cartan subspace. For example,
one can use $X_1, X_{14}, X_{24}, X_{27}$ or $X_1, X_{14}, X_{25}, X_{26}$, or
$X_1, X_{15}, X_{20}, X_{28}$ and so on. }
We rearrange the $SO(8)$ generators in (\ref{SO8originalgenerators}) as follows:
\bea
\mathbf{\widetilde{X}} =\{ X_2, X_3, \cdots, X_{12},X_{13}, X_{15}, \cdots, X_{21},X_{22}, X_{24},X_{25},X_{26},X_{27},X_{1},X_{14},X_{23},X_{28}\}.
\nonu
\eea
The last four components of $\mathbf{\widetilde{X}}$ form the Cartan subspace.
Then we find the transformation matrix $S^{'}$ that diagonalizes
$X_{1},X_{14},X_{23}$ and $X_{28}$ simultaneously. It is expressed as
\bea
S^{'} &=&\mbox{diag($C,C,C,C$)},
\nonu
\eea
where
\bea
C &=& \frac{1}{\sqrt{2}}\left(
\begin{array}{cc}
i & -i \\
1 & 1 \\
\end{array} \right).
\nonu
\eea
Then we transform $\mathbf{\widetilde{X}}_i$ by the transformation $\mathbf{\widetilde{X}}_i\rightarrow Y_i=(S^{'})^{ \dagger}\mathbf{\widetilde{X}}_i S^{'}$
and find the structure constants with $Y_i$.
Next we find the adjoint representation  of $Y_i$. 
Then we write the  most general vector in the Cartan subspace as follows:
\bea
Z=aY^{adj}_{25}+bY^{adj}_{26}+cY^{adj}_{27}+dY^{adj}_{28}.
\nonu
\eea
Then we find the  matrix $S$ that makes $Z$ diagonalized by the transformation $Z^{'}=S^{\dagger}Z S$.
The $28 \times 28$ matrix $S$ is given by
\bea
S & = &
\left(
\begin{array}{ccccccc}
A_{1} & A_{2}  & 0  & 0 & 0 & 0 & 0   \\
A_{3}  & 0 & A_{4}   & 0 & 0 & 0 & 0   \\
0 & 0   & 0 & A_{5}  & A_{6} & 0 & 0   \\
0 & A_{7}  & A_{8}   & 0  & 0   & 0   & 0 \\
0 & 0   & 0 & A_{9}  & A_{10} & 0 & 0   \\
0 & 0   & 0 & 0  & 0 & A_{11} & 0   \\
0 & 0   & 0 & 0  & 0 & 0 & A_{12}   \\
\end{array} \right),
\label{diagonalizer2}
\eea
where the zeros $0$ stand for 
$4 \times 4$ zero matrices and 
the nonzero $4 \times 4$ matrix elements are given by
\bea
A_{1} & = &
\left(
\begin{array}{cccc}
-\frac{i}{2} & -\frac{1}{2}& 0 & 0   \\
\frac{i}{2} & -\frac{1}{2}& 0 & 0   \\
\frac{i}{2} & \frac{1}{2}& 0 & 0   \\
-\frac{i}{2} & \frac{1}{2}& 0 & 0   \\
\end{array} \right)= A_{5}= A_{7},
\,\,\,A_{2} =
\left(
\begin{array}{cccc}
 0 & 0 &-\frac{1}{2} & \frac{i}{2}  \\
 0 & 0 &-\frac{1}{2} & -\frac{i}{2}  \\
 0 & 0 &-\frac{1}{2} & \frac{i}{2}  \\
 0 & 0 &-\frac{1}{2} & -\frac{i}{2}  \\
\end{array} \right)=A_{8}=A_{10},
\nonu \\
A_{3} &=&
\left(
\begin{array}{cccc}
 0 & 0 &-\frac{i}{2} & -\frac{1}{2}  \\
 0 & 0 &\frac{i}{2} & -\frac{1}{2}  \\
 0 & 0 &\frac{i}{2} & \frac{1}{2}  \\
 0 & 0 &-\frac{i}{2} & \frac{1}{2}  \\
\end{array} \right) = A_{9},
\,\,\,\,\,\,\,\,\,\,\,\,\,\,\,\,\,A_{4} =
\left(
\begin{array}{cccc}
-\frac{1}{2} & \frac{i}{2}& 0 & 0   \\
-\frac{1}{2} & -\frac{i}{2}& 0 & 0   \\
-\frac{1}{2} & \frac{i}{2}& 0 & 0   \\
-\frac{1}{2} & -\frac{i}{2}& 0 & 0   \\
\end{array} \right) = A_{6},
\nonu \\
A_{11} &=&
\left(
\begin{array}{cccc}
 -\frac{i}{2} & -\frac{1}{2}& -\frac{1}{2} & \frac{i}{2}   \\
 \frac{i}{2} & -\frac{1}{2}& -\frac{1}{2} & -\frac{i}{2}   \\
 \frac{i}{2} & \frac{1}{2}& -\frac{1}{2} & \frac{i}{2}   \\
 -\frac{i}{2} & \frac{1}{2}& -\frac{1}{2} & -\frac{i}{2}   \\
\end{array} \right),
\,\,\,\,\,\,\,\,\,\,\,\,\,\,\,\,\,\,\,\,A_{12} =
\left(
\begin{array}{cccc}
1 & 0 & 0 & 0   \\
0 &1 & 0 & 0   \\
0 & 0 & 1 & 0   \\
0 &0 & 0 & 1   \\
\end{array} \right).
\nonu
\eea

The diagonalized matrix $Z^{'}$ with (\ref{diagonalizer2}) is expressed as
\bea
Z^{'}&=& S^{\dag} Z S
      = \mbox{diag} (B_1,B_2,B_3,B_4,B_5,B_6,B_7),
\nonu
\eea
where each matrix element is given by
\bea
B_1 &=& \mbox{diag} (a+b,-(a+b),-a+b,-(-a+b)),\nonu \\
B_2 &=& \mbox{diag} (a+c,-(a+c),-a+c,-(-a+c)),\nonu \\
B_3 &=& \mbox{diag} (b+c,-(b+c),-b+c,-(-b+c)),\nonu \\
B_4 &=& \mbox{diag} (a+d,-(a+d),-a+d,-(-a+d)),\nonu \\
B_5 &=& \mbox{diag} (b+d,-(b+d),-b+d,-(-b+d)),\nonu \\
B_6 &=& \mbox{diag} (c+d,-(c+d),-c+d,-(-c+d)),\nonu \\
B_7 &=& \mbox{diag} (0,0,0,0).
\label{zdiag}
\eea

From (\ref{zdiag}), one can find all the roots (eigenvalues) as follows
\footnote{ Usually, positive roots are roots whose the first non-vanishing element is positive.
In our case, positive roots are roots whose the last non-vanishing element is positive. }:
\bea
\alpha_1&=& a+b\equiv (1,1,0,0),\,\,\,\,\,\,\,\,\,\,\,\,\,\,-\alpha_1 = -a+b\equiv (-1,-1,0,0),
\nonu \\
\alpha_2&=& -a+b\equiv (-1,1,0,0),\,\,\,\,\,-\alpha_2 = -(-a+b)\equiv (1,-1,0,0),
\nonu \\
\alpha_3&=&  a+c\equiv (1,0,1,0),\,\,\,\,\,\,\,\,\,\,\,\,\,\,-\alpha_3 = -(a+c)\equiv (-1,0,-1,0),
\nonu \\
\alpha_4&=&  -a+c\equiv (-1,0,1,0),\,\,\,\,\,-\alpha_4 = -(-a+c)\equiv (1,0,-1,0),
\nonu \\
\alpha_5 &=&  b+c\equiv (0,1,1,0),\,\,\,\,\,\,\,\,\,\,\,\,\,\,-\alpha_5 = -(b+c)\equiv (0,-1,-1,0),
\nonu \\
\alpha_6 &=&  -b+c\equiv (0,-1,1,0),\,\,\,\,\,-\alpha_6 = -(-b+c)\equiv (0,1,-1,0),
\nonu \\
\alpha_7 &=&  a+d\equiv (1,0,0,1),\,\,\,\,\,\,\,\,\,\,\,\,\,\,-\alpha_7 = -(a+d)\equiv (-1,0,0,-1),
\nonu \\
\alpha_8 &=&  -a+d\equiv (-1,0,0,1),\,\,\,\,\,-\alpha_8 = -(-a+d)\equiv (1,0,0,-1),
\nonu \\
\alpha_9 &=&  b+d\equiv (0,1,0,1),\,\,\,\,\,\,\,\,\,\,\,\,\,\,-\alpha_9 =-( b+d)\equiv (0,-1,0,-1),
\nonu \\
\alpha_{10} &=&  -b+d\equiv (0,-1,0,1),\,\,\,\,\,-\alpha_{10} = -(-b+d)\equiv (0,1,0,-1),
\nonu \\
\alpha_{11} &=&  c+d\equiv (0,0,1,1),\,\,\,\,\,\,\,\,\,\,\,\,\,\,-\alpha_{11} = -(c+d)\equiv (0,0,-1,-1),
\nonu \\
\alpha_{12} &=&  -c+d\equiv (0,0,-1,1),\,\,\,\,\,-\alpha_{12} = -(-c+d)\equiv (0,0,1,-1),
\nonu \\
\alpha_{13} &=& -\alpha_{13}=\alpha_{14}=-\alpha_{14} = 0 \equiv (0,0,0,0).
\label{SO8roots}
\eea
Finally we find $E_{\pm \alpha}$ and $H_i$ by the following linear transformation:
\bea
\mathbf{E} &=&  S \,\,\mathbf{\widetilde{X}},
\nonu
\eea
where the left hand side is
\bea
\mathbf{E} &=& \{ E_{\alpha_1},E_{-\alpha_1},E_{\alpha_2},E_{-\alpha_2},\cdots,E_{\alpha_{12}},E_{-\alpha_{12}},H_1,H_2,H_3,H_4 \}.
\nonu 
\eea
Then we construct the $SO(8)$ generators, for our purpose,
in the complex basis as follows:
\bea
T_{A}&=& \{ E_{\alpha_1},E_{\alpha_2},\frac{1}{\sqrt{2}}(H_1+iH_2),E_{\alpha_{12}},\frac{1}{\sqrt{2}}(H_3+iH_4),E_{\alpha_{11}},E_{\alpha_{3}},E_{\alpha_{4}}
\nonu \\
&& E_{\alpha_{9}},E_{\alpha_{10}},E_{\alpha_{5}},E_{\alpha_{6}},E_{\alpha_{7}},E_{\alpha_{8}}\},
\nonu \\
T_{\bar{A}} &=& T_{A}^{\dagger}.
\nonu
\eea
Note that $T_3= \frac{1}{\sqrt{2}}( H_1 +i H_2)$ and $T_5= \frac{1}{\sqrt{2}}( H_3 +i H_4)$ where the four
Cartan generators are given \cite{Georgi}
\bea
H_1 & = &  \mbox{diag}(1,-1,0,0,0,0,0,0), \nonu \\
H_2  & = &
\mbox{diag}(0,0,1,-1,0,0,0,0), \nonu \\
H_3 & = &
\mbox{diag}(0,0,0,0,1,-1,0,0), \nonu \\
H_4  & = & 
\mbox{diag}(0,0,0,0,0,0,1,-1).
\nonu
\eea

We present the 14 generators $T_{A}$ of $SO(8)$ (the other 
$14$ generators are given by $T_{\bar{A}}=T_{A}^{\dagger}$) 
as follows:
\bea
T_1 & = &
\left(
\begin{array}{cccc|cccccc}
0 & 0 &0 &-1 &0 &0 &0 &0  \\
0& 0 &0 &0 &0 &0 &0 &0 \\
0 & 1 &0 &0 &0 &0 &0 &0\\
0 & 0 &0 & 0 & 0  &0 &0 &0\\
\hline
0 & 0 &0 & 0 & 0 &0 &0 &0\\
0 & 0 &0 & 0 & 0 &0 &0 &0\\
0 & 0 &0 & 0 & 0 &0 &0 &0\\
0 & 0 &0 & 0 & 0 &0 &0 &0\\
\end{array} \right),
T_2 =
\left(
\begin{array}{cccc|cccccc}
0 & 0 &0 &0 &0 &0 &0 &0  \\
0 & 0 &0 &-1 &0 &0 &0 &0 \\
1 & 0 &0 &0 &0 &0 &0 &0\\
0 & 0 &0 & 0 & 0  &0 &0 &0\\
\hline
0 & 0 &0 & 0 & 0 &0 &0 &0\\
0 & 0 &0 & 0 & 0 &0 &0 &0\\
0 & 0 &0 & 0 & 0 &0 &0 &0\\
0 & 0 &0 & 0 & 0 &0 &0 &0\\
\end{array} \right),
\nonu \\
T_3  & = &
\left(
\begin{array}{cccc|cccccc}
\frac{1}{\sqrt{2}} & 0 &0 &0 &0 &0 &0 &0  \\
0 & -\frac{1}{\sqrt{2}} &0 &0 &0 &0 &0 &0 \\
0 & 0 &\frac{i}{\sqrt{2}} &0 &0 &0 &0 &0\\
0 & 0 &0 & -\frac{i}{\sqrt{2}} & 0  &0 &0 &0\\
\hline
0 & 0 &0 & 0 & 0 &0 &0 &0\\
0 & 0 &0 & 0 & 0 &0 &0 &0\\
0 & 0 &0 & 0 & 0 &0 &0 &0\\
0 & 0 &0 & 0 & 0 &0 &0 &0\\
\end{array} \right),
T_4 =
\left(
\begin{array}{cccc|cccccc}
0 & 0 &0 &0 &0 &0 &0 &0  \\
0 & 0 &0 &0 &0 &0 &0 &0 \\
0 & 0 &0 &0 &0 &0 &0 &0\\
0 & 0 &0 & 0 & 0  &0 &0 &0\\
\hline
0 & 0 &0 & 0 & 0 &0 &0 &0\\
0 & 0 &0 & 0 & 0 &0 &0 &-1\\
0 & 0 &0 & 0 & 1 &0 &0 &0\\
0 & 0 &0 & 0 & 0 &0 &0 &0\\
\end{array} \right),
\nonu \\
T_5  & = &
\left(
\begin{array}{cccc|cccccc}
0 & 0 &0 &0 &0 &0 &0 &0  \\
0 & 0 &0 &0 &0 &0 &0 &0 \\
0 & 0 &0 &0 &0 &0 &0 &0\\
0 & 0 &0 & 0 & 0  &0 &0 &0\\
\hline
0 & 0 &0 & 0 & \frac{1}{\sqrt{2}} &0 &0 &0\\
0 & 0 &0 & 0 & 0 &-\frac{1}{\sqrt{2}} &0 &0\\
0 & 0 &0 & 0 & 0 &0 &\frac{i}{\sqrt{2}} &0\\
0 & 0 &0 & 0 & 0 &0 &0 &-\frac{i}{\sqrt{2}}\\
\end{array} \right),
T_6  =
\left(
\begin{array}{cccc|cccccc}
0 & 0 &0 &0 &0 &0 &0 &0  \\
0 & 0 &0 &0 &0 &0 &0 &0 \\
0 & 0 &0 &0 &0 &0 &0 &0\\
0 & 0 &0 & 0 & 0  &0 &0 &0\\
\hline
0 & 0 &0 & 0 & 0 &0 &0 &-1\\
0 & 0 &0 & 0 & 0 &0 &0 &0\\
0 & 0 &0 & 0 & 0 &1 &0 &0\\
0 & 0 &0 & 0 & 0 &0 &0 &0\\
\end{array} \right),
\nonu \\
T_7 & = &
\left(
\begin{array}{cccc|cccccc}
0 & 0 &0 &0 &0 &-1 &0 &0  \\
0 & 0 &0 &0 &0 &0 &0 &0 \\
0 & 0 &0 &0 &0 &0 &0 &0\\
0 & 0 &0 & 0 & 0  &0 &0 &0\\
\hline
0 & 1 &0 & 0 & 0 &0 &0 &0\\
0 & 0 &0 & 0 & 0 &0 &0 &0\\
0 & 0 &0 & 0 & 0 &0 &0 &0\\
0 & 0 &0 & 0 & 0 &0 &0 &0\\
\end{array} \right),
T_{8} =
\left(
\begin{array}{cccc|cccccc}
0 & 0 &0 &0 &0 &0 &0 &0  \\
0 & 0 &0 &0 &0 &-1 &0 &0 \\
0 & 0 &0 &0 &0 &0 &0 &0\\
0 & 0 &0 & 0 & 0  &0 &0 &0\\
\hline
1 & 0 &0 & 0 & 0 &0 &0 &0\\
0 & 0 &0 & 0 & 0 &0 &0 &0\\
0 & 0 &0 & 0 & 0 &0 &0 &0\\
0 & 0 &0 & 0 & 0 &0 &0 &0\\
\end{array} \right),
\nonu \\
T_9 & = &
\left(
\begin{array}{cccc|cccccc}
0 & 0 &0 &0 &0 &0 &0 &0  \\
0 & 0 &0 &0 &0 &0 &0 &0 \\
0 & 0 &0 &0 &0 &0 &0 &-1\\
0 & 0 &0 & 0 & 0  &0 &0 &0\\
\hline
0 & 0 &0 & 0 & 0 &0 &0 &0\\
0 & 0 &0 & 0 & 0 &0 &0 &0\\
0 & 0 &0 & 1 & 0 &0 &0 &0\\
0 & 0 &0 & 0 & 0 &0 &0 &0\\
\end{array} \right),
T_{10} =
\left(
\begin{array}{cccc|cccccc}
0 & 0 &0 &0 &0 &0 &0 &0  \\
0 & 0 &0 &0 &0 &0 &0 &0 \\
0 & 0 &0 &0 &0 &0 &0 &0\\
0 & 0 &0 & 0 & 0  &0 &0 &-1\\
\hline
0 & 0 &0 & 0 & 0 &0 &0 &0\\
0 & 0 &0 & 0 & 0 &0 &0 &0\\
0 & 0 &1 & 0 & 0 &0 &0 &0\\
0 & 0 &0 & 0 & 0 &0 &0 &0\\
\end{array} \right),
\nonu \\
T_{11} & = &
\left(
\begin{array}{cccc|cccccc}
0 & 0 &0 &0 &0 &0 &0 &0  \\
0 & 0 &0 &0 &0 &0 &0 &0 \\
0 & 0 &0 &0 &0 &-1 &0 &0\\
0 & 0 &0 & 0 & 0  &0 &0 &0\\
\hline
0 & 0 &0 & 1 & 0 &0 &0 &0\\
0 & 0 &0 & 0 & 0 &0 &0 &0\\
0 & 0 &0 & 0 & 0 &0 &0 &0\\
0 & 0 &0 & 0 & 0 &0 &0 &0\\
\end{array} \right),
T_{12} =
\left(
\begin{array}{cccc|cccccc}
0 & 0 &0 &0 &0 &0 &0 &0  \\
0 & 0 &0 &0 &0 &0 &0 &0 \\
0 & 0 &0 &0 &0 &0 &0 &0\\
0 & 0 &0 & 0 & 0  &-1 &0 &0\\
\hline
0 & 0 &1 & 0 & 0 &0 &0 &0\\
0 & 0 &0 & 0 & 0 &0 &0 &0\\
0 & 0 &0 & 0 & 0 &0 &0 &0\\
0 & 0 &0 & 0 & 0 &0 &0 &0\\
\end{array} \right),
\nonu \\
T_{13} & = &
\left(
\begin{array}{cccc|cccccc}
0 & 0 &0 &0 &0 &0 &0 &-1  \\
0 & 0 &0 &0 &0 &0 &0 &0 \\
0 & 0 &0 &0 &0 &0 &0 &0\\
0 & 0 &0 & 0 & 0  &0 &0 &0\\
\hline
0 & 0 &0 & 0 & 0 &0 &0 &0\\
0 & 0 &0 & 0 & 0 &0 &0 &0\\
0 & 1 &0 & 0 & 0 &0 &0 &0\\
0 & 0 &0 & 0 & 0 &0 &0 &0\\
\end{array} \right),
T_{14} =
\left(
\begin{array}{cccc|cccccc}
0 & 0 &0 &0 &0 &0 &0 &0  \\
0 & 0 &0 &0 &0 &0 &0 &-1 \\
0 & 0 &0 &0 &0 &0 &0 &0\\
0 & 0 &0 & 0 & 0  &0 &0 &0\\
\hline
0 & 0 &0 & 0 & 0 &0 &0 &0\\
0 & 0 &0 & 0 & 0 &0 &0 &0\\
1 & 0 &0 & 0 & 0 &0 &0 &0\\
0 & 0 &0 & 0 & 0 &0 &0 &0\\
\end{array} \right).
\label{Generators}
\eea
The 6 generators $T_m$ where $m=1, 2, \cdots, 6$ (and their conjugated ones) 
belong to the 
Wolf space subgroup $SO(4) \times SU(2) \times SU(2)$  of the $SO(8)$.
Among them, the ones with $1,2,3$ (and conjugated ones)
are located at the first $4 \times 4$ matrices
and the remaining ones with $4,5,6$ (and conjugated ones)
are in the last $4 \times 4$ matrices inside $8 \times 8$ matrices.
The 8 generators $T_a$ where $a=7, 8, \cdots, 14$ belong to the 
Wolf space coset $\frac{SO(8)}{SO(4) \times SU(2) \times SU(2)}$
and are located off diagonal $4 \times 4$ matrices.
Then the $8 \times 8$ matrices are classified as 
the first $4 \times 4$ matrices, the last $4 \times 4$ matrices and the
remaining two off  diagonal $4 \times 4$ matrices. 
For general $N$, the subgroup $SO(N)$ lives in the first 
$N \times N$ matrices and the remaining $SU(2) \times SU(2)$ lives in
the last $4 \times 4$ matrices.

The commutators between $T_A$, and $T_{\bar{A}}$ are given by 
\footnote{ Other commutators can be obtained from (\ref{Generators}).}
\bea
&&[T_{1},\,\,\,T_{\bar{1}}]=  [E_{\alpha_{1}},\,\,\,E_{-\alpha_{1}}]=H_{1}+ H_{2}=\frac{(1-i)}{\sqrt{2}} (T_3 + iT_{\bar{3}}),
\nonu \\
&&[T_{2},\,\,\,T_{\bar{2}}]= [E_{\alpha_{2}},\,\,\,E_{-\alpha_{2}}]= -H_{1}+ H_{2}=-\frac{(1+i)}{\sqrt{2}} (T_3- i T_{\bar{3}}),
\nonu \\
&&[T_{4},\,\,\,T_{\bar{4}}] = [E_{\alpha_{12}},\,\,\,E_{-\alpha_{12}}]= -H_{3}+ H_{4}=-\frac{(1+i)}{\sqrt{2}} (T_5- i T_{\bar{5}}),
\nonu \\
&&[T_{6},\,\,\,T_{\bar{6}}] =[E_{\alpha_{11}},\,\,\,E_{-\alpha_{11}}]=  H_{3}+ H_{4}=\frac{(1-i)}{\sqrt{2}} (T_5 + iT_{\bar{5}}),
\nonu \\
&&[T_{7},\,\,\,T_{\bar{7}}] = [E_{\alpha_{3}},\,\,\,E_{-\alpha_{3}}]= H_{1}+ H_{3}=\frac{1}{\sqrt{2}} (T_3 +  T_{\bar{3}}+T_5+ T_{\bar{5}}),
\nonu \\
&&[T_{8},\,\,\,T_{\bar{8}}]= [E_{\alpha_{4}},\,\,\,E_{-\alpha_{4}}]=- H_{1}+ H_{3}=\frac{1}{\sqrt{2}} (-T_3 - T_{\bar{3}}+T_5+  T_{\bar{5}}),
\nonu \\
&&[T_{9},\,\,\,T_{\bar{9}}] = [E_{\alpha_{9}},\,\,\,E_{-\alpha_{9}}]= H_{2}+ H_{4}=-\frac{i}{\sqrt{2}} (T_3 - T_{\bar{3}}+T_5-T_{\bar{5}}),
\nonu \\
&&[T_{10},\,\,\,T_{\bar{10}}] =[E_{\alpha_{10}},\,\,\,E_{-\alpha_{10}}]= - H_{2}+ H_{4}=\frac{i}{\sqrt{2}} (T_3 -  T_{\bar{3}}-T_5+ T_{\bar{5}}),
\nonu \\
&&[T_{11},\,\,\,T_{\bar{11}}] =[E_{\alpha_{5}},\,\,\,E_{-\alpha_{5}}]=  H_{2}+ H_{3}=\frac{1}{\sqrt{2}} (-iT_3 + i T_{\bar{3}}+T_5+  T_{\bar{5}}),
\nonu \\
&&[T_{12},\,\,\,T_{\bar{12}}]= [E_{\alpha_{6}},\,\,\,E_{-\alpha_{6}}]= -H_{2}+ H_{3}=\frac{1}{\sqrt{2}} (iT_3 - i T_{\bar{3}}+T_5+  T_{\bar{5}}),
\nonu \\
&&[T_{13},\,\,\,T_{\bar{13}}] = [E_{\alpha_{7}},\,\,\,E_{-\alpha_{7}}]= H_{1}+ H_{4}=\frac{1}{\sqrt{2}} (T_3 +  T_{\bar{3}}-iT_5+ i T_{\bar{5}}),
\nonu \\
&&[T_{14},\,\,\,T_{\bar{14}}] = [E_{\alpha_{8}},\,\,\,E_{-\alpha_{8}}]= -H_{1}+ H_{4}=\frac{1}{\sqrt{2}} (-T_3 - T_{\bar{3}}-iT_5+ iT_{\bar{5}}).
\label{su2algebra}
\eea
Each commutator in (\ref{su2algebra}) represents a $SU(2)$ subgroup of $SO(8)$ and the coefficients of $H_{i}$
in each algebra  gives the corresponding root vector (\ref{SO8roots}) 
\cite{Georgi}.
Moreover, $T_{1}, T_{\bar{1}}, T_{2}, T_{\bar{2}}$
with $\frac{(1-i)}{\sqrt{2}} (T_3 + iT_{\bar{3}})$, and $-\frac{(1+i)}{\sqrt{2}} (T_3- i T_{\bar{3}})$ forms the $SO(4)$ subgroup and
$T_{4}, T_{\bar{4}},T_{6}, T_{\bar{6}}$ with $-\frac{(1+i)}{\sqrt{2}} (T_5- i T_{\bar{5}})$ and $\frac{(1-i)}{\sqrt{2}} (T_5 + iT_{\bar{5}})$
 forms the $SU(2) \times SU(2)$ subgroup in the coset.

We present the nonzero structure constants of $SO(8)$ in Table 11
as follows.

\begin{table}[ht]
\centering 
\begin{tabular}{|c||c| } 
\hline 
 & \\
\mbox{ values}  
& $(A,B,C)$ indices of $f_{AB}^{\;\;\;\;C}$ \\ [1ex] 
\hline \hline 
$1$
& $(1,8,11),\,(1,14,9),\,(2,7,11),\,(2,13,9),\,(4,7,13),\,(4,8,14)\,(4,11,9)\,(4,12,10),$
\\
& $(7,14,6),\,(8,13,6),\,(10,1,13)\,(10,2,14),\,(11,10,6),\,(12,1,7),\,(12,2,8)\,(12,9,6),$
\\
& $(6,\bar{9},12),\,(6,\bar{10},11),\,(6,\bar{13},18),\,(6,\bar{14},7),\,(7,\bar{1},12)\,(8,\bar{2},12)\,(9,\bar{11},4),\,(9,\bar{13},2),$
\\
& $(9,\bar{14},1),\,(10,\bar{12},4)\,(11,\bar{7},2),\,(11,\bar{8},1),\,(13,\bar{1},10),\,(13,\bar{7},4)\,(14,\bar{2},10)\,(14,\bar{8},4),$
\\
& $(1,\bar{7},\bar{12}),\,(1,\bar{3},\bar{10}),\,(2,\bar{8},\bar{12}),\,(2,\bar{14},\bar{10}),\,(7,\bar{11},\bar{2}),\,(7,\bar{13},\bar{4}),\,(8,\bar{11},\bar{1})\,(8,\bar{14},\bar{4}),$
\\
& $(9,\bar{6},\bar{12}),(10,\bar{6},\bar{11}),\,(11,\bar{9},\bar{4}),\,(12,\bar{10},\bar{4})\,(13,\bar{6},\bar{8}),\,(13,\bar{9},\bar{2}),\,(14,\bar{6},\bar{7}),\,(14,\bar{9},\bar{1}),$
\\
& $(\bar{1},\bar{10},\bar{13}),\,(\bar{1},\bar{12},\bar{7}),\,(\bar{2},\bar{10},\bar{14}),\,(\bar{12},\bar{12},\bar{8}),\,(\bar{7},\bar{2},\bar{11}),\,(\bar{7},\bar{4},\bar{13})\,(\bar{8},\bar{1},\bar{11})\,(\bar{8},\bar{4},\bar{14}),$
\\
&  $(\bar{9},\bar{12},\bar{6}),\,(\bar{10},\bar{11},\bar{6})\,(\bar{11},\bar{4},\bar{9}),\,(\bar{12},\bar{4},\bar{10}),\,(\bar{13},\bar{2},\bar{9}),\,(\bar{13},\bar{8},\bar{6})\,(\bar{14},\bar{1},\bar{9})\,(\bar{14},\bar{7},\bar{6}),$
\\
& $(\bar{6},7,\bar{14}),\,(\bar{6},8,\bar{13}),\,(\bar{6},11,\bar{10}),\,(\bar{6},12,\bar{9}),\,(\bar{7},12,\bar{1}),\,(\bar{8},12,\bar{2})\,(\bar{9},1,\bar{14})\,(\bar{9},2,\bar{13}),$
\\
& $(\bar{9},4,\bar{11}),\,(\bar{10},4,\bar{12})\,(\bar{11},1,\bar{8}),\,(\bar{11},2,\bar{7}),\,(\bar{13},4,\bar{7}),\,(\bar{13},10,\bar{1})\,(\bar{14},4,\bar{8})\,(\bar{14},10,\bar{2}),$
\\
& $(\bar{1},9,14),\,(\bar{1},11,8),\,(\bar{2},9,13),\,(\bar{2},11,7),\,(\bar{4},9,11),\,(\bar{4},10,12)\,(\bar{4},13,7)\,(\bar{4},14,8),$
\\
& $(\bar{7},6,14),\,(\bar{8},6,13)\,(\bar{10},13,1),\,(\bar{10},14,2),\,(\bar{11},6,10),\,(\bar{12},6,9)\,(\bar{12},7,1)\,(\bar{12},8,2)$
\\ 
\hline
$\frac{1}{\sqrt{2}}$
& $(3,7,7),\,(3,13,13)\,(5,7,7)\,(5,8,8)\,(5,11,11),\,(5,12,12),\,(8,3,8),\,(14,3,14),$
\\
& $(7,\bar{7},3),\,(7,\bar{7},5)\,(8,\bar{3},8)\,(8,\bar{8},5)\,(11,\bar{11},5),\,(12,\bar{12},5),\,(13,\bar{13},3),\,(14,\bar{3},14),$
\\
& $(3,\bar{8},\bar{8}),\,(3,\bar{14},\bar{14}),\,(7,\bar{7},\bar{3})\,(7,\bar{7},\bar{5}),\,(8,\bar{8},\bar{5}),\,(11,\bar{11},\bar{5}),\,(12,\bar{12},\bar{5})\,(13,\bar{13},\bar{3}),$
\\
&  $(\bar{3},\bar{8},\bar{8}),\,(\bar{3},\bar{14},\bar{14})\,(\bar{7},\bar{3},\bar{7}),\,(\bar{7},\bar{5},\bar{7}),\,(\bar{8},\bar{5},\bar{8}),\,(\bar{11},\bar{5},\bar{11})\,(\bar{12},\bar{5},\bar{12})\,(\bar{13},\bar{3},\bar{13}),$
\\
& $(\bar{7},3,\bar{7}),\,(\bar{7},5,\bar{7})\,(\bar{8},5,\bar{8}),\,(\bar{8},8,\bar{3}),\,(\bar{11},5,\bar{11}),\,(\bar{12},5,\bar{12})\,(\bar{13},3,\bar{13})\,(\bar{14},14,\bar{3}),$
\\
& $(\bar{3},7,7),\,(\bar{3},13,13)\,(\bar{5},7,7),\,(\bar{5},8,8),\,(\bar{5},11,11),\,(\bar{5},12,12)\,(\bar{8},8,3)\,(\bar{14},14,3)$
\\
\hline
$\frac{i}{\sqrt{2}}$
& $(3,9,9),\,(3,11,11)\,(5,9,9)\,(5,10,10)\,(5,13,13),\,(5,14,14),\,(10,3,10),\,(12,3,12),$
\\
& $(9,\bar{3},9),\,(9,\bar{5},9)\,(10,\bar{5},10)\,(10,\bar{10},3)\,(11,\bar{3},11),\,(12,\bar{12},3),\,(13,\bar{5},13),\,(14,\bar{5},14),$
\\
& $(3,\bar{10},\bar{10}),\,(3,\bar{12},\bar{12})\,(9,\bar{9},\bar{3})\,(9,\bar{9},\bar{5})\,(10,\bar{10},\bar{5}),\,(11,\bar{11},\bar{3}),\,(13,\bar{13},\bar{5}),\,(14,\bar{14},\bar{5}),$
\\
&  $(\bar{3},\bar{9},\bar{9}),\,(\bar{3},\bar{11},\bar{11})\,(\bar{5},\bar{9},\bar{9}),\,(\bar{5},\bar{10},\bar{10}),\,(\bar{5},\bar{13},\bar{13}),\,(\bar{5},\bar{14},\bar{14})\,(\bar{10},\bar{3},\bar{10})\,(\bar{12},\bar{3},\bar{12}),$
\\
& $(\bar{9},3,\bar{9}),\,(\bar{9},5,\bar{9})\,(\bar{10},5,\bar{10}),\,(\bar{10},10,\bar{3}),\,(\bar{11},3,\bar{11}),\,(\bar{12},12,\bar{3})\,(\bar{13},5,\bar{13})\,(\bar{14},5,\bar{14}),$
\\
& $(\bar{3},10,10),\,(\bar{3},12,12)\,(\bar{9},9,3),\,(\bar{9},9,5),\,(\bar{10},10,5),\,(\bar{11},11,3)\,(\bar{13},13,5)\,(\bar{14},14,5)$
\\
\hline
$\frac{(1+i)}{\sqrt{2}} $
& $(3,1,1),\,(5,6,6),  \, (2,\bar{3},2),\,(4,\bar{5},4), \, 
(1,\bar{1},\bar{3}),\,(6,\bar{6},\bar{5}), 
(\bar{3},\bar{2},\bar{2}),\,(\bar{5},\bar{4},\bar{4}),
 $
\\
& $(\bar{1},3,\bar{1}),\,(\bar{6},5,\bar{6}), \,
(\bar{2},2,3),\,(\bar{4},4,5)
$
\\
\hline
$\frac{(1-i)}{\sqrt{2}}$
& $(2,3,2),\,(4,5,4), \, 
(1,\bar{1},3),\,(6,\bar{6},5),
\,
(3,\bar{2},\bar{2}),\,(5,\bar{4},\bar{4}),
\,
(\bar{1},\bar{3},\bar{1}),\,(\bar{6},\bar{5},\bar{6}),
$
\\
& $(\bar{2},2,\bar{3}),\,(\bar{4},4,\bar{5}), \, 
(\bar{3},1,1),\,(\bar{5},6,6)
$
\\
[1ex] 
\hline 
\end{tabular}
\caption{The nonzero structure constants of $SO(8)$.
  } 
\end{table}

\section{The current algebra for the ${\cal N}=2$ supersymmetric
WZW model on a group $SO(N+4)$ in terms of components of spin 
$s=\frac{1}{2}$ and $1$}
Most of the OPEs between the component fields 
for the ${\cal N}=2$ supersymmetric
WZW model on a group $SO(N+4)$ take the same 
forms for those OPEs of \cite{Ahn1206}.
Let us describe the 
 following OPEs which have the same behaviors to  the ones in 
\cite{Ahn1206}
\bea
K^m(z) \, \overline{D} K^n(w) & = & -\frac{1}{(z-w)} f_{\bar{m}
\bar{n}}^{\;\;\;\;\bar{p}} K^p(w) + \cdots,
\nonu \\
\overline{D} K^m (z) \, \overline{D} K^n(w) & =&   -\frac{1}{(z-w)} f_{\bar{m}
  \bar{n}}^
{\;\;\;\;\bar{p}} \overline{D} K^p(w) + \cdots,
\nonu \\
K^m(z) \, \overline{D} J^a(w) & = & -\frac{1}{(z-w)} f_{\bar{m}
  \bar{a}}^{\;\;\;\;\bar{b}}
J^b(w) + \cdots,
\nonu \\
\overline{D} K^m (z) \, J^a(w) & = & -\frac{1}{(z-w)} f_{\bar{m} \bar{a}}^
{\;\;\;\;\bar{b}} J^b(w) +\cdots,
\nonu \\
\overline{D} K^m(z) \, \overline{D} J^a(w) & = & -\frac{1}{(z-w)} f_{\bar{m}
  \bar{a}}^{\;\;\;\; \bar{b}} \overline{D} J^b(w) +\cdots,
\nonu \\
K^{\bar{m}} (z) \, D K^{\bar{n}} (w) & = & -\frac{1}{(z-w)} f_{m
n}^{\;\;\;\; p} K^{\bar{p}}(w) + \cdots,
\nonu \\
D K^{\bar{m}} (z) \, D K^{\bar{n}} (w) & = & -
\frac{1}{(z-w)} f_{m
 n}^
{\;\;\;\; p} D K^{\bar{p}} (w) + \cdots,
\nonu \\
K^{\bar{m}} (z) \, D J^{\bar{a}} (w) & = & -\frac{1}{(z-w)}
f_{m a}^{\;\;\;\; b}
J^{\bar{b}}(w) + \cdots,
\nonu \\
D K^{\bar{m}} (z) \, J^{\bar{a}} (w) & = &
-\frac{1}{(z-w)} f_{m a}^
{\;\;\;\; b} J^{\bar{b}} (w) +\cdots,
\nonu \\
D K^{\bar{m}} (z) \, D J^{\bar{a}} (w) & = &
-\frac{1}{(z-w)} f_{m
  a}^{\;\;\;\; b} D J^{\bar{b}} (w) +\cdots,
\nonu \\
K^m(z) \, K^{\bar{n}}(w) & = & -\frac{1}{(z-w)} (k+N+2) \delta^{m\bar{n}}+
\cdots,
\nonu \\
K^{m} (z) \, D K^{\bar{n}} (w) & = & -\frac{1}{(z-w)} f_{\bar{m}
n}^{\;\;\;\; \bar{p}} K^{p}(w) + \cdots,
\nonu \\
\overline{D} K^{m} (z) \, K^{\bar{n}} (w) & = & -
\frac{1}{(z-w)} f_{\bar{m}
 n}^
{\;\;\;\; p}  K^{\bar{p}} (w) + \cdots,
\nonu \\
\overline{D} J^{a} (z) \, D J^{\bar{b}} (w) & = & \frac{1}{(z-w)^2}
\frac{1}{2} \left[2(k+N+2) \delta^{a\bar{b}} +
f_{\bar{a} m}^{\;\;\;\;
\bar{c}} f_{b\bar{m}}^{\;\;\;\;c} + f_{\bar{a} c}^{\;\;\;\;\bar{m}}
f_{b\bar{c}}^{\;\;\;\; m}\right] \nonu \\
&  - &
\frac{1}{(z-w)} \left[ f_{\bar{a}
 b}^
{\;\;\;\; \bar{m}} \overline{D} K^{m} +f_{\bar{a}
 b}^
{\;\;\;\; m} D K^{\bar{m}}
 \right. \nonu \\
& + & \left.  \frac{1}{(k+N+2)} \left(
f_{\bar{a} m}^{\;\;\;\;\bar{c}} f_{b\bar{m}
}^{\;\;\;\;d} J^c J^{\bar{d}}  + f_{\bar{a}c}^{\;\;\;\;\bar{m}}
f_{b\bar{c}}^{\;\;\;\;n} K^m K^{\bar{n}} \right) \right] (w)  + \cdots,
\nonu \\
J^a(z) \, J^{\bar{b}}(w)
& = & -\frac{1}{(z-w)} (k+N+2) \delta^{a\bar{b}}+
\cdots,
\nonu \\
J^{a} (z) \, D J^{\bar{b}} (w) & = & -\frac{1}{(z-w)}
f_{\bar{a} b}^{\;\;\;\; \bar{m}}
K^{m}(w) + \cdots,
\nonu \\
\overline{D} J^{a} (z) \, J^{\bar{b}} (w) & = &
-\frac{1}{(z-w)} f_{\bar{a} b}^
{\;\;\;\; m} K^{\bar{m}} (w) +\cdots,
\nonu \\
\overline{D} K^{m} (z) \, J^{\bar{a}} (w) & = &
-\frac{1}{(z-w)} f_{\bar{m} a}^
{\;\;\;\; b} J^{\bar{b}} (w) +\cdots,
\nonu \\
J^{a} (z) \, D K^{\bar{m}} (w) & = & -\frac{1}{(z-w)}
f_{\bar{a} m}^{\;\;\;\; \bar{b}}
J^{b}(w) + \cdots.
\label{compexpression}
\eea
The following  OPEs have different behaviors from the ones in \cite{Ahn1206}
because, for the present case, the nonzero structure constants 
$f_{ab}^{\,\,\,\,m}$ and $f_{\bar{a} \bar{b}}^{\,\,\,\,\bar{m}}$ occur
\bea
\overline{D} K^{m} (z) \, D K^{\bar{n}} (w) & = & \frac{1}{(z-w)^2}
\frac{1}{2} \left[2(k+N+2) \delta^{m\bar{n}} + f_{\bar{m} p}^{\;\;\;\;
\bar{q}} f_{n\bar{p}}^{\;\;\;\;q}
+f_{\bar{m} a}^{\;\;\;\;
\bar{b}} f_{n\bar{a}}^{\;\;\;\;b}\right] \nonu \\
&  - &
\frac{1}{(z-w)} \left[ f_{\bar{m}
 n}^
{\;\;\;\; p} D K^{\bar{p}} +f_{\bar{m}
 n}^
{\;\;\;\; \bar{p}} \overline{D} K^{p} \right.
 \nonu \\
&  + & \left.
\frac{1}{(k+N+2)}( f_{\bar{m} p}^{\;\;\;\;\bar{q}} f_{n\bar{p}
}^{\;\;\;\;r} K^q K^{\bar{r}}
+f_{\bar{m} a}^{\;\;\;\;\bar{b}} f_{n\bar{a}
}^{\;\;\;\;c} J^b J^{\bar{c}})\right] (w)  + \cdots,
\nonu \\
\overline{D} K^{m} (z) \, D J^{\bar{a}} (w) & = &-
\frac{1}{(z-w)} \left[  f_{\bar{m}
 a}^
{\;\;\;\; \bar{b}} \overline{D} J^{b}+f_{\bar{m}
 a}^
{\;\;\;\; b} D J^{\bar{b}}
\right.
 \nonu \\
&  + & \left.
\frac{1}{(k+N+2)} (f_{\bar{m} p}^{\;\;\;\;\bar{q}} f_{a \bar{p}
}^{\;\;\;\;b} K^q J^{\bar{b}}
+f_{\bar{m} b}^{\;\;\;\;\bar{c}} f_{a \bar{b}
}^{\;\;\;\;p} J^c K^{\bar{p}}) \right] (w)  + \cdots,
\nonu \\
\overline{D} J^{a} (z) \, D K^{\bar{m}} (w) & = &   -
\frac{1}{(z-w)} \left[
f_{\bar{a}
 m}^
{\;\;\;\; \bar{b}} \overline{D} J^{b}
 +
 f_{\bar{a}
 m}^
{\;\;\;\; b} D J^{\bar{b}}
\right.
 \nonu \\
&  + & \left.
\frac{1}{(k+N+2)} (f_{\bar{a} b}^{\;\;\;\;\bar{n}} f_{m \bar{b}
}^{\;\;\;\;c} K^n J^{\bar{c}}
+f_{\bar{a} p}^{\;\;\;\;\bar{b}} f_{m \bar{p}
}^{\;\;\;\;n} J^b K^{\bar{n}}) \right] (w)  + \cdots,
\nonu \\
K^m(z) \, D J^{\bar{a}}(w) & = & -\frac{1}{(z-w)} f_{\bar{m}
a}^{\;\;\;\;\bar{b}} J^b(w) + \cdots,
\nonu \\
\overline{D} J^a (z) \, \overline{D} J^b(w) & =&   -\frac{1}{(z-w)} f_{\bar{a}
  \bar{b}}^
{\;\;\;\;\bar{m}} \overline{D} K^m(w) + \cdots,
\nonu \\
\overline{D} J^a (z) \, K^{\bar{m}}(w) & =&   -\frac{1}{(z-w)} f_{\bar{a}
  m}^
{\;\;\;\;b} J^{\bar{b}}(w) + \cdots,
\nonu \\
J^{\bar{a}} (z) \, D J^{\bar{b}}(w) & =&   -\frac{1}{(z-w)} f_{a
  b}^
{\;\;\;\;m}  K^{\bar{m}}(w) + \cdots,
\nonu \\
D J^{\bar{a}} (z) \,D J^{\bar{b}}(w) & =&   -\frac{1}{(z-w)} f_{a
  b}^
{\;\;\;\;m} D K^{\bar{m}}(w) + \cdots.
\label{compexpressionpaeng}
\eea
For simplicity, the notation
$|_{\theta=\bar{\theta}=0}$  acting on the  ${\cal N}=2$ superfields
\cite{Ahn1208} are ignored in (\ref{compexpression}) and 
(\ref{compexpressionpaeng}).
That is, $ D K^{\bar{m}}|_{\theta=\bar{\theta}=0}(z) \equiv
 D K^{\bar{m}}(z), \overline{D} K^m|_{\theta=\bar{\theta}=0}(z) \equiv
 \overline{D} K^m(z),  D J^{\bar{a}}|_{\theta=\bar{\theta}=0}(z) \equiv
 D J^{\bar{a}}(z)$, and $ \overline{D} J^a|_{\theta=\bar{\theta}=0}(z) \equiv
 \overline{D} J^a(z)$.

Therefore, the complete OPEs between the WZW affine currents 
with $s=\frac{1}{2}, 1$ are summarized by (\ref{compexpression}) and 
(\ref{compexpressionpaeng})
\footnote{
 Other spin-$1$ currents are not independent because 
of the following constraints. 
The nonlinear constraints, by taking $\theta, \bar{\theta}$ independent
terms \cite{HS,Ahn1206}, are given by
\bea
D K^m|_{\theta=\bar{\theta}=0}(z) & = &
-\frac{1}{2(k+N+2)} (f_{\bar{m} n}^{\;\;\;\;\bar{p}} K^n
K^p+f_{\bar{m} a}^{\;\;\;\;\bar{b}} J^a
J^b)|_{\theta=\bar{\theta}=0}(z),
\nonu \\
D J^a|_{\theta=\bar{\theta}=0}(z) & = &
-\frac{1}{(k+N+2)} f_{\bar{a} b}^{\;\;\;\;\bar{m}} (J^b K^m)|_{\theta=\bar{\theta}=0}(z),
\nonu \\
\overline{D} K^{\bar{m}}|_{\theta=\bar{\theta}=0}
(z)& =& -\frac{1}{2(k+N+2)} (f_{m \bar{n}}^{\;\;\;\;p} K^{\bar{n}}
K^{\bar{p}}+f_{m \bar{a}}^{\;\;\;\;b} J^{\bar{a}}
J^{\bar{b}})|_{\theta=\bar{\theta}=0}(z),
\nonu \\
\overline{D} J^{\bar{a}}|_{\theta=\bar{\theta}=0}(z) & = &
-\frac{1}{(k+N+2)}
f_{a \bar{b}}^{\;\;\;\;m} (J^{\bar{b}} K^{\bar{m}})|_{\theta=\bar{\theta}=0}(z).
\label{Constraints}
\eea
Compared to the unitary case, there exist the extra terms in the right hand 
side of first and third terms of (\ref{Constraints})
because the structure constants $f_{ab}^{\,\,\, m}$ and $f_{\bar{a} 
\bar{b}}^{\,\,\,\bar{m}}$
are nonvanishing from Appendix $A$. 
Furthermore, the spin-$\frac{3}{2}$ currents, 
the last components,  
can be expressed in terms of known 
independent spin-$\frac{1}{2}$ currents (and their
spin-$1$ superpartners) as in footnote $12$ of \cite{Ahn1311}.}.

\section{The OPEs between the six spin-$1$
currents and the higher spin currents}
In this section, we present the OPEs between the six spin-$1$ currents
of large ${\cal N}=4$ nonlinear superconformal algebra  
in section $2$ and the $16$ higher spin currents in section $3$.

\subsection{ The OPEs between six spin-$1$ currents and the higher spin
current of spins $(2, \frac{5}{2}, \frac{5}{2}, 3)$
}
The nontrivial OPEs between six spin-$1$ currents (\ref{nona1a2a3}) and 
(\ref{b1b2b3nonlinear}) 
and the higher spin
current $T_{+}^{(\frac{5}{2})}(w)$ (\ref{t+}) are
\bea
\left(
\begin{array}{c}
\hat{A}_{+}\\
\hat{B}_{+}\\
 \end{array} \right)(z)
 \, T_{+}^{(\frac{5}{2})}(w)
& = & -\frac{1}{(z-w)} \, i \,
\left(
\begin{array}{c}
U^{(\frac{5}{2})}\\
V^{(\frac{5}{2})}\\
\end{array} \right)(w) +\cdots,
\label{a+t+5half} \\
\hat{A}_3(z) \, T_{+}^{(\frac{5}{2})}(w)
& = &\frac{1}{(z-w)} \, \frac{i}{2} \,
T_{+}^{(\frac{5}{2})}(w) +\cdots=\hat{B}_3(z) \, T_{+}^{(\frac{5}{2})}(w).
\label{a3t+5half}
\eea
One can check the $U(1)$ charge conservation in these OPEs from Table $2$
or $4$.
There are no nonlinear terms in (\ref{a+t+5half}) or (\ref{a3t+5half}).

The nontrivial OPEs between six spin-$1$ currents (\ref{nona1a2a3}) 
and (\ref{b1b2b3nonlinear}) and the higher spin
current $T_{-}^{(\frac{5}{2})}(w)$ (\ref{t-})
as follows:
\bea
\left(
\begin{array}{c}
\hat{A}_{-}\\
\hat{B}_{-}\\
 \end{array} \right)(z) \, T_{-}^{(\frac{5}{2})}(w)
& = & \frac{1}{(z-w)} \, i \,
\left(
\begin{array}{c}
V^{(\frac{5}{2})}   \\
U^{(\frac{5}{2})}   \\
\end{array} \right)(w) +\cdots,
\label{a+t-5half} \\
\hat{A}_3(z) \, T_{-}^{(\frac{5}{2})}(w)
& = &-\frac{1}{(z-w)} \, \frac{i}{2} \,
T_{-}^{(\frac{5}{2})}(w) +\cdots=\hat{B}_3(z) \, T_{-}^{(\frac{5}{2})}(w).
\label{a3t-5half}
\eea
The $U(1)$ charge conservation in these OPEs can be 
checked from Table $2$ or $4$ also. 
No nonlinear terms in (\ref{a+t-5half}) or (\ref{a3t-5half}).

The OPEs between six spin-$1$ currents (\ref{nona1a2a3}) and 
(\ref{b1b2b3nonlinear}) and the higher spin
current $T^{(3)}(w)$ (\ref{t3}) are
as follows:
\bea
\hat{A}_{\pm}(z)   \, T^{(3)}(w) & = &
 \frac{1}{(z-w)}\, i
\left(
\begin{array}{c}
U^{(3)}_{-} \\
 V^{(3)}_{+} \\
\end{array} \right)(w)
+ \cdots,
\label{a+t3} \\
\hat{A}_3(z) \, T^{(3)}(w) & = &
\frac{1}{(z-w)^2} \, 2iT^{(2)}(w) +\cdots=\hat{B}_3(z) \, T^{(3)}(w),
\label{a3t3} \\
\hat{B}_{\pm} (z) \, T^{(3)}(w) & = &
 \frac{1}{(z-w)}\,
 i
\left(
\begin{array}{c}
V_{-}^{(3)} \\
U_{+}^{(3)} \\
\end{array} \right)(w) +\cdots.
\label{b+t3}
\eea
The $U(1)$ charge conservation in these OPEs are checked by Table $2$ or $5$.
There are no nonlinear terms in (\ref{a+t3}), (\ref{a3t3}) or (\ref{b+t3}).
There is no descendant field for the spin-$2$ field 
in the first order pole of (\ref{a3t3}).

Therefore, one obtains 
 the OPEs between six spin-$1$ currents and the higher spin
current of spins $(2, \frac{5}{2}, \frac{5}{2}, 3)$ explicitly.
There are no nonlinear terms in the right hand side of these OPEs.
Some of the 
higher spin currents corresponding to three ${\cal N}=2$ multiplets
(\ref{new16comp})
occur in the right hand sides of the OPEs.

\subsection{The OPEs between six spin $1$ currents and two higher spin
currents of spins $(\frac{5}{2}, 3, 3, \frac{7}{2})$}
One has  the following nontrivial OPEs between the spin-$1$ currents and the
spin-$\frac{5}{2}$ currents $U^{(\frac{5}{2})}(w)$ (\ref{u5half}) 
and $V^{(\frac{5}{2})}(w)$ (\ref{v5half}):
\bea
\hat{A}_{\mp} (z)   \,
\left(
\begin{array}{c}
U^{(\frac{5}{2})} \\
V^{(\frac{5}{2})} \\
\end{array} \right)(w) & = &
\frac{1}{(z-w)} \, i \, \left(
\begin{array}{c}
-T_{+}^{(\frac{5}{2})}\\
T_{-}^{(\frac{5}{2})} \\
\end{array} \right)(w)+\cdots,
\label{a+5half} \\
\hat{A}_3(z) \,
\left(
\begin{array}{c}
U^{(\frac{5}{2})} \\
V^{(\frac{5}{2})} \\
\end{array} \right)(w)
 & = &
\frac{1}{(z-w)} \, \frac{i}{2} \,
\left(
\begin{array}{c}
-U^{(\frac{5}{2})} \\
V^{(\frac{5}{2})} \\
\end{array} \right)(w) +\cdots =
-\hat{B}_3(z) \,
\left(
\begin{array}{c}
U^{(\frac{5}{2})} \\
V^{(\frac{5}{2})} \\
\end{array} \right)(w),
\label{a35half} \\
\hat{B}_{\pm} (z) \,
\left(
\begin{array}{c}
U^{(\frac{5}{2})} \\
V^{(\frac{5}{2})} \\
\end{array} \right)(w)
 & = &
\frac{1}{(z-w)} \, i \, \left(
\begin{array}{c}
T_{-}^{(\frac{5}{2})} \\
-T_{+}^{(\frac{5}{2})}\\
\end{array} \right)(w)+\cdots.
\label{b+5half}
\eea
The $U(1)$ charge conservation in these OPEs
(\ref{a+5half}), (\ref{a35half}) and (\ref{b+5half})
can be checked from Table $2$ or $4$.

As performed before, one has the nontrivial OPEs between the spin-$1$ currents and the
spin-$3$ currents, $U_{\pm}^{(3)}(w)$ and $V_{\pm}^{(3)}(w)$ in (\ref{u3+}), 
(\ref{u3-}), (\ref{v3+}) and (\ref{v3-}) as follows:
\bea
\hat{B}_{\pm} (z) \,
\left(
\begin{array}{c}
U_{+}^{(3)} \\
V_{-}^{(3)}
\end{array} \right)
(w) & = &
\mp\frac{1}{(z-w)^2}\,4i\,T^{(2)}(w)
+\frac{1}{(z-w)}\,i(T^{(3)}-W^{(3)})(w)+\cdots,
\label{b+u+3} \\
\hat{B}_3(z) \,
\left(
\begin{array}{c}
U_{+}^{(3)} \\
V_{-}^{(3)}
\end{array} \right)(w)
 & = &
\frac{1}{(z-w)}\,i \left(
\begin{array}{c}
U_{+}^{(3)} \\
-V_{-}^{(3)}
\end{array} \right)(w) +\cdots,
\label{b3u+3}
\\
\hat{A}_{\mp} (z) \,
\left(
\begin{array}{c}
U_{-}^{(3)} \\
V_{+}^{(3)}
\end{array} \right)
(w) & = &
\pm\frac{1}{(z-w)^2}\,4i\,T^{(2)}(w)
+\frac{1}{(z-w)}\,i(T^{(3)}+W^{(3)})(w)\cdots,
\label{a-u-3} \\
\hat{A}_3(z) \,
\left(
\begin{array}{c}
U_{-}^{(3)} \\
V_{+}^{(3)}
\end{array} \right)(w)
 & = &
\frac{1}{(z-w)}\,i \left(
\begin{array}{c}
-U_{-}^{(3)} \\
V_{+}^{(3)}
\end{array} \right)(w) +\cdots.
\label{a+u-3}
\eea
One can check the $U(1)$ charge conservation in these OPEs 
(\ref{b+u+3}), (\ref{b3u+3}), (\ref{a-u-3}) and (\ref{a+u-3}) from Table
$2$ or $5$.
There is no descendant field for the spin-$2$ field 
in the first order pole of (\ref{b+u+3}) or (\ref{a-u-3}).

The following nontrivial OPEs between the spin-$1$ currents and the spin-$\frac{7}{2}$
current $U^{(\frac{7}{2})}(w)$ (\ref{u7half}) can be obtained as
\bea
\hat{A}_{+} (z) \, U^{(\frac{7}{2})}(w) & = &
\frac{1}{(z-w)} \left[
\frac{2}{(6+k)} \hat{A}_{+}U^{(\frac{5}{2})} \right](w)
+\cdots,
\label{a+u7half} \\
\hat{A}_{-} (z) \, U^{(\frac{7}{2})}(w) & = &
\frac{1}{(z-w)^2} \left[
\frac{2i(71+13k)}{5(6+k)} T_{+}^{(\frac{5}{2})} \right](w)
\label{a-u7half}  \\
& + & \frac{1}{(z-w)} \left[ i W_{+}^{(\frac{7}{2})}
- \frac{2}{(6+k)} \left(  \hat{B}_{-} V^{(\frac{5}{2})}+U^{(\frac{5}{2})}\hat{A}_{-}
-\frac{3i}{5} \pa T_{+}^{(\frac{5}{2})}\right)\right](w) +  \cdots,
\nonu \\
\hat{A}_3(z) \, U^{(\frac{7}{2})}(w) & = &
\frac{1}{(z-w)^2} \left[ \frac{6i(11+3k)}{5(6+k)} U^{(\frac{5}{2})}
\right](w)
 -  \frac{1}{(z-w)} \, \frac{i}{2}
\, U^{(\frac{7}{2})}(w) +\cdots,
\label{a3u7half} \\
\hat{B}_{+} (z) \, U^{(\frac{7}{2})}(w) & = &
\frac{1}{(z-w)^2} \left[ \frac{8i(11+3k)}{5(6+k)} T_{-}^{(\frac{5}{2})} \right](w)
\label{b+u7half}  \\
& + & \frac{1}{(z-w)} \left[ i W_{-}^{(\frac{7}{2})}
+ \frac{2}{(6+k)} \left( 2 \hat{A}_{3} T_{-}^{(\frac{5}{2})}-U^{(\frac{5}{2})}\hat{B}_{+}
-\frac{3i}{5} \pa T_{-}^{(\frac{5}{2})}\right)\right](w) +  \cdots,
\nonu \\
\hat{B}_3(z) \, U^{(\frac{7}{2})}(w) & = &
\frac{1}{(z-w)^2} \left[\frac{4i(11+3k)}{5(6+k)} U^{(\frac{5}{2})} \right](w)
+ \frac{1}{(z-w)} \, \frac{i}{2}
\, U^{(\frac{7}{2})}(w) +\cdots.
\label{b3u7half}
\eea
One can check the $U(1)$ charge conservation in these OPEs from Table
$2$ or $6$. Nonlinear terms in (\ref{a+u7half}), (\ref{a-u7half}), 
(\ref{b+u7half}) (linear in the higher spin currents) appear.
In (\ref{a3u7half}) and (\ref{b3u7half}), only linear terms appear.
There is no descendant field for the spin-$\frac{5}{2}$ field 
in the first order pole of (\ref{a-u7half}), (\ref{a3u7half}), 
(\ref{b+u7half}) or (\ref{b3u7half}).
One can check  
in (\ref{a-u7half}) or (\ref{b+u7half}) whether 
the last terms (containing the composite fields)  
in the first order poles 
are  (quasi)primary fields. 

The nontrivial OPEs between the spin-$1$ currents and the spin-$\frac{7}{2}$
current $V^{(\frac{7}{2})}(w)$ (\ref{v7half}) are given by
\bea
\hat{A}_{+} (z) \, V^{(\frac{7}{2})}(w) & = &
\frac{1}{(z-w)^2} \left[
\frac{8i(15+2k)}{5(6+k)} T_{-}^{(\frac{5}{2})} \right](w)
\label{a+v7half}  \\
& + & \frac{1}{(z-w)} \left[ -i W_{-}^{(\frac{7}{2})}
+ \frac{2}{(6+k)} \left(  2\hat{B}_{3} T_{-}^{(\frac{5}{2})}-\hat{A}_{+}V^{(\frac{5}{2})}
+\frac{2i}{5} \pa T_{-}^{(\frac{5}{2})}\right)\right](w) +  \cdots,
\nonu \\
\hat{A}_3(z) \, V^{(\frac{7}{2})}(w) & = &
\frac{1}{(z-w)^2} \left[ \frac{4i(15+2k)}{5(6+k)} V^{(\frac{5}{2})} \right](w)
 + \frac{1}{(z-w)} \, \frac{i}{2}
\, V^{(\frac{7}{2})}(w) +\cdots,
\label{a3v7half} \\
\hat{B}_{+} (z) \, V^{(\frac{7}{2})}(w) & = &
\frac{1}{(z-w)} \left[
\frac{2}{(6+k)} V^{(\frac{5}{2})}\hat{B}_{+} \right](w) +\cdots,
\label{b+v7half}  \\
\hat{B}_{-} (z) \, V^{(\frac{7}{2})}(w) & = &
\frac{1}{(z-w)^2} \left[
\frac{6i(25+4k)}{5(6+k)} T_{+}^{(\frac{5}{2})} \right](w)
\label{b-v7half}  \\
& - & \frac{1}{(z-w)} \left[ i W_{+}^{(\frac{7}{2})}
+ \frac{2}{(6+k)} \left(  \hat{B}_{-} V^{(\frac{5}{2})}+U^{(\frac{5}{2})}\hat{A}_{-}
-\frac{3i}{5} \pa T_{+}^{(\frac{5}{2})}\right)\right](w) +  \cdots,
\nonu \\
\hat{B}_3(z) \, V^{(\frac{7}{2})}(w) & = &
\frac{1}{(z-w)^2} \left[
\frac{6i(15+2k)}{5(6+k)} V^{(\frac{5}{2})} \right](w)
- \frac{1}{(z-w)} \, \frac{i}{2}
\, V^{(\frac{7}{2})}(w) +\cdots.
\label{b3v7half}
\eea
One can check the $U(1)$ charge conservation in these OPEs from Table
$2$ or $6$. In (\ref{a+v7half}), (\ref{b+v7half}) and (\ref{b-v7half}),
the nonlinear terms (still linear in higher spin currents) 
occur while, in (\ref{a3v7half}) and (\ref{b3v7half}),
the linear terms occur.
There is no descendant field for the spin-$\frac{5}{2}$ field 
in the first order pole of (\ref{a+v7half}), (\ref{a3v7half}), 
(\ref{b-v7half}) or (\ref{b3v7half}).
One can check  
in (\ref{a+v7half}) or (\ref{b-v7half}) whether 
the last terms (containing the composite fields)  
in the first order poles 
are  (quasi)primary fields. 

Therefore, one obtains 
 the OPEs between six spin-$1$ currents and the higher spin
current of spins $(\frac{5}{2}, 3, 3, \frac{7}{2})$ explicitly.
Most of the 
higher spin currents corresponding to four ${\cal N}=2$ multiplets
(\ref{new16comp})
occur in the right hand sides of the OPEs.

\subsection{ The OPEs between six spin $1$ currents and the higher spin
current of spins $(3, \frac{7}{2}, \frac{7}{2}, 4)$}
The OPEs between the spin-$1$ currents and the spin-$3$ current $W^{(3)}(w)$
(\ref{w3})
can be described as
\bea
\hat{A}_{\pm}(z) \, W^{(3)}(w) & = &
\frac{1}{(z-w)} i
\left(
\begin{array}{c}
U_{-}^{(3)} \\
V_{+}^{(3)}
\end{array} \right)
(w) +  \cdots,
\label{a+w3} \\
\hat{A}_3(z) \, W^{(3)}(w) & = &
\frac{1}{(z-w)^2} \,2i T^{(2)}(w)
+\cdots = -\hat{B}_3(z) \, W^{(3)}(w),
\label{a3w3} \\
\hat{B}_{\pm}(z) \, W^{(3)}(w) & = &
-\frac{1}{(z-w)} i
\left(
\begin{array}{c}
V_{-}^{(3)} \\
U_{+}^{(3)}
\end{array} \right)(w)
+  \cdots.
\label{b+w3}
\eea
From Table $2$ or $5$, one can check the $U(1)$ charges in (\ref{a+w3}), 
(\ref{a3w3}) 
and (\ref{b+w3}).

The OPEs between the spin-$1$ currents and the
spin-$\frac{7}{2}$ current $W_{+}^{(\frac{7}{2})}(w)$ (\ref{w7+}) are as follows:
\bea
\hat{A}_{+}(z) \, W_{+}^{(\frac{7}{2})}(w) & = &
-\frac{1}{(z-w)^2} \left[
\frac{2i(63+13k)}{5(6+k)} U^{(\frac{5}{2})} \right](w)
\label{a+w+7half}  \\
& + & \frac{1}{(z-w)} \left[ i U^{(\frac{7}{2})}
+ \frac{2}{(6+k)} \left(  \hat{B}_{-} T_{-}^{(\frac{5}{2})}-T_{+}^{(\frac{5}{2})}\hat{A}_{+}
+\frac{3i}{5} \pa U^{(\frac{5}{2})}\right)\right](w) +  \cdots,
\nonu \\
\hat{A}_{-}(z) \, W_{+}^{(\frac{7}{2})}(w) & = &
\frac{1}{(z-w)} \left[
\frac{2}{(6+k)} T_{+}^{(\frac{5}{2})}\hat{A}_{-}\right](w)+\cdots,
\label{a-w+7half} \\
\hat{A}_3(z) \, W_{+}^{(\frac{7}{2})}(w) & = &
\frac{1}{(z-w)^2} \left[
\frac{4i(17+2k)}{5(6+k)} T_{+}^{(\frac{5}{2})}\right](w) +
\frac{1}{(z-w)}\frac{i}{2} W_{+}^{(\frac{7}{2})}(w)+\cdots,
\label{a3w+7half} \\
\hat{B}_{+}(z) \, W_{+}^{(\frac{7}{2})}(w) & = &
\frac{1}{(z-w)^2} \left[
\frac{2i(67+12k)}{5(6+k)}V^{(\frac{5}{2})}\right](w)
\label{b+w+7half}  \\
& + & \frac{1}{(z-w)} \left[ -i V^{(\frac{7}{2})}
+ \frac{2}{(6+k)} \left(  T_{+}^{(\frac{5}{2})}\hat{B}_{+} -\hat{A}_{-}T_{-}^{(\frac{5}{2})}
-\frac{3i}{5} \pa V^{(\frac{5}{2})}\right)\right](w) +  \cdots,
\nonu \\
\hat{B}_{-}(z) \, W_{+}^{(\frac{7}{2})}(w) & = &
-\frac{1}{(z-w)} \left[
\frac{2}{(6+k)} \hat{B}_{-}T_{+}^{(\frac{5}{2})} \right](w)+\cdots,
\label{b-w+7half} \\
\hat{B}_3(z) \, W_{+}^{(\frac{7}{2})}(w) & = &
-\frac{1}{(z-w)^2} \left[
\frac{4i(13+3k)}{5(6+k)} T_{+}^{(\frac{5}{2})}\right](w) +
\frac{1}{(z-w)}\frac{i}{2} W_{+}^{(\frac{7}{2})}(w)+\cdots.
\label{b3w+7half}
\eea
The $U(1)$ charge conservation in these OPEs can be checked by Table
$2$ or $6$. 
The nonlinear terms (still linear in higher spin currents) 
appear in (\ref{a+w+7half}), 
(\ref{a-w+7half}), (\ref{b+w+7half}) and (\ref{b-w+7half})
while the linear terms appear in (\ref{a3w+7half}) and (\ref{b3w+7half}).
There is no descendant field for the spin-$\frac{5}{2}$ field 
in the first order pole of (\ref{a+w+7half}), (\ref{a3w+7half}), 
(\ref{b+w+7half}) or (\ref{b3w+7half}).
One can check  
in (\ref{a+w+7half}) or (\ref{b+w+7half}) whether 
the last terms (containing the composite fields)  
in the first order poles 
are (quasi)primary fields. 

The OPEs between the spin-$1$ currents and the
spin-$\frac{7}{2}$ current $W_{-}^{(\frac{7}{2})}(w)$ (\ref{w7-}) 
are expressed as
\bea
\hat{A}_{+}(z) \, W_{-}^{(\frac{7}{2})}(w) & = &
-\frac{1}{(z-w)} \left[
\frac{2}{(6+k)}\hat{A}_{-} T_{-}^{(\frac{5}{2})}\right](w)+\cdots,
\label{a+w-7half} \\
\hat{A}_{-}(z) \, W_{-}^{(\frac{7}{2})}(w) & = &
-\frac{1}{(z-w)^2} \left[\frac{2i(63+13k)}{5(6+k)} V^{(\frac{5}{2})} \right](w)
\label{a-w-7half}  \\
& + & \frac{1}{(z-w)} \left[ -i V^{(\frac{7}{2})}
+ \frac{4}{(6+k)} \left(  \hat{B}_{3}V^{(\frac{5}{2})} -\hat{A}_{3}V^{(\frac{5}{2})}
+\frac{i}{5} \pa V^{(\frac{5}{2})}\right)\right](w) +  \cdots,
\nonu \\
\hat{A}_3(z) \, W_{-}^{(\frac{7}{2})}(w) & = &
\frac{1}{(z-w)^2} \left[\frac{4i(17+2k)}{5(6+k)} T_{-}^{(\frac{5}{2})} \right](w)
-\frac{1}{(z-w)}\frac{i}{2} W_{-}^{(\frac{7}{2})}(w)+\cdots,
\label{a3w-7half} \\
\hat{B}_{+}(z) \, W_{-}^{(\frac{7}{2})}(w) & = &
\frac{1}{(z-w)} \left[
\frac{2}{(6+k)} T_{-}^{(\frac{5}{2})}\hat{B}_{+}\right](w)+\cdots,
\label{b+w-7half} \\
\hat{B}_{-}(z) \, W_{-}^{(\frac{7}{2})}(w) & = &
\frac{1}{(z-w)^2} \left[
\frac{2i(67+12k)}{5(6+k)}U^{(\frac{5}{2})}\right](w)
\label{b-w-7half}  \\
& + & \frac{1}{(z-w)} \left[ i U^{(\frac{7}{2})}
+ \frac{4}{(6+k)} \left(  U^{(\frac{5}{2})}\hat{B}_{3} -U^{(\frac{5}{2})}\hat{A}_{3}
+\frac{4i}{5} \pa U^{(\frac{5}{2})}\right)\right](w) +  \cdots,
\nonu \\
\hat{B}_3(z) \, W_{-}^{(\frac{7}{2})}(w) & = &
-\frac{1}{(z-w)^2} \left[\frac{4i(13+3k)}{5(6+k)} T_{-}^{(\frac{5}{2})} 
\right](w)
-\frac{1}{(z-w)}\frac{i}{2} W_{-}^{(\frac{7}{2})}(w)+\cdots.
\label{b3w-7half}
\eea
The $U(1)$ charge conservation in these OPEs can be checked by Table
$2$ or $6$. 
The linear terms appear in (\ref{a3w-7half}) and (\ref{b3w-7half})
and the nonlinear terms (linear in higher spin currents) 
appear in (\ref{a+w-7half}), (\ref{a-w-7half}), 
(\ref{b+w-7half}) and (\ref{b-w-7half}).
There is no descendant field for the spin-$\frac{5}{2}$ field 
in the first order pole of (\ref{a-w-7half}), (\ref{a3w-7half}), 
(\ref{b-w-7half}) or (\ref{b3w-7half}).
One can check  
in (\ref{a-w-7half}) or (\ref{b3w-7half}) whether 
the last terms (containing the composite fields)  
in the first order poles 
are  (quasi)primary fields. 

Finally the six nontrivial OPEs between the spin-$1$ currents and the
spin-$4$ current  $ W^{(4)}(w)$ (\ref{w4}) can be described as
\bea
\hat{A}_{+}(z) \, W^{(4)}(w) & = &
\frac{1}{(z-w)^2}\left[
 \frac{2i(43+6k)}{3(6+k)}  U_{-}^{(3)}
+\frac{16(-972+k(-125+27k))}{3(324+179k)(6+k)} T^{(2)}\hat{A}_{+}
\right](w) \nonu \\
& + &
\frac{1}{(z-w)}
\frac{1}{(6+k)}\left[
\frac{8}{3}\hat{A}_{+} \pa T^{(2)}+4 \hat{B}_{3} U_{-}^{(3)}-
\frac{16}{3} T^{(2)} \pa \hat{A}_{+}
 \right. \nonu \\
& -&
\left.
 4 T^{(3)} \hat{A}_{+}-\frac{10i}{3}\pa U_{-}^{(3)}\right](w) +\cdots,
\label{A+w4} \\
\hat{A}_{-}(z) \, W^{(4)}(w) & = &
\frac{1}{(z-w)^2}\left[
 -\frac{2i(47+9k)}{3(6+k)}  V_{+}^{(3)}
+\frac{8(-1620-71k+54k^2)}{3(324+179k)(6+k)}  \hat{A}_{-} T^{(2)}
\right](w) \nonu \\
& + &
\frac{1}{(z-w)}
\frac{1}{(6+k)}\left[
2\hat{A}_{-} T^{(3)}-2 \hat{A}_{-} W^{(3)}-\frac{8}{3} \hat{A}_{-}\pa T^{(2)} +\frac{16}{3}T^{(2)} \pa \hat{A}_{-}
 \right. \nonu \\
& -&
\left.
 4 V_{+}^{(3)} \hat{A}_{3} +4 V_{+}^{(3)} \hat{B}_{3}-\frac{10i}{3}\pa V_{+}^{(3)}\right](w) +\cdots,
\label{A-w4} \\
\hat{A}_{3}(z) \, W^{(4)}(w) & = &
-\frac{1}{(z-w)^3}
 \left[ \frac{16i(1+k)}{3(6+k)}  T^{(2)} \right](w) \nonu \\
& + &
\frac{1}{(z-w)^2}
\left[\frac{i(16+k)}{(6+k)}T^{(3)}+\frac{i(16+3k)}{(6+k)}W^{(3)}
+\frac{8}{(6+k)}\hat{B}_3 T^{(2)}
 \right. \nonu \\
& +&
\left.
\frac{8(-756-143k+18k^2)}{(324+179k)(6+k)}\hat{A}_{3}T^{(2)}
+\frac{4i(1+k)}{3(6+k)} \pa T^{(2)}\right](w) +\cdots,
\label{A3w4} \\
\hat{B}_{+}(z) \, W^{(4)}(w) & = &
\frac{1}{(z-w)^2}\left[
 -\frac{2i(31+9k)}{3(6+k)}  V_{-}^{(3)}
+\frac{16(-324+233k +27k^2)}{3(324+179k)(6+k)} T^{(2)}\hat{B}_{+}
\right](w) \nonu \\
& + &
\frac{1}{(z-w)}
\frac{1}{(6+k)}\left[
-4 \hat{A}_{3} V_{-}^{(3)}-\frac{8}{3}\hat{B}_{+} \pa T^{(2)}+\frac{16}{3} T^{(2)} \pa \hat{B}_{+}
 \right. \nonu \\
& +&
\left.
 4 T^{(3)} \hat{B}_{+}+\frac{10i}{3}\pa V_{-}^{(3)}\right](w) +\cdots,
\label{B+w4} \\
\hat{B}_{-}(z) \, W^{(4)}(w) & = &
\frac{1}{(z-w)^2}\left[
 \frac{2i(47+9k)}{3(6+k)}  U_{+}^{(3)}
+\frac{8(-972+287k+54k^2)}{3(324+179k)(6+k)}  \hat{B}_{-} T^{(2)}
\right](w) \nonu \\
& + &
\frac{1}{(z-w)}
\frac{1}{(6+k)}\left[
-2\hat{B}_{-} T^{(3)}-2 \hat{B}_{-} W^{(3)}+\frac{8}{3} \hat{B}_{-}\pa T^{(2)} +\frac{16}{3}T^{(2)} \pa \hat{B}_{-}
 \right. \nonu \\
& -&
\left.
 4 U_{+}^{(3)} \hat{A}_{3} +4 U_{+}^{(3)} \hat{B}_{3}+\frac{10i}{3}\pa U_{+}^{(3)}\right](w) +\cdots,
\label{B-w4} \\
\hat{B}_{3}(z) \, W^{(4)}(w) & = &
\frac{1}{(z-w)^3}
 \left[ \frac{80i}{3(6+k)} i T^{(2)} \right](w) \nonu \\
& + &
\frac{1}{(z-w)^2}
\left[-\frac{i(8+3k)}{(6+k)}T^{(3)}+\frac{i(16+3k)}{(6+k)}W^{(3)}
-\frac{8}{(6+k)}\hat{A}_3 T^{(2)}
 \right. \nonu \\
& +&
\left.
\frac{8(-108+215k+18k^2)}{(324+179k)(6+k)}\hat{B}_{3}T^{(2)}
-\frac{20i}{3(6+k)} \pa T^{(2)}\right](w) +\cdots.
\label{B3w4}
\eea
The $U(1)$ charge conservation in these OPEs (\ref{A+w4}), (\ref{A-w4}), 
(\ref{A3w4}), (\ref{B+w4}), (\ref{B-w4}) and (\ref{B3w4}) 
can be checked by Table
$2$ or $5$.
There is no descendant field for the spin-$3$ field 
in the first order pole of (\ref{A+w4}), (\ref{A-w4}), (\ref{A3w4}), 
(\ref{B+w4}), (\ref{B-w4}) or (\ref{B3w4}).
In (\ref{A3w4}) or (\ref{B3w4}), each term (except the last term) 
in the second order pole 
is a primary field. 

Therefore, one obtains 
 the OPEs between six spin-$1$ currents and the higher spin
current of spins $(3, \frac{7}{2}, \frac{7}{2}, 4)$ explicitly.
Most of the 
higher spin currents corresponding to four ${\cal N}=2$ multiplets
(\ref{new16comp})
occur in the right hand sides of the OPEs.
The nonlinear terms in this Appendix are quadratic terms.
Note that, for unitary case \cite{Ahn1311}, 
there are also cubic terms in their
corresponding OPEs.   

One obtains the following results between the spin-$1$
currents and the $16$ higher spin currents by focusing on the linear terms of
higher spin currents in the right hand side of the OPEs as follows:
\bea
\hat{A}_{+} & \times &
\left(
\begin{array}{cccc}
T^{(2)}, & T_{+}^{(\frac{5}{2})}, & T_{-}^{(\frac{5}{2})}, & T^{(3)}  \\
U^{(\frac{5}{2})}, & U_{+}^{(3)}, & U_{-}^{(3)}, & U^{(\frac{7}{2})}  \\
V^{(\frac{5}{2})}, & V^{(3)}_{+}, & V^{(3)}_{-}, & V^{(\frac{7}{2})}  \\
W^{(3)}, & W_{+}^{(\frac{7}{2})}, & W_{-}^{(\frac{7}{2})}, & W^{(4)}
\end{array} \right)
 \rightarrow
\left(
\begin{array}{cccc}
0, & U^{(\frac{5}{2})}, & 0, & U_{-}^{(3)} \\
0,  & 0, & 0,  & 0  \\
T_{-}^{(\frac{5}{2})}, & T^{(2)},T^{(3)},W^{(3)}, & 0, & T_{-}^{(\frac{5}{2})},W_{-}^{(\frac{7}{2})}   \\
U_{-}^{(3)}, & U^{(\frac{5}{2})},U^{(\frac{7}{2})}, & 0, & U_{-}^{(3)}
\end{array} \right),
\nonu \\
\hat{A}_{-} & \times &
\left(
\begin{array}{cccc}
T^{(2)}, & T_{+}^{(\frac{5}{2})}, & T_{-}^{(\frac{5}{2})}, & T^{(3)}  \\
U^{(\frac{5}{2})}, & U_{+}^{(3)}, & U_{-}^{(3)}, & U^{(\frac{7}{2})}  \\
V^{(\frac{5}{2})}, & V^{(3)}_{+}, & V^{(3)}_{-}, & V^{(\frac{7}{2})}  \\
W^{(3)}, & W_{+}^{(\frac{7}{2})}, & W_{-}^{(\frac{7}{2})}, & W^{(4)}
\end{array} \right)
 \rightarrow
\left(
\begin{array}{cccc}
0, & 0, & V^{(\frac{5}{2})}, & V_{+}^{(3)} \\
T_{+}^{(\frac{5}{2})},  & 0, & T^{(2)},T^{(3)},W^{(3)},  & T_{+}^{(\frac{5}{2})},W_{+}^{(\frac{7}{2})}  \\
0, & 0, & 0, & 0   \\
V_{+}^{(3)}, & 0, & V^{(\frac{5}{2})},V^{(\frac{7}{2})}, & V_{+}^{(3)}
\end{array} \right),
\nonu \\
\hat{A}_{3} & \times &
\left(
\begin{array}{cccc}
T^{(2)}, & T_{+}^{(\frac{5}{2})}, & T_{-}^{(\frac{5}{2})}, & T^{(3)}  \\
U^{(\frac{5}{2})}, & U_{+}^{(3)}, & U_{-}^{(3)}, & U^{(\frac{7}{2})}  \\
V^{(\frac{5}{2})}, & V^{(3)}_{+}, & V^{(3)}_{-}, & V^{(\frac{7}{2})}  \\
W^{(3)}, & W_{+}^{(\frac{7}{2})}, & W_{-}^{(\frac{7}{2})}, & W^{(4)}
\end{array} \right)
 \rightarrow
\left(
\begin{array}{cccc}
0, & T_{+}^{(\frac{5}{2})}, & T_{-}^{(\frac{5}{2})}, &T^{(2)} \\
U^{(\frac{5}{2})},  & 0, & U_{-}^{(3)},  & U^{(\frac{5}{2})},U^{(\frac{7}{2})}  \\
V^{(\frac{5}{2})} & V_{+}^{(3)}, & 0, & V^{(\frac{5}{2})},V^{(\frac{7}{2})}   \\
T^{(2)} & T_{+}^{(\frac{5}{2})},W_{+}^{(\frac{7}{2})}, & T_{-}^{(\frac{5}{2})},W_{-}^{(\frac{7}{2})}, & T^{(2)},T^{(3)},W^{(3)}
\end{array} \right),
\nonu  \\
\hat{B}_{+} & \times &
\left(
\begin{array}{cccc}
T^{(2)}, & T_{+}^{(\frac{5}{2})}, & T_{-}^{(\frac{5}{2})}, & T^{(3)}  \\
U^{(\frac{5}{2})}, & U_{+}^{(3)}, & U_{-}^{(3)}, & U^{(\frac{7}{2})}  \\
V^{(\frac{5}{2})}, & V^{(3)}_{+}, & V^{(3)}_{-}, & V^{(\frac{7}{2})}  \\
W^{(3)}, & W_{+}^{(\frac{7}{2})}, & W_{-}^{(\frac{7}{2})}, & W^{(4)}
\end{array} \right)
 \rightarrow
\left(
\begin{array}{cccc}
0, & V^{(\frac{5}{2})}, & 0, & V_{-}^{(3)} \\
T_{-}^{(\frac{5}{2})},  & T^{(2)},T^{(3)},W^{(3)}, & 0,  & T_{-}^{(\frac{5}{2})},W_{-}^{(\frac{7}{2})}  \\
0 & 0, & 0, & 0   \\
V_{-}^{(3)}, & V^{(\frac{5}{2})},V^{(\frac{7}{2})}, & 0, & V_{-}^{(3)}
\end{array} \right),
\nonu \\
\hat{B}_{-} & \times &
\left(
\begin{array}{cccc}
T^{(2)}, & T_{+}^{(\frac{5}{2})}, & T_{-}^{(\frac{5}{2})}, & T^{(3)}  \\
U^{(\frac{5}{2})}, & U_{+}^{(3)}, & U_{-}^{(3)}, & U^{(\frac{7}{2})}  \\
V^{(\frac{5}{2})}, & V^{(3)}_{+}, & V^{(3)}_{-}, & V^{(\frac{7}{2})}  \\
W^{(3)}, & W_{+}^{(\frac{7}{2})}, & W_{-}^{(\frac{7}{2})}, & W^{(4)}
\end{array} \right)
 \rightarrow
\left(
\begin{array}{cccc}
0, & 0, & U^{(\frac{5}{2})}, &U_{+}^{(3)} \\
0,  & 0, & 0,  & 0  \\
T_{+}^{(\frac{5}{2})}, & 0, & T^{(2)},T^{(3)},W^{(3)}, & T_{+}^{(\frac{5}{2})},W_{+}^{(\frac{7}{2})}  \\
U_{+}^{(3)}, & 0, & U^{(\frac{5}{2})},U^{(\frac{7}{2})}, & U_{+}^{(3)}
\end{array} \right),
\nonu \\
\hat{B}_{3} & \times &
\left(
\begin{array}{cccc}
T^{(2)}, & T_{+}^{(\frac{5}{2})}, & T_{-}^{(\frac{5}{2})}, & T^{(3)}  \\
U^{(\frac{5}{2})}, & U_{+}^{(3)}, & U_{-}^{(3)}, & U^{(\frac{7}{2})}  \\
V^{(\frac{5}{2})}, & V^{(3)}_{+}, & V^{(3)}_{-}, & V^{(\frac{7}{2})}  \\
W^{(3)}, & W_{+}^{(\frac{7}{2})}, & W_{-}^{(\frac{7}{2})}, & W^{(4)}
\end{array} \right)
 \rightarrow
\left(
\begin{array}{cccc}
0 & T_{+}^{(\frac{5}{2})}, & T_{-}^{(\frac{5}{2})}, &T^{(2)} \\
U^{(\frac{5}{2})},  & U_{+}^{(3)}, & 0,  & U^{(\frac{5}{2})},U^{(\frac{7}{2})} \\
V^{(\frac{5}{2})}, & 0, & V_{-}^{(3)}, & V^{(\frac{5}{2})},V^{(\frac{7}{2})}   \\
T^{(2)} & T_{+}^{(\frac{5}{2})},W_{+}^{(\frac{7}{2})}, & T_{-}^{(\frac{5}{2})},W_{-}^{(\frac{7}{2})}, & T^{(2)},T^{(3)},W^{(3)}
\end{array} \right).
\nonu
\eea
Note that the zeros of the right hand sides for the OPEs between 
$\hat{A}_{\pm}$ and $\hat{B}_{\pm}$ acting on the $16$ higher spin currents
are located at 
the row and column containing the corresponding spin-$1$ current in
$(2.44)$ of \cite{Ahn1311}.
The zeros from the OPEs 
between these four spin-$1$ currents and spin-$2$ current 
$T^{(2)}(w)$ also occur.

\section{The OPEs between the four spin-$\frac{3}{2}$ currents and 
the higher spin currents}
In this Appendix, we present the OPEs between the four
spin-$\frac{3}{2}$ currents
in section $2$ and the $16$ higher spin currents in section $3$.

\subsection{The OPEs between four spin $\frac{3}{2}$ currents
and the higher spin
current of spins $(2, \frac{5}{2}, \frac{5}{2}, 3)$
}
The OPEs between the
spin-$\frac{3}{2}$ currents (\ref{nong11}), (\ref{nong22}), (\ref{nong12}) and 
(\ref{nong21}) and the
spin-$2$ current $T^{(2)}(w)$ are derived as follows:
\bea
\left(
\begin{array}{c}
\hat{G}_{11} \\
\hat{G}_{22}
\end{array} \right)(z) \, T^{(2)}(w) & = &
\frac{1}{(z-w)} \left(
\begin{array}{c}
  U^{(\frac{5}{2})}
\\
 V^{(\frac{5}{2})}
\end{array}
\right)(w) +
\cdots,
\label{G11t1} \\
\left(
\begin{array}{c}
\hat{G}_{12} \\
\hat{G}_{21}
\end{array} \right) (z) \, T^{(2)}(w) & = &
\frac{1}{(z-w)} \left(
\begin{array}{c}
T_{-}^{(\frac{5}{2})} \\
 T_{+}^{(\frac{5}{2})}
\end{array}
\right)(w) +
\cdots.
\label{G12t2}
\eea
One can check the $U(1)$ charge conservation in these OPEs (\ref{G11t1}) 
and (\ref{G12t2}) from Table
$2$ or $4$.

The following nontrivial OPEs between the spin-$\frac{3}{2}$ currents and the
spin-$\frac{5}{2}$ currents $T_{\pm}^{(\frac{5}{2})}(w)$  in (\ref{t+}) 
and (\ref{t-})
are as follows:
\bea
\left(
\begin{array}{c}
\hat{G}_{11} \\
\hat{G}_{22}
\end{array} \right)(z) \,
T_{+}^{(\frac{5}{2})} (w) & = &
-\frac{1}{(z-w)} \left(
\begin{array}{c}
  U_{+}^{(3)}
\\
 V_{+}^{(3)}
\end{array}
\right)(w) +
\cdots,
\label{G11t+5half}
\\
\left(
\begin{array}{c}
\hat{G}_{12} \\
\hat{G}_{21}
\end{array} \right)(z) \, T_{\pm}^{(\frac{5}{2})}(w) & = &
\frac{1}{(z-w)}  4T^{(2)}(w)
+
\frac{1}{(z-w)^2} (\pa T^{(2)}\mp T^{(3)})(w)
+\cdots,
\label{G12t+5half}\\
\left(
\begin{array}{c}
\hat{G}_{11} \\
\hat{G}_{22}
\end{array} \right)(z) \,
T_{-}^{(\frac{5}{2})} (w) & = &
-\frac{1}{(z-w)} \left(
\begin{array}{c}
  U_{-}^{(3)}
\\
 V_{-}^{(3)}
\end{array}
\right)(w) +
\cdots.
\label{G11t-5half}
\eea
The $U(1)$ charge conservation in these OPEs can be checked from Tables
$2$ or $5$. No nonlinear terms in (\ref{G11t+5half}), (\ref{G12t+5half}) 
and (\ref{G11t-5half}) occur.

One calculates the following OPEs between the spin-$\frac{3}{2}$ currents
and the spin-$3$ current $T^{(3)}(w)$ (\ref{t3}) as follows:
\bea
\left(
\begin{array}{c}
\hat{G}_{11} \\
\hat{G}_{22} \\
\end{array} \right) (z) \, T^{(3)}(w) & = &
\frac{1}{(z-w)^2}\frac{(-4+k)}{(6+k)}
\left(
\begin{array}{c}
U^{(\frac{5}{2})} \\
-V^{(\frac{5}{2})} \\
\end{array} \right)
\label{G11t3}
\\
& + & \frac{1}{(z-w)} \left[
\frac{(-4+k)}{5(6+k)} \pa
\left(
\begin{array}{c}
U^{(\frac{5}{2})} \\
-V^{(\frac{5}{2})} \\
\end{array} \right)
+\left(
\begin{array}{c}
U^{(\frac{7}{2})} \\
V^{(\frac{7}{2})} \\
\end{array} \right)
\right. \nonu \\
& + & \left. \frac{2}{(6+k)}
\left(
\begin{array}{c}
2i U^{(\frac{5}{2})}\hat{A}_{3}-i \hat{B}_{-}T_{-}^{(\frac{5}{2})}+\frac{3}{5} \pa U^{(\frac{5}{2})}\\
2i \hat{B}_3 V^{(\frac{5}{2})}-i \hat{A}_{-}T_{-}^{(\frac{5}{2})}-\frac{2}{5} \pa V^{(\frac{5}{2})}\\
\end{array} \right) \right](w)+\cdots,
\nonu \\
\left(
\begin{array}{c}
\hat{G}_{12} \\
\hat{G}_{21} \\
\end{array} \right) (z) \, T^{(3)}(w) & = &
\frac{1}{(z-w)^2}\frac{(26+5k)}{(6+k)}
\left(
\begin{array}{c}
T_{-}^{(\frac{5}{2})} \\
-T_{+}^{(\frac{5}{2})} \\
\end{array} \right)
\label{G12t3}
\\
& + & \frac{1}{(z-w)} \left[
\frac{(26+5k)}{5(6+k)} \pa
\left(
\begin{array}{c}
T_{-}^{(\frac{5}{2})} \\
-T_{+}^{(\frac{5}{2})} \\
\end{array} \right)
\right. \nonu \\
& + & \left. \frac{2}{(6+k)}
\left(
\begin{array}{c}
i \hat{A}_{+}V^{(\frac{5}{2})}+i U^{(\frac{5}{2})}\hat{B}_{+}-\frac{3}{5} \pa T_{-}^{(\frac{5}{2})}\\
i \hat{B}_{-}V^{(\frac{5}{2})}+i U^{(\frac{5}{2})}\hat{A}_{-}+\frac{3}{5} \pa T_{+}^{(\frac{5}{2})}\\
\end{array} \right) \right](w)+\cdots.
\nonu
\eea
The $U(1)$ charge conservation in these OPEs can be checked from Tables
$2$ or $6$. The quadratic terms (still linear in higher spin currents) 
in (\ref{G11t3}) and (\ref{G12t3}) 
occur. The factor $(k-4)$ appears in (\ref{G11t3}). 

Therefore, one obtains 
 the OPEs between six spin-$1$ currents and the higher spin
current of spins $(2, \frac{5}{2}, \frac{5}{2}, 3)$ explicitly.
The 
higher spin currents corresponding to three ${\cal N}=2$ multiplets
(\ref{new16comp})
occur in the right hand sides of the OPEs.

\subsection{The OPEs between four spin $\frac{3}{2}$ currents
and two higher spin
currents of spins $(\frac{5}{2}, 3, 3, \frac{7}{2})$}
The nontrivial OPEs between the spin-$\frac{3}{2}$ currents and the
spin-$\frac{5}{2}$ currents, $U^{(\frac{5}{2})}(z)$ (\ref{u5half}) 
and $V^{(\frac{5}{2})}(z)$ (\ref{v5half}), can be obtained as follows:
\bea
\left(
\begin{array}{c}
\hat{G}_{22} \\
\hat{G}_{11} \\
\end{array} \right) (z)
 \left(
\begin{array}{c}
U^{(\frac{5}{2})} \\
V^{(\frac{5}{2})}\\
\end{array} \right)(z) \, & = &
\frac{1}{(z-w)^2}4T^{(2)}(w) +
\frac{1}{(z-w)}(\pa T^{(2)}\pm W^{(3)})(w)+
\cdots, 
\label{G22u5half}
\\
\hat{G}_{12}(z)\,
 \left(
\begin{array}{c}
U^{(\frac{5}{2})} \\
V^{(\frac{5}{2})}\\
\end{array} \right)(z) \, & = &
\frac{1}{(z-w)} \left(
\begin{array}{c}
U_{-}^{(3)}\\
V_{-}^{(3)}\\
\end{array}
\right)(w) +\cdots,
\label{G12u5half} \\
\hat{G}_{21}(z)\,
 \left(
\begin{array}{c}
U^{(\frac{5}{2})} \\
V^{(\frac{5}{2})}\\
\end{array} \right)(z) \, & = &
\frac{1}{(z-w)} \left(
\begin{array}{c}
U_{+}^{(3)}\\
V_{+}^{(3)}\\
\end{array}
\right)(w) +\cdots.
\label{G11u5half}
\eea
The $U(1)$ charge conservation in these OPEs (\ref{G22u5half}), 
(\ref{G12u5half}) and (\ref{G11u5half})
can be checked from Table
$2$ or $5$.

The OPEs between the spin-$\frac{3}{2}$ currents
and the spin-$3$ currents $U_{\pm}^{(3)}(w)$ in (\ref{u3+}) and (\ref{u3-}) 
are as follows:
\bea
\left(
\begin{array}{c}
\hat{G}_{11} \\
\hat{G}_{21}
\end{array} \right)(z) \, U_{+}^{(3)}(w) & = &
-\frac{1}{(z-w)}\frac{2i}{(6+k)}\hat{B}_{-} \left(
\begin{array}{c}
  U^{(\frac{5}{2})}
\\
T_{+}^{(\frac{5}{2})}
\end{array}
\right)(w) +
\cdots,
\label{G11u+3}\\
\left(
\begin{array}{c}
\hat{G}_{22} \\
\hat{G}_{12} \\
\end{array} \right) (z) \, U_{+}^{(3)}(w) & = &
\frac{1}{(z-w)^2}\frac{(30+4k)}{(6+k)}
\left(
\begin{array}{c}
-T_{+}^{(\frac{5}{2})} \\
U^{(\frac{5}{2})} \\
\end{array} \right)
\label{G22u+3}
\\
& + & \frac{1}{(z-w)} \left[
\frac{(30+4k)}{5(6+k)} \pa
\left(
\begin{array}{c}
-T_{+}^{(\frac{5}{2})} \\
U^{(\frac{5}{2})} \\
\end{array} \right)
-\left(
\begin{array}{c}
W_{+}^{(\frac{7}{2})} \\
U^{(\frac{7}{2})} \\
\end{array} \right)
\right. \nonu \\
& + & \left. \frac{2}{(6+k)}
\left(
\begin{array}{c}
i U^{(\frac{5}{2})}\hat{A}_{-}-2i T_{+}^{(\frac{5}{2})}\hat{B}_{3}+\frac{8}{5} \pa T_{+}^{(\frac{5}{2})}\\
i \hat{B}_{-}T_{-}^{(\frac{5}{2})}-i T_{+}^{(\frac{5}{2})}\hat{A}_{+}
-2iU^{(\frac{5}{2})}(\hat{A}_{3}-\hat{B}_{3})-\frac{11}{5} \pa U^{(\frac{5}{2})}\\
\end{array} \right) \right](w)
\nonu \\
&+&\cdots,
\nonu \\
\left(
\begin{array}{c}
\hat{G}_{11} \\
\hat{G}_{12}
\end{array} \right)(z) \, U_{-}^{(3)}(w) & = &
\frac{1}{(z-w)}\frac{2i}{(6+k)} \left(
\begin{array}{c}
  U^{(\frac{5}{2})}\hat{A}_{+}
\\
\hat{A}_{+}T_{-}^{(\frac{5}{2})}
\end{array}
\right)(w) +
\cdots,
\label{G11u-3}\\
\left(
\begin{array}{c}
\hat{G}_{22} \\
\hat{G}_{21} \\
\end{array} \right) (z) \, U_{-}^{(3)}(w) & = &
\frac{1}{(z-w)^2}\frac{(22+6k)}{(6+k)}
\left(
\begin{array}{c}
-T_{-}^{(\frac{5}{2})} \\
U^{(\frac{5}{2})} \\
\end{array} \right)
\label{G22u-3}
\\
& + & \frac{1}{(z-w)} \left[
\frac{(22+6k)}{5(6+k)} \pa
\left(
\begin{array}{c}
-T_{-}^{(\frac{5}{2})} \\
U^{(\frac{5}{2})} \\
\end{array} \right)
+\left(
\begin{array}{c}
-W_{-}^{(\frac{7}{2})} \\
U^{(\frac{7}{2})} \\
\end{array} \right)
\right. \nonu \\
& + & \left. \frac{2}{(6+k)}
\left(
\begin{array}{c}
2i\hat{A}_{3} T_{-}^{(\frac{5}{2})}-i U^{(\frac{5}{2})}\hat{B}_{+}+\frac{3}{5} \pa T_{-}^{(\frac{5}{2})}\\
0\\
\end{array} \right) \right](w)
+\cdots.
\nonu
\eea
The $U(1)$ charge conservation in these OPEs can be checked from Table
$2$ or $6$. The nonlinear terms (linear in higher spin currents) 
occur in (\ref{G11u+3}), 
(\ref{G22u+3}), (\ref{G11u-3}) and 
(\ref{G22u-3}).

The OPEs between the spin-$\frac{3}{2}$ currents
and the spin-$3$ currents $V_{\pm}^{(3)}(w)$ in (\ref{v3+}) and (\ref{v3-}) 
are expressed as
\bea
\left(
\begin{array}{c}
\hat{G}_{22} \\
\hat{G}_{21}
\end{array} \right)(z) \, V_{+}^{(3)}(w) & = &
-\frac{1}{(z-w)}\frac{2i}{(6+k)} \left(
\begin{array}{c}
 \hat{A}_{-} V^{(\frac{5}{2})}
\\
T_{+}^{(\frac{5}{2})}\hat{A}_{-}
\end{array}
\right)(w) +
\cdots,
\label{G22uv+3}\\
\left(
\begin{array}{c}
\hat{G}_{11} \\
\hat{G}_{12} \\
\end{array} \right) (z) \, V_{+}^{(3)}(w) & = &
\frac{1}{(z-w)^2}\frac{(22+6k)}{(6+k)}
\left(
\begin{array}{c}
-T_{+}^{(\frac{5}{2})} \\
V^{(\frac{5}{2})} \\
\end{array} \right)
\label{G11v+3}
\\
& + & \frac{1}{(z-w)} \left[
\frac{(22+6k)}{5(6+k)} \pa
\left(
\begin{array}{c}
-T_{+}^{(\frac{5}{2})} \\
V^{(\frac{5}{2})} \\
\end{array} \right)
+\left(
\begin{array}{c}
W_{+}^{(\frac{7}{2})} \\
-V^{(\frac{7}{2})} \\
\end{array} \right)
\right. \nonu \\
& + & \left. \frac{2}{(6+k)}
\left(
\begin{array}{c}
i \hat{B}_{-}V^{(\frac{5}{2})}-2i T_{+}^{(\frac{5}{2})}\hat{A}_{3}+\frac{3}{5} \pa T_{+}^{(\frac{5}{2})}\\
i \hat{A}_{-}T_{-}^{(\frac{5}{2})}+i T_{+}^{(\frac{5}{2})}\hat{B}_{+}
+2i(\hat{A}_{3}-\hat{B}_{3})V^{(\frac{5}{2})}-\frac{1}{5} \pa V^{(\frac{5}{2})}\\
\end{array} \right) \right](w)
\nonu \\
&+&\cdots,
\nonu \\
\left(
\begin{array}{c}
\hat{G}_{22} \\
\hat{G}_{12}
\end{array} \right)(z) \, V_{-}^{(3)}(w) & = &
\frac{1}{(z-w)}\frac{2i}{(6+k)} \left(
\begin{array}{c}
  V^{(\frac{5}{2})}
\\
T_{-}^{(\frac{5}{2})}
\end{array}
\right)\hat{B}_{+}(w) +
\cdots,
\label{G22v-3}\\
\left(
\begin{array}{c}
\hat{G}_{11} \\
\hat{G}_{21} \\
\end{array} \right) (z) \, V_{-}^{(3)}(w) & = &
\frac{1}{(z-w)^2}\frac{(30+4k)}{(6+k)}
\left(
\begin{array}{c}
-T_{-}^{(\frac{5}{2})} \\
V^{(\frac{5}{2})} \\
\end{array} \right)
\label{G11v-3}
\\
& + & \frac{1}{(z-w)} \left[
\frac{(30+4k)}{5(6+k)} \pa
\left(
\begin{array}{c}
-T_{-}^{(\frac{5}{2})} \\
V^{(\frac{5}{2})} \\
\end{array} \right)
+\left(
\begin{array}{c}
W_{-}^{(\frac{7}{2})} \\
V^{(\frac{7}{2})} \\
\end{array} \right)
\right. \nonu \\
& + & \left. \frac{2}{(6+k)}
\left(
\begin{array}{c}
-i\hat{A}_{+} V^{(\frac{5}{2})}+2i \hat{B}_{3}T_{-}^{(\frac{5}{2})}-\frac{2}{5} \pa T_{-}^{(\frac{5}{2})}\\
0\\
\end{array} \right) \right](w)
+\cdots.
\nonu
\eea
The $U(1)$ charge conservation in these OPEs can be checked from Table
$2$ or $6$. In this case also, the nonlinear terms (linear in higher 
spin currents) appear in (\ref{G22uv+3}), 
(\ref{G11v+3}), (\ref{G22v-3}) and (\ref{G11v-3}).

The OPEs between the spin-$\frac{3}{2}$ currents and the
spin-$\frac{7}{2}$ currents $U^{(\frac{7}{2})}(w)$ (\ref{u7half}) are given by
\bea
\hat{G}_{11}(z) \, U^{(\frac{7}{2})}(w)
& = &\frac{1}{(z-w)}
\left[-\frac{4i}{(6+k)}\hat{B}_{-}U_{-}^{(3)}+\frac{2}{(6+k)}\hat{G_{11}}U^{(\frac{5}{2})}
-\frac{2i}{(6+k)}U_{+}^{(3)}\hat{A}_{+}\right](w)
\nonu \\
&+&\cdots,
\label{g11u7half}
\\
\hat{G}_{22}(z) \, U^{(\frac{7}{2})}(w)
& = &
-\frac{1}{(z-w)^3} \left[ \frac{48(-3+k)}{5(6+k)}
T^{(2)} \right](w)
\label{g22u7half}
\\
&+& \frac{1}{(z-w)^2}
\left[\frac{6(5+k)}{(6+k)}T^{(3)} -\frac{6(-3+k)}{5(6+k)}W^{(3)} +\frac{16i}{(6+k)}\hat{B}_{3}T^{(2)}\right](w)
\nonu \\
&+& \frac{1}{(z-w)}
\left[\frac{1}{6}\pa \{\hat{G}_{22} \, U^{(\frac{7}{2})}\}_{-2}+W^{(4)}
-\frac{4i}{3(6+k)}(\hat{A}_{3}+4\hat{B}_{3})\pa T^{(2)}
\right. \nonu \\
& + & \left.
\frac{4i}{(6+k)}\hat{B}_{3}(T^{(3)}+W^{(3)})
-\frac{4i}{(6+k)}\hat{A}_{3}T^{(3)}
-\frac{2}{(6+k)}\hat{G}_{21}T_{-}^{(\frac{5}{2})}
\right. \nonu \\
& + & \left.
\frac{8i}{3(6+k)}T^{(2)}\pa (\hat{A}_{3}+4\hat{B}_{3})
-\frac{144(-4+k)}{(324+179k)}\hat{T}T^{(2)}
+\frac{2i}{(6+k)}U_{+}^{(3)}\hat{B}_{+}
\right. \nonu \\
&+& \left.
\frac{2}{(6+k)}U^{(\frac{5}{2})}\hat{G}_{22} +\frac{1}{3(6+k)}\pa (W^{(3)}-3T^{(3)})
\right. \nonu \\
&+& \left.
\frac{6(-1188-107k+36k^2)}{5(324+179k)(6+k)}\pa^2 T^{(2)}\right](w)
+\cdots,
\nonu \\
\hat{G}_{12}(z) \, U^{(\frac{7}{2})}(w)
& = &
\frac{1}{(z-w)^2}
\left[ \frac{2(89+12k)}{5(6+k)}U_{-}^{(3)}
+\frac{8i}{(6+k)}T^{(2)}\hat{A}_{+}\right](w)+
\label{g12u7half}
\\
&+& \frac{1}{(z-w)}
\left[\frac{1}{6}\pa \{\hat{G}_{12} \, U^{(\frac{7}{2})}\}_{-2}
-\frac{4i}{(6+k)}(\hat{A}_{3}-\hat{B}_{3})U_{-}^{(3)}
\right. \nonu \\
& - & \left.
\frac{10i}{3(6+k)}\hat{A}_{+}\pa T^{(2)}
+\frac{2}{(6+k)}\hat{G}_{11}T_{-}^{(\frac{5}{2})}
+\frac{2i}{(6+k)}W^{(3)}\hat{A}_{+}
\right. \nonu \\
& + & \left.
\frac{20i}{3(6+k)}T^{(2)}\pa \hat{A}_{+}
-\frac{1}{3(6+k)}\pa U_{-}^{(3)}\right](w)
+\cdots,
\nonu \\
\hat{G}_{21}(z) \, U^{(\frac{7}{2})}(w)
& = &
\frac{1}{(z-w)^2}
\left[ -\frac{12(11+3k)}{5(6+k)}U_{+}^{(3)}
+\frac{16i}{(6+k)}\hat{B}_{-}T^{(2)}\right](w)
\label{g21u7half}
\\
&+& \frac{1}{(z-w)}
\left[\frac{1}{6}\pa \{\hat{G}_{21} \, U^{(\frac{7}{2})}\}_{-2}
+\frac{2i}{(6+k)}\hat{B}_{-}(T^{(3)}+W^{(3)})
\right. \nonu \\
& - & \left.
\frac{8i}{3(6+k)}\hat{B}_{-}\pa T^{(2)}
+\frac{16i}{3(6+k)}T^{(2)}\pa \hat{B}_{-}\right](w)
+\cdots.
\nonu
\eea
The $U(1)$ charge conservation in these OPEs (\ref{g11u7half}), 
(\ref{g22u7half}), (\ref{g12u7half}) 
and (\ref{g21u7half})
can be checked from Table
$2$, $7$ or $8$.
Each term in the second order pole of (\ref{g22u7half}), (\ref{g12u7half}), 
or (\ref{g21u7half}) is a primary field.
There is no descendant field for the spin-$2$ field $T^{(2)}(w)$ 
in the second order pole of (\ref{g22u7half}).

The OPEs between the spin-$\frac{3}{2}$ currents and the
spin-$\frac{7}{2}$ current $V^{(\frac{7}{2})}(w)$ (\ref{v7half}) are as follows:
\bea
\hat{G}_{11}(z) \, V^{(\frac{7}{2})}(w)
& = &
\frac{1}{(z-w)^3} \left[ \frac{48(-5+k)}{5(6+k)}
T^{(2)} \right](w)
\label{g11v7half}
\\
&+& \frac{1}{(z-w)^2}
\left[\frac{6(5+k)}{(6+k)}T^{(3)}
-\frac{6(-5+k)}{5(6+k)}W^{(3)}
+\frac{16i}{(6+k)}\hat{A}_{3}T^{(2)}\right](w)
\nonu \\
&+& \frac{1}{(z-w)}
\left[\frac{1}{6}\pa \{\hat{G}_{11}\, V^{(\frac{7}{2})}\}_{-2}
-W^{(4)}
+\frac{4i}{(6+k)}\hat{A}_{3}(T^{(3)}-W^{(3)})
\right. \nonu \\
& - & \left.
\frac{4i}{3(6+k)}(4\hat{A}_{3}+\hat{B}_{3})\pa T^{(2)}
-\frac{4i}{(6+k)}\hat{B}_{3}T^{(3)}
-\frac{2}{(6+k)}\hat{G}_{11}V^{(\frac{5}{2})}
\right. \nonu \\
& - & \left.
\frac{2}{(6+k)}\hat{G}_{21}T_{-}^{(\frac{5}{2})}
+\frac{8i}{3(6+k)}T^{(2)}\pa (4\hat{A}_{3}+\hat{B}_{3})
+\frac{144(-4+k)}{(324+179k)}\hat{T}T^{(2)}
\right. \nonu \\
&+& \left.
\frac{2i}{(6+k)}V_{+}^{(3)}\hat{A}_{+}
-\frac{1}{3(6+k)}\pa (7W^{(3)}+3T^{(3)})
\right. \nonu \\
&-& \left.
\frac{8k(412+27k)}{5(324+179k)(6+k)}\pa^2 T^{(2)}\right](w)
+\cdots,
\nonu \\
\hat{G}_{22}(z) \, V^{(\frac{7}{2})}(w)
& = &\frac{1}{(z-w)}
\left[-\frac{4i}{(6+k)}\hat{A}_{-}V_{-}^{(3)}
+\frac{2}{(6+k)}\hat{G_{22}}V^{(\frac{5}{2})}
-\frac{2i}{(6+k)}V_{+}^{(3)}\hat{B}_{+}\right](w)
\nonu \\
&+&\cdots,
\label{g22v7half}
\\
\hat{G}_{12}(z) \, V^{(\frac{7}{2})}(w)
& = &
\frac{1}{(z-w)^2}
\left[ \frac{2(65+18k)}{5(6+k)}V_{-}^{(3)}
+\frac{8i}{(6+k)}T^{(2)}\hat{B}_{+}\right](w)
\label{g12v7half}
\\
&+& \frac{1}{(z-w)}
\left[\frac{1}{6}\pa \{\hat{G}_{12} \, V^{(\frac{7}{2})}\}_{-2}
+\frac{4i}{(6+k)}(\hat{A}_{3}-\hat{B}_{3})V_{-}^{(3)}
\right. \nonu \\
& - & \left.
\frac{10i}{3(6+k)}\hat{B}_{+}\pa T^{(2)}
+\frac{2}{(6+k)}\hat{G}_{22}T_{-}^{(\frac{5}{2})}
-\frac{2i}{(6+k)}W^{(3)}\hat{B}_{+}
\right. \nonu \\
& + & \left.
\frac{20i}{3(6+k)}T^{(2)}\pa \hat{B}_{+}
-\frac{1}{3(6+k)}\pa V_{-}^{(3)}\right](w)
+\cdots,
\nonu \\
\hat{G}_{21}(z) \, V^{(\frac{7}{2})}(w)
& = &
\frac{1}{(z-w)^2}
\left[ -\frac{12(15+2k)}{5(6+k)}V_{+}^{(3)}
+\frac{16i}{(6+k)}\hat{A}_{-}T^{(2)}\right](w)
\label{g21v7half}
\\
&+& \frac{1}{(z-w)}
\left[\frac{1}{6}\pa \{\hat{G}_{21} \, V^{(\frac{7}{2})}\}_{-2}
+\frac{2i}{(6+k)}\hat{A}_{-}(T^{(3)}-W^{(3)})
\right. \nonu \\
& - & \left.
\frac{8i}{3(6+k)}\hat{A}_{-}\pa T^{(2)}
+\frac{16i}{3(6+k)}T^{(2)}\pa \hat{A}_{-}\right](w)
+\cdots.
\nonu
\eea
The $U(1)$ charge conservation in these OPEs (\ref{g11v7half}), 
(\ref{g22v7half}), 
(\ref{g12v7half}) and (\ref{g21v7half})
can be checked from Table
$2$, $7$ or $8$.
Each term in the second order pole of (\ref{g11v7half}), (\ref{g12v7half}), 
or (\ref{g21v7half}) is a primary field.
There is no descendant field for the spin-$2$ field $T^{(2)}(w)$ 
in the second order pole of (\ref{g11v7half}). 

Therefore, one obtains 
 the OPEs between six spin-$1$ currents and the higher spin
current of spins $(\frac{5}{2}, 3, 3, \frac{7}{2})$ explicitly.
The 
higher spin currents corresponding to four ${\cal N}=2$ multiplets
(\ref{new16comp})
occur in the right hand sides of the OPEs.

\subsection{The OPEs between four spin $\frac{3}{2}$ currents
and the higher spin
current of spins $(3, \frac{7}{2}, \frac{7}{2}, 4)$}
The following four OPEs between the spin-$\frac{3}{2}$ currents and the
spin-$3$ current $W^{(3)}(w)$ (\ref{w3}) can be obtained
\bea
\left(
\begin{array}{c}
\hat{G}_{11} \\
\hat{G}_{22} \\
\end{array} \right) (z) \, W^{(3)}(w)  & = &
\frac{1}{(z-w)^2} \frac{(26+5k)}{(6+k)}
\left(
\begin{array}{c}
U^{(\frac{5}{2})} \\
-V^{(\frac{5}{2})} \\
\end{array} \right)(w)
\nonu \\
& + &
\frac{1}{(z-w)} \frac{1}{(6+k)} \left[
 (4+k) \pa
\left(
\begin{array}{c}
U^{(\frac{5}{2})} \\
-V^{(\frac{5}{2})} \\
\end{array} \right)
\right. \nonu \\
& +& \left.
 2i\left(
\begin{array}{c}
 \hat{B}_{-}T_{-}^{(\frac{5}{2})}-T_{+}^{(\frac{5}{2})}\hat{A}_{+} \\
-\hat{A}_{-}T_{-}^{(\frac{5}{2})}+T_{+}^{(\frac{5}{2})}\hat{B}_{+} \\
\end{array} \right) \right](w)+\cdots,
\label{G11w3}
\\
\left(
\begin{array}{c}
\hat{G}_{12} \\
\hat{G}_{21} \\
\end{array} \right) (z) \, W^{(3)}(w)  & = &
\frac{1}{(z-w)^2} \frac{(-4+k)}{(6+k)}
\left(
\begin{array}{c}
T_{-}^{(\frac{5}{2})} \\
-T_{+}^{(\frac{5}{2})} \\
\end{array} \right)(w)
\label{G22w3}
\\
& + &
\frac{1}{(z-w)}  \left[
 \frac{(-4+k)}{5(6+k)}
\pa \left(
\begin{array}{c}
T_{-}^{(\frac{5}{2})} \\
-T_{+}^{(\frac{5}{2})} \\
\end{array} \right)
+\left(
\begin{array}{c}
W_{-}^{(\frac{7}{2})} \\
W_{+}^{(\frac{7}{2})} \\
\end{array} \right)
 \right](w)+\cdots.
\nonu
\eea
The $U(1)$ charge conservation in these OPEs (\ref{G11w3}) and (\ref{G22w3}) 
can be checked from Table
$2$ or $6$.
 The factor $(k-4)$ appears in (\ref{G22w3}). 

The OPEs between the spin-$\frac{3}{2}$ currents and the
spin-$\frac{7}{2}$  current $W_{+}^{(\frac{7}{2})}(w)$ (\ref{w7+}) 
are expressed as
\bea
\hat{G}_{11}(z) \, W_{+}^{(\frac{7}{2})}(w)
& = & \frac{1}{(z-w)^2} \left[
-\frac{18(7+2k)}{5(6+k)} U_{+}^{(3)}
+\frac{8i}{(6+k)}  \hat{B}_{-} \,  \hat{T}^{(2)}
\right](w)
\nonu \\
& + & \frac{1}{(z-w)}
\left[ \frac{1}{6}\pa \{\hat{G}_{11} \, W_{+}^{(\frac{7}{2})}\}_{-2}
-\frac{2i}{(6+k)} \hat{B}_{-} \, T^{(3)}
- \frac{10i}{3(6+k)} \hat{B}_{-}   \pa T^{(2)}
\right. \nonu \\
& + & \left.
\frac{20i}{3(6+k)}T^{(2)} \pa \hat{B}_{-}
+\frac{2}{(6+k)} T_{+}^{(\frac{5}{2})} \hat{G}_{11}
-\frac{4i}{(6+k)} U_{+}^{(3)}\hat{A}_{3}
\right. \nonu \\
&+& \left.
\frac{1}{(6+k)} \pa U_{+}^{(3)}\right](w)+\cdots,
\label{g11w+7half}
\\
\hat{G}_{22}(z) \, W_{+}^{(\frac{7}{2})}(w)
& = & \frac{1}{(z-w)^2} \left[
\frac{6(29+4k)}{5(6+k)} V_{+}^{(3)}
-\frac{8i}{(6+k)}  \hat{A}_{-} \,  \hat{T}^{(2)}
\right](w)
\nonu \\
& + & \frac{1}{(z-w)}
\left[ \frac{1}{6}\pa \{\hat{G}_{22} \, W_{+}^{(\frac{7}{2})}\}_{-2}
+\frac{2i}{(6+k)} \hat{A}_{-} \, T^{(3)}
+\frac{10i}{3(6+k)} \hat{A}_{-}   \pa T^{(2)}
\right. \nonu \\
& -& \left.
\frac{20i}{3(6+k)} T^{(2)} \pa \hat{A}_{-}
-\frac{2}{(6+k)} T_{+}^{(\frac{5}{2})} \hat{G}_{22}
+\frac{4i}{(6+k)} V_{+}^{(3)}\hat{B}_{3}
\right. \nonu \\
&-& \left.
\frac{1}{(6+k)} \pa V_{+}^{(3)}\right](w)+\cdots,
\label{g22w+7half}
\\
\hat{G}_{12}(z) \, W_{+}^{(\frac{7}{2})}(w)
& = & \frac{1}{(z-w)^3}
\left[ \frac{48(-4+k)}{5(6+k)} T^{(2)} \right](w)
\nonu \\
& + & \frac{1}{(z-w)^2} \left[
-\frac{6(-4+k)}{5(6+k)} T^{(3)}
+\frac{(32+6k)}{(6+k)}  W^{(3)}
+\frac{16i}{(6+k)} (\hat{A}_{3}-\hat{B}_{3})T^{(2)}
\right](w)
\nonu \\
& + & \frac{1}{(z-w)}
\left[ \frac{1}{6}\pa \{\hat{G}_{12} \, W_{+}^{(\frac{7}{2})}\}_{-2}
-W^{(4)}
-\frac{16i}{3(6+k)} (\hat{A}_{3}-\hat{B}_3) \pa T^{(2)}
\right. \nonu \\
& +& \left.
\frac{2i}{(6+k)} \hat{A}_{-}  U_{-}^{(3)}
-\frac{2i}{(6+k)} \hat{B}_{-} V_{-}^{(3)}
+\frac{32i}{3(6+k)} T^{(2)}\pa (\hat{A}_{3}-\hat{B}_{3})
\right. \nonu \\
&+& \left.
\frac{144(-4+k)}{(324+179k)} \hat{T} T^{(2)}
-\frac{2i}{(6+k)} U_{+}^{(3)}\hat{B}_{+}
+\frac{2i}{(6+k)} V_{+}^{(3)}\hat{A}_{+}
\right. \nonu \\
&-& \left.
\frac{8}{3(6+k)} \pa W^{(3)}-\frac{216(-4+k)}{(1620+895k)} \pa^2 T^{(2)}\right](w)+\cdots,
\label{g12w+7half}
\\
\hat{G}_{21}(z) \, W_{+}^{(\frac{7}{2})}(w)
& = &
\frac{1}{(z-w)} \left[
\frac{2i}{(6+k)}\hat{B}_{-} V_{+}^{(3)}
-\frac{2i}{(6+k)}  U_{+}^{(3)}\hat{A}_{-}
\right](w)+\cdots.
\label{g21w+7half}
\eea
The $U(1)$ charge conservation in these OPEs can be checked from Table
$2$, $7$ or $8$. 
The nonlinear terms appear in (\ref{g11w+7half}), (\ref{g22w+7half}), 
(\ref{g12w+7half}) and (\ref{g21w+7half}).
Each term in the second order pole of (\ref{g11w+7half}), (\ref{g22w+7half}), 
or (\ref{g12w+7half}) is a primary field.
There is no descendant field for the spin-$2$ field $T^{(2)}(w)$ 
in the second order pole of (\ref{g12w+7half}).

The OPEs between the spin-$\frac{3}{2}$ currents and the
spin-$\frac{7}{2}$  current $W_{-}^{(\frac{7}{2})}(w)$ (\ref{w7-}) are given by
\bea
\hat{G}_{11}(z) \, W_{-}^{(\frac{7}{2})}(w)
& = & \frac{1}{(z-w)^2} \left[
-\frac{6(29+4k)}{5(6+k)} U_{-}^{(3)}
-\frac{8i}{(6+k)}  \hat{T}^{(2)}\hat{A}_{+}
\right](w)
\nonu \\
& + & \frac{1}{(z-w)}
\left[ \frac{1}{6}\pa \{\hat{G}_{11} \, W_{-}^{(\frac{7}{2})}\}_{-2}
+\frac{10i}{3(6+k)} \hat{A}_{+}   \pa T^{(2)}
+\frac{4i}{(6+k)} \hat{B}_{3} \, U_{-}^{(3)}
\right. \nonu \\
& - & \left.
\frac{2}{(6+k)} \hat{G}_{11}T_{-}^{(\frac{5}{2})}
-\frac{20i}{3(6+k)}T^{(2)} \pa \hat{A}_{+}
-\frac{2i}{(6+k)} T^{(3)}\hat{A}_{+}
\right. \nonu \\
&+& \left.
\frac{1}{(6+k)} \pa U_{-}^{(3)}\right](w)+\cdots,
\label{g11w-7half}
\\
\hat{G}_{22}(z) \, W_{-}^{(\frac{7}{2})}(w)
& = & \frac{1}{(z-w)^2} \left[
\frac{18(7+2k)}{5(6+k)} V_{-}^{(3)}
+\frac{8i}{(6+k)}  \hat{T}^{(2)} \hat{B}_{+}
\right](w)
\nonu \\
& + & \frac{1}{(z-w)}
\left[ \frac{1}{6}\pa \{\hat{G}_{22} \, W_{-}^{(\frac{7}{2})}\}_{-2}
-\frac{4i}{(6+k)} \hat{A}_{3} \, V_{-}^{(3)}
-\frac{10i}{3(6+k)} \hat{B}_{+}   \pa T^{(2)}
\right. \nonu \\
& +& \left.
\frac{2}{(6+k)}\hat{G}_{22} T_{-}^{(\frac{5}{2})}
+\frac{20i}{3(6+k)} T^{(2)} \pa \hat{B}_{+}
+\frac{2i}{(6+k)} T^{(3)}\hat{B}_{+}
\right. \nonu \\
&-& \left.
\frac{1}{(6+k)} \pa V_{-}^{(3)}\right](w)+\cdots,
\label{g22w-7half}
\\
\hat{G}_{12}(z) \, W_{-}^{(\frac{7}{2})}(w)
& = &
\frac{1}{(z-w)} \left[
-\frac{2i}{(6+k)}\hat{A}_{+} V_{-}^{(3)}
+\frac{2i}{(6+k)}  U_{-}^{(3)}\hat{B}_{+}
\right](w)+\cdots,
\label{g12w-7half}
\\
\hat{G}_{21}(z) \, W_{-}^{(\frac{7}{2})}(w)
& = & -\frac{1}{(z-w)^3}
\left[\frac{48(-4+k)}{5(6+k)} T^{(2)} \right](w)
\nonu \\
& + & \frac{1}{(z-w)^2} \left[
-\frac{6(-4+k)}{5(6+k)} T^{(3)}
+\frac{(32+6k)}{(6+k)}  W^{(3)}
+\frac{16i}{(6+k)} (\hat{A}_{3}-\hat{B}_{3})T^{(2)}
\right](w)
\nonu \\
& + & \frac{1}{(z-w)}
\left[ \frac{1}{6}\pa \{\hat{G}_{21} \, W_{+}^{(\frac{7}{2})}\}_{-2}
+W^{(4)}
\right. \nonu \\
&-& \left.
\frac{144(-4+k)}{(324+179k)} \left( \hat{T} T^{(2)}
-\frac{3}{10} \pa^2 T^{(2)} \right) \right](w)+\cdots.
\label{g21w-7half}
\eea
The $U(1)$ charge conservation in these OPEs (\ref{g11w-7half}), 
(\ref{g22w-7half}), 
(\ref{g12w-7half}) and (\ref{g21w-7half})
can be checked from Table
$2$, $7$ or $8$.
 The factor $(k-4)$ appears in (\ref{g21w-7half}). 
Each term in the second order pole of (\ref{g11w-7half}), (\ref{g22w-7half}), 
or (\ref{g21w-7half}) is a primary field.
There is no descendant field for the spin-$2$ field $T^{(2)}(w)$ 
in the second order pole of (\ref{g21w-7half}).

The OPEs between the spin-$\frac{3}{2}$ currents and the
spin-$4$  current $W^{(4)}(w)$ (\ref{w4}) is
\bea
\hat{G}_{11}(z) \, W^{(4)}(w)  & = &
\frac{1}{(z-w)^3} \left[
\frac{4(-87804+k(-32897+9k(549+166k)))}{3(6+k)^2(324+179k)}
U^{(\frac{5}{2})} \right](w)
\nonu \\
& + &
\frac{1}{(z-w)^2} \left[
\frac{(36+7k)}{(6+k)} U^{(\frac{7}{2})}
+ \frac{4i(34+3k)}{3(6+k)^2} \hat{B}_{-} T_{-}^{(\frac{5}{2})}
\right.
\nonu \\
&+ & \left.
\frac{8(-2592-196k+81k^2)}{3(6+k)(324+179k)} \hat{G}_{11}T^{(2)}
+ \frac{2i(88+15k)}{3(6+k)^2}  T_{+}^{(\frac{5}{2})}\hat{A}_{+}
\right.
\nonu \\
&+ & \left.
\frac{4i(72+19k)}{3(6+k)^2} U^{(\frac{5}{2})}\hat{A}_{3}
- \frac{8i(-3+2k)}{3(6+k)^2} U^{(\frac{5}{2})}\hat{B}_{3}
\right.
\nonu \\
&+ & \left.
\frac{4(57672+33962k+4327k^2-108k^3)}{5(6+k)^2(324+179k)} \pa U^{(\frac{5}{2})}
\right](w)+\cdots,
\nonu \\
& + &
\frac{1}{(z-w)}  \left[
\frac{1}{7}\{\hat{G}_{11}\, W^{(4)}\}_{-2}
-\frac{8i(53+24k)}{35(6+k)^2}\hat{A}_{3} \pa U^{(\frac{5}{2})}
-\frac{4i(31+6k)}{7(6+k)^2}\hat{A}_{+} \pa T_{+}^{(\frac{5}{2})}
\right.
\nonu \\
&-& \frac{8i(47+6k)}{35(6+k)^2} \hat{B}_{3}\pa U^{(\frac{5}{2})}
- \frac{2i}{(6+k)} \hat{B}_{-} W_{-}^{(\frac{7}{2})}
+\frac{8}{(6+k)^2}\hat{B}_{-}\hat{A}_{3} T_{-}^{(\frac{5}{2})}
\nonu \\
&-&
\frac{4}{(6+k)^2}\hat{B}_{-}\hat{A}_{+} V^{(\frac{5}{2})}
- \frac{8i(68+13k)}{35(6+k)^2} \hat{B}_{-} \pa T_{-}^{(\frac{5}{2})}
- \frac{4}{(6+k)} \hat{G}_{11}T^{(3)}
\nonu \\
& -& \frac{24(-540-161k+9k^2)}{7(6+k)(324+179k)}\hat{G}_{11}\pa T^{(2)}
-\frac{2}{(6+k)}  \hat{G}_{21} U_{-}^{(3)}
\nonu \\
& +& \frac{32(-540-161k+9k^2)}{7(6+k)(324+179k)}T^{(2)} \pa\hat{G}_{11}
+\frac{4i(61+13k)}{7(6+k)^2}  T_{-}^{(\frac{5}{2})} \pa \hat{B}_{-}
\nonu \\
& +& \frac{4i(88+15k)}{7(6+k)^2}T_{+}^{(\frac{5}{2})} \pa\hat{A}_{+}
+\frac{144(-4+k)}{(324+179k)}  U^{(\frac{5}{2})} \hat{T}
-\frac{16}{(6+k)^2}  U^{(\frac{5}{2})} \hat{A}_{3}\hat{A}_{3}
 \nonu \\
& + &
\frac{16}{(6+k)^2}U^{(\frac{5}{2})} \hat{A}_{3}\hat{B}_{3}
-\frac{4}{(6+k)^2}U^{(\frac{5}{2})} \hat{A}_{-}\hat{A}_{+}
+\frac{8i(79+12k)}{7(6+k)^2}U^{(\frac{5}{2})}\pa \hat{A}_{3}
\nonu \\
& + &  \frac{8i (13+3k)}{7(6+k)^2} U^{(\frac{5}{2})}\pa \hat{B}_{3}
+\frac{4i}{(6+k)} U^{(\frac{7}{2})}( \hat{A}_3 -\hat{B}_3)
+\frac{34}{7(6+k)}\pa U^{(\frac{7}{2})}
\nonu \\
&+& \left.
\frac{(239760+90124k+5112k^2-972k^3)}{21(6+k)^2(324+179k)} \pa^2 U^{(\frac{5}{2})}
\right](w) +\cdots,
\label{g11w4}
\\
\hat{G}_{22}(z) \, W^{(4)}(w)  & = &
\frac{1}{(z-w)^3} \left[
\frac{4(-71604+k(-21031+18k(364+83k)))}{3(6+k)^2(324+179k)}
V^{(\frac{5}{2})} \right](w)
\nonu \\
& + &
\frac{1}{(z-w)^2} \left[
-\frac{(36+7k)}{(6+k)} V^{(\frac{7}{2})}
- \frac{8i(-9+k)}{3(6+k)^2} \hat{A}_{3} V^{(\frac{5}{2})}
\right.
\nonu \\
&- & \left.
 \frac{8i(11+3k)}{3(6+k)^2} \hat{A}_{-} T_{-}^{(\frac{5}{2})}
 -\frac{4i(96+13k)}{3(6+k)^2} \hat{B}_{3} V^{(\frac{5}{2})}
\right.
\nonu \\
&+& \left.
\frac{8(-1296+520k+81k^2)}{3(6+k)(324+179)}  T^{(2)}\hat{G}_{22}
-\frac{2i(88+15k)}{3(6+k)^2} T_{+}^{(\frac{5}{2})}\hat{B}_{+}
\right.
\nonu \\
&+ & \left.
\frac{2(-184680-48354k+1153k^2+972k^3)}{15(6+k)^2(324+179k)} \pa V^{(\frac{5}{2})}
\right](w)
\nonu \\
& + &
\frac{1}{(z-w)}  \left[
\frac{1}{7}\{\hat{G}_{22}\, W^{(4)}\}_{-2}
+\frac{4i}{(6+k)}\hat{A}_{3} V^{(\frac{7}{2})}
-\frac{16}{(6+k)^2}\hat{A}_{3} \hat{B}_{3}V^{(\frac{5}{2})}
\right.
\nonu \\
&+&
\frac{32i(5+k)}{35(6+k)^2}\hat{A}_{3}\pa V^{(\frac{5}{2})}
- \frac{2i}{(6+k)} \hat{A}_{-} W_{-}^{(\frac{7}{2})}
-\frac{8}{(6+k)^2}\hat{A}_{-} \hat{B}_{3}T_{-}^{(\frac{5}{2})}
\nonu \\
& +& \frac{4i(109+24k)}{35(6+k)^2}\hat{A}_{-}\pa T_{-}^{(\frac{5}{2})}
-\frac{4i}{(6+k)}  \hat{B}_{3} V^{(\frac{7}{2})}
+\frac{16}{(6+k)^2}\hat{B}_{3} \hat{B}_{3}V^{(\frac{5}{2})}
\nonu \\
& +& \frac{16i(110+13k)}{35(6+k)^2}\hat{B}_{3}\pa V^{(\frac{5}{2})}
+\frac{4}{(6+k)^2}\hat{B}_{-}V^{(\frac{5}{2})} \hat{B}_{+}
+\frac{4i(31+6k)}{7(6+k)^2}\hat{B}_{+}\pa T_{+}^{(\frac{5}{2})}
\nonu \\
& +& \frac{2}{(6+k)}\hat{G}_{21} V_{-}^{(3)}
-\frac{24(108+197k+9k^2)}{7(6+k)(324+179k)}\hat{G}_{22} \pa T^{(2)}
 \nonu \\
& + &
\frac{4}{(6+k)}  T^{(3)} \hat{G}_{22}
+\frac{32(108+197k+9k^2)}{7(6+k)(324+179k)}T^{(2)} \pa \hat{G}_{22}
-\frac{4i(79+12k)}{7(6+k)^2}T_{-}^{(\frac{5}{2})} \pa \hat{A}_{-}
\nonu \\
& - &  \frac{4i (88+15k)}{7(6+k)^2} T_{+}^{(\frac{5}{2})}\pa \hat{B}_{+}
+\frac{144(-4+k)}{(324+179k)} \hat{T} V^{(\frac{5}{2})}
+\frac{4}{(6+k)^2}U^{(\frac{5}{2})}\hat{A}_{-} \hat{B}_{+}
\nonu \\
& - &  \frac{8i (17+2k)}{7(6+k)^2} V^{(\frac{5}{2})}\pa \hat{A}_{3}
-\frac{8i (61+13k)}{7(6+k)^2} V^{(\frac{5}{2})}\pa \hat{B}_{3}
+\frac{22}{7(6+k)}\pa V^{(\frac{7}{2})}
\nonu \\
&-& \left.
\frac{4(27540+36371k+3771k^2+270k^3)}{105(6+k)^2(324+179k)} \pa^2 V^{(\frac{5}{2})}
\right](w) +\cdots,
\label{g22w4}
\\
\hat{G}_{12}(z) \, W^{(4)}(w)  & = &
\frac{1}{(z-w)^3} \left[
\frac{8(-41472-13972k+3097k^2+747k^3)}{3(6+k)^2(324+179k)}
T_{-}^{(\frac{5}{2})} \right](w)
\nonu \\
& + &
\frac{1}{(z-w)^2} \left[
\frac{(116+21k)}{3(6+k)} W_{-}^{(\frac{7}{2})}
+\frac{8i(27+5k)}{3(6+k)^2} \hat{A}_{3} T_{-}^{(\frac{5}{2})}
- \frac{2i(28+5k)}{(6+k)^2} \hat{A}_{+}V^{(\frac{5}{2})}
\right.
\nonu \\
&- & \left.
 \frac{8i(27+5k)}{3(6+k)^2} \hat{B}_{3}T_{-}^{(\frac{5}{2})}
+ \frac{216(-4+k)}{(324+179k)} T^{(2)}\hat{G}_{12}
+\frac{2i(28+5k)}{(6+k)^2} U^{(\frac{5}{2})}\hat{B}_{+}
\right.
\nonu \\
&+ & \left.
\frac{2(-92016-37048k-1883k^2+324k^3)}{5(6+k)^2(324+179k)} \pa T_{-}^{(\frac{5}{2})}
\right](w)
\nonu \\
& + &
\frac{1}{(z-w)}  \left[
\frac{1}{7}\{\hat{G}_{12}\, W^{(4)}\}_{-2}
-\frac{8}{(6+k)^2}\hat{A}_{3}\hat{A}_{+} V^{(\frac{5}{2})}
-\frac{8i(23+4k)}{7(6+k)^2}\hat{A}_{3} \pa T_{-}^{(\frac{5}{2})}
\right.
\nonu \\
&-&
\frac{4}{(6+k)^2}\hat{A}_{-}\hat{A}_{+} T_{-}^{(\frac{5}{2})}
+\frac{4i(35+6k)}{7(6+k)^2} \hat{A}_{+} \pa V^{(\frac{5}{2})}
+\frac{8}{(6+k)^2}\hat{B}_{3} \hat{A}_{+}V^{(\frac{5}{2})}
\nonu \\
& +& \frac{16i(15+2k)}{7(6+k)^2}\hat{B}_{3}\pa T_{-}^{(\frac{5}{2})}
+\frac{4}{(6+k)^2}  \hat{B}_{-} T_{-}^{(\frac{5}{2})} \hat{B}_{+}
-\frac{8i(14+3k)}{7(6+k)^2}\hat{B}_{+} \pa U^{(\frac{5}{2})}
\nonu \\
& -& \frac{2}{(6+k)}\hat{G}_{11}V_{-}^{(3)}
-\frac{216(-4+k)}{7(324+179k)}\hat{G}_{12} \pa T^{(2)}
+\frac{288(-4+k)}{7(324+179k)}T^{(2)}\pa \hat{G}_{12}
\nonu \\
& +& \frac{16i(34+5k)}{7(6+k)^2}T_{-}^{(\frac{5}{2})}\pa \hat{A}_{3}
-\frac{8i(61+10k)}{7(6+k)^2}T_{-}^{(\frac{5}{2})} \pa \hat{B}_{3}
+\frac{144(-4+k)}{(324+179k)}\hat{T} T_{-}^{(\frac{5}{2})}
 \nonu \\
& + &
\frac{2}{(6+k)}  U_{-}^{(3)} \hat{G}_{22}
-\frac{8}{(6+k)^2}U^{(\frac{5}{2})} (\hat{A}_{3}-\hat{B}_{3}) \hat{B}_{+}
+\frac{12i(28+5k)}{7(6+k)^2}U^{(\frac{5}{2})} \pa \hat{B}_{+}
\nonu \\
& - &  \frac{4i (91+15k)}{7(6+k)^2} V^{(\frac{5}{2})}\pa \hat{A}_{+}
-\frac{6}{7(6+k)}\pa  W_{-}^{(\frac{7}{2})}
\nonu \\
&+& \left.
\frac{(804168+498326k+61528k^2-1080k^3)}{105(6+k)^2(324+179k)} \pa^2 T_{-}^{(\frac{5}{2})}
\right](w) +\cdots,
\label{g12w4}
\\
\hat{G}_{21}(z) \, W^{(4)}(w)  & = &
\frac{1}{(z-w)^3} \left[
\frac{4(-84240-28336k+6373k^2+1494k^3)}{3(6+k)^2(324+179k)}
T_{+}^{(\frac{5}{2})} \right](w)
\nonu \\
& + &
\frac{1}{(z-w)^2} \left[
-\frac{(124+21k)}{3(6+k)} W_{+}^{(\frac{7}{2})}
-\frac{6i(20+3k)}{(6+k)^2} \hat{B}_{-} V^{(\frac{5}{2})}
\right.
\nonu \\
&+ & \left.
 \frac{216(-4+k)}{(324+179k)} \hat{G}_{21}T^{(2)}
 -\frac{8i(36+7k)}{3(6+k)^2} T_{+}^{(\frac{5}{2})}(\hat{A}_{3}-\hat{B}_{3})
+ \frac{2i(52+11k)}{(6+k)^2} U^{(\frac{5}{2})}\hat{A}_{-}
\right.
\nonu \\
&+ & \left.
\frac{(235872+135472k+15518k^2-432k^3)}{5(6+k)^2(324+179k)} \pa T_{+}^{(\frac{5}{2})}
\right](w)
\nonu \\
& + &
\frac{1}{(z-w)}  \left[
\frac{1}{7}\{\hat{G}_{21}\, W^{(4)}\}_{-2}
+\frac{40i}{7(6+k)^2}\hat{A}_{3} \pa T_{+}^{(\frac{5}{2})}
-\frac{12i(41+8k)}{35(6+k)^2}\hat{A}_{-} \pa U^{(\frac{5}{2})}
\right.
\nonu \\
&-&
\frac{40i}{7(6+k)^2}\hat{B}_{3}\pa T_{+}^{(\frac{5}{2})}
-\frac{2i}{(6+k)} \hat{B}_{-} V^{(\frac{7}{2})}
-\frac{8}{(6+k)^2}\hat{B}_{-} (\hat{A}_{3}-\hat{B}_{3})V^{(\frac{5}{2})}
\nonu \\
& +&
\frac{4i(115+26k)}{35(6+k)^2}  \hat{B}_{-} \pa V^{(\frac{5}{2})}
-\frac{216(-4+k)}{7(324+179k)}\hat{G}_{21} \pa T^{(2)}
\nonu \\
& +& \frac{288(-4+k)}{7(324+179k)} T^{(2)}\pa \hat{G}_{21}
+\frac{144(-4+k)}{(324+179k)}T_{+}^{(\frac{5}{2})}\hat{T}
-\frac{16}{7(6+k)^2}T_{+}^{(\frac{5}{2})}\pa \hat{A}_{3}
\nonu \\
& +& \frac{16i}{7(6+k)^2}T_{+}^{(\frac{5}{2})}\pa \hat{B}_{3}
-\frac{8i}{(6+k)^2}U^{(\frac{5}{2})}\hat{A}_{-} (\hat{A}_{3}-\hat{B}_{3})
+\frac{8i(29+6k)}{7(6+k)^2 }U^{(\frac{5}{2})} \pa \hat{A}_{-}
 \nonu \\
& + &
\frac{2i}{(6+k)}  U^{(\frac{7}{2})} \hat{A}_{-}
-\frac{4i(68+13k)}{7(6+k)^2}V^{(\frac{5}{2})} \pa \hat{B}_{-}
-\frac{10}{7(6+k)}\pa W_{+}^{(\frac{7}{2})}
\nonu \\
&+& \left.
\frac{(1625832+736438k+53126k^2-4860k^3)}{105(6+k)^2(324+179k)} \pa^2 T_{+}^{(\frac{5}{2})}
\right](w) +\cdots.
\label{g21w4}
\eea
One can check the $U(1)$ charges from Table $2$, $9$ or $10$ 
and the nonlinear terms (linear in higher spin currents) appear in 
(\ref{g11w4}), (\ref{g22w4}), (\ref{g12w4}) and (\ref{g21w4}).
There are no descendant fields for the spin-$\frac{5}{2}$ fields  
in the second  order pole of (\ref{g11w4}), (\ref{g22w4}), (\ref{g12w4}) 
or (\ref{g21w4}).
One can analyze each second-order pole in these OPEs 
whether it is written in terms of
(quasi)primary fields. 

Therefore, one obtains 
 the OPEs between six spin-$1$ currents and the higher spin
current of spins $(3, \frac{7}{2}, \frac{7}{2}, 4)$ explicitly.
The 
higher spin currents corresponding to three ${\cal N}=2$ multiplets
(\ref{new16comp})
occur in the right hand sides of the OPEs.

As described in the 
previous Appendix, one can analyze the above OPEs similarly.
The first result is as follows:
\bea
\hat{G}_{11} & \times &
\left(
\begin{array}{cccc}
T^{(2)}, & T_{+}^{(\frac{5}{2})}, & T_{-}^{(\frac{5}{2})}, & T^{(3)}  \\
U^{(\frac{5}{2})}, & U_{+}^{(3)}, & U_{-}^{(3)}, & U^{(\frac{7}{2})}  \\
V^{(\frac{5}{2})}, & V^{(3)}_{+}, & V^{(3)}_{-}, & V^{(\frac{7}{2})}  \\
W^{(3)}, & W_{+}^{(\frac{7}{2})}, & W_{-}^{(\frac{7}{2})}, & W^{(4)}
\end{array} \right)
\nonu \\
& \rightarrow &
\left(
\begin{array}{cccc}
U^{(\frac{5}{2})}, & U_{+}^{(3)}, & U_{-}^{(3)}, & U^{(\frac{5}{2})},U^{(\frac{7}{2})}  \\
0,  & 0, & 0,  & 0  \\
T^{(2)}, W^{(3)}, & T_{+}^{(\frac{5}{2})},  W_{+}^{(\frac{7}{2})},
& T_{-}^{(\frac{5}{2})}, W_{-}^{(\frac{7}{2})}, & T^{(2)},
T^{(3)},  W^{(3)},  W^{(4)}    \\
U^{(\frac{5}{2})}, & U_{+}^{(3)}, & U_{-}^{(3)}, & U^{(\frac{5}{2})},U^{(\frac{7}{2})}
\end{array} \right).
\label{operesult}
\eea
There are no higher spin currents, denoted by zeros in (\ref{operesult})
and the four currents living in the second multiplet of (\ref{new16comp}) 
appear in the first and last rows 
in (\ref{operesult}).  
Also the current $U^{(\frac{5}{2})}(w)$ appears in the last columns of these rows.
The first and last multiplets of (\ref{new16comp}) appear in the 
third row in (\ref{operesult}).
The currents $T^{(2)}(w)$ and $W^{(3)}(w)$ can appear in the last 
column of this row.

Similarly, it turns out that the following result for the
spin-$\frac{3}{2}$ current and $16$ currents is given by
\bea
\hat{G}_{12} & \times &
\left(
\begin{array}{cccc}
T^{(2)}, & T_{+}^{(\frac{5}{2})}, & T_{-}^{(\frac{5}{2})}, & T^{(3)}  \\
U^{(\frac{5}{2})}, & U_{+}^{(3)}, & U_{-}^{(3)}, & U^{(\frac{7}{2})}  \\
V^{(\frac{5}{2})}, & V^{(3)}_{+}, & V^{(3)}_{-}, & V^{(\frac{7}{2})}  \\
W^{(3)}, & W_{+}^{(\frac{7}{2})}, & W_{-}^{(\frac{7}{2})}, & W^{(4)}
\end{array} \right)
\nonu \\
& \rightarrow&
\left(
\begin{array}{cccc}
T_{-}^{(\frac{5}{2})}, & T^{(2)}, T^{(3)}, & 0, & T_{-}^{(\frac{5}{2})}   \\
U_{-}^{(3)}, & U^{(\frac{5}{2})},U^{(\frac{7}{2})}, & 0, & U_{-}^{(3)}  \\
V_{-}^{(3)}, & V^{(\frac{5}{2})}, V^{(\frac{7}{2})}, & 0, & V_{-}^{(3)}  \\
T_{-}^{(\frac{5}{2})}, W_{-}^{(\frac{7}{2})},  & T^{(2)}, T^{(3)}, W^{(3)}, W^{(4)},
& 0, & T_{-}^{(\frac{5}{2})}, W_{-}^{(\frac{7}{2})}
\end{array} \right).
\label{operesult1}
\eea
The four currents living in the third component of each  
multiplet of (\ref{new16comp}) appear in the first and the last columns 
in (\ref{operesult1}).  
Also the current $T_{-}^{(\frac{5}{2})}(w)$ 
appears in the last rows of these columns.
The first and last components in each multiplet 
of (\ref{new16comp}) appear in the 
second column in (\ref{operesult1}).
The currents $T^{(2)}(w)$ and $W^{(3)}(w)$ can appear in the last 
row of this column.

One has the following result
\bea
\hat{G}_{21} & \times &
\left(
\begin{array}{cccc}
T^{(2)}, & T_{+}^{(\frac{5}{2})}, & T_{-}^{(\frac{5}{2})}, & T^{(3)}  \\
U^{(\frac{5}{2})}, & U_{+}^{(3)}, & U_{-}^{(3)}, & U^{(\frac{7}{2})}  \\
V^{(\frac{5}{2})}, & V^{(3)}_{+}, & V^{(3)}_{-}, & V^{(\frac{7}{2})}  \\
W^{(3)}, & W_{+}^{(\frac{7}{2})}, & W_{-}^{(\frac{7}{2})}, & W^{(4)}
\end{array} \right)
\nonu \\
& \rightarrow&
\left(
\begin{array}{cccc}
T_{+}^{(\frac{5}{2})}, & 0, & T^{(2)}, T^{(3)},  & T_{+}^{(\frac{5}{2})}   \\
U_{+}^{(3)}, & 0, & U^{(\frac{5}{2})},U^{(\frac{7}{2})}, & U_{+}^{(3)}  \\
V_{+}^{(3)}, & 0,  & V^{(\frac{5}{2})}, V^{(\frac{7}{2})},  & V_{+}^{(3)}  \\
T_{+}^{(\frac{5}{2})}, W_{+}^{(\frac{7}{2})},  & 0, & T^{(2)}, T^{(3)}, W^{(3)}, W^{(4)},
 & T_{+}^{(\frac{5}{2})}, W_{+}^{(\frac{7}{2})}
\end{array} \right).
\label{operesult2}
\eea
The four currents living in the second component of each  
multiplet of (\ref{new16comp}) appear in the first and the last columns 
in (\ref{operesult2}).  
Also the current $T_{+}^{(\frac{5}{2})}(w)$ 
appears in the last rows of these columns.
The first and last components in each multiplet 
of (\ref{new16comp}) appear in the 
third column in (\ref{operesult2}).
The currents $T^{(2)}(w)$ and $W^{(3)}(w)$ can appear in the last 
row of this column.

Finally, one obtains the following result
\bea
\hat{G}_{22} & \times &
\left(
\begin{array}{cccc}
T^{(2)}, & T_{+}^{(\frac{5}{2})}, & T_{-}^{(\frac{5}{2})}, & T^{(3)}  \\
U^{(\frac{5}{2})}, & U_{+}^{(3)}, & U_{-}^{(3)}, & U^{(\frac{7}{2})}  \\
V^{(\frac{5}{2})}, & V^{(3)}_{+}, & V^{(3)}_{-}, & V^{(\frac{7}{2})}  \\
W^{(3)}, & W_{+}^{(\frac{7}{2})}, & W_{-}^{(\frac{7}{2})}, & W^{(4)}
\end{array} \right)
\nonu \\
& \rightarrow &
\left(
\begin{array}{cccc}
V^{(\frac{5}{2})}, & V_{+}^{(3)}, & V_{-}^{(3)}, & V^{(\frac{5}{2})},V^{(\frac{7}{2})}  \\
T^{(2)}, W^{(3)}, & T_{+}^{(\frac{5}{2})},  W_{+}^{(\frac{7}{2})},
& T_{-}^{(\frac{5}{2})}, W_{-}^{(\frac{7}{2})}, & T^{(2)},T^{(3)},  W^{(3)},  W^{(4)}    \\
0,  & 0, & 0,  & 0  \\
V^{(\frac{5}{2})}, & V_{+}^{(3)}, & V_{-}^{(3)}, & V^{(\frac{5}{2})},V^{(\frac{7}{2})}
\end{array} \right).
\label{operesult3}
\eea
The four currents living in the third multiplet of (\ref{new16comp}) 
appear in the first and last rows in  
 (\ref{operesult3}).  
Also the current $V^{(\frac{5}{2})}(w)$ appears in the last columns of these rows.
The first and last multiplets of (\ref{new16comp}) appear in the 
second row in (\ref{operesult}).
The currents $T^{(2)}(w)$ and $W^{(3)}(w)$ can appear in the last 
column of this row.


\end{document}